\documentclass[%
preprintnumbers,
nofootinbib,
 amsmath,amssymb,
 aps, physrev,
 superscriptaddress
]{revtex4-2}

\usepackage{graphicx}
\usepackage{dcolumn}
\usepackage{bm}
\usepackage[hidelinks]{hyperref}

\usepackage{mleftright,mathtools,mycommands,mathrsfs}

\begin{document}

\preprint{NU-QG-22}

\title{\textbf{Odd-parity ringdown gravitational waves of a spherically symmetric 
black hole \\ with perfect fluid accretion} 
}%

\author{Rikuto Ohashi}\email{ohashi.rikuto.z3@s.mail.nagoya-u.ac.jp}
\affiliation{%
 Graduate School of Science, Nagoya University, Nagoya 464-8602, Japan}
\author{Yasutaka Koga}
\affiliation{%
Department of Physics, College of Humanities and Sciences, Nihon University, Tokyo 156-8550, Japan}

\author{Hiroyuki Nakano}

\affiliation{%
Faculty of Law, Ryukoku University, Kyoto 612-8577, Japan}

\author{Kota Ogasawara}

\affiliation{%
Department of Physics, School of Science and Technology, Meiji University, Kanagawa 214-8571, Japan}

\author{Chul-Moon Yoo}%
\affiliation{%
 Graduate School of Science, Nagoya University, Nagoya 464-8602, Japan}
\affiliation{Kobayashi-Maskawa Institute for the Origin of Particles and the Universe (KMI), Nagoya 464-8602, Japan}
 

\begin{abstract}
The ringdown waves from 
a black hole offer a clean probe of strong-field gravity, but a matter distribution that may be present around a realistic black hole renders the background spacetime dynamical and the ringdown frequencies time dependent. 
We study the odd-parity ringdown of a Schwarzschild black hole that grows through the dilute, steady, spherically symmetric accretion of a perfect fluid. 
Working to first order in the accretion rate, we compute the ringdown waveform directly in the time domain on this dynamical background. Since the odd-parity matter perturbation decouples from the metric perturbation, 
 the wave mode can be described by a purely tensorial mode on the accreting background.
In particular, the ratio of the imaginary to the real part of the frequency cancels both the secular variation caused by the growth of the black hole and the redshift factor, so that its deviation from the Schwarzschild value purely reflects the surrounding environment. The time dependence of the frequency, on the other hand, reflects the accretion rate and allows us to define a second observable tied to it. 
We argue that measuring these observables
across multiple modes may provide significant information to constrain the surrounding environment of the black hole.
\end{abstract}

\maketitle


\section{Introduction}\label{1.Intro}
The direct detections of gravitational waves (GWs) from compact binary mergers~\cite{LIGOGW150914, LIGOGW170817} have firmly established GW astronomy as a quantitative probe of the strong-field regime of general relativity (GR). The signal of a binary black hole (BH) merger consists of inspiral, merger, and ringdown phases, the last of which is well described by a superposition of damped sinusoids known as the quasinormal modes (QNMs) of the remnant BH~\cite{KokkotasSchmidt, BertiCardosoStarinets, KonoplyaZhidenko}. Within GR and under the assumption of vacuum, the QNM spectrum of the final state is uniquely determined by the mass and spin of a Kerr BH; the systematic measurement of multiple QNMs—often referred to as ``BH spectroscopy''—therefore offers a stringent test of the Kerr hypothesis~\cite{Dreyer, BertiSpec, MGtestII} (see, e.g., Refs.~\cite{LIGOScientific:2025wao, LIGOScientific:2026wpt} for tests of GR with the BH spectroscopy).\par

Astrophysical BHs, however, are generically immersed in matter: accretion flows and possibly dark-matter halos contribute to the local geometry and can shift the QNM spectrum at a level relevant for high-precision spectroscopy~\cite{Barausse, CardosoMaselli,ZhaoPani}. Once the background (BG) spacetime is nonstationary, as is typically the case for an accreting BH, the very notion of a frequency-domain QNM ceases to be sharply defined and the ringdown signal acquires an explicit time dependence that encodes the dynamics of the matter distribution~\cite{Xue:2003vs, Shao:2004ws, Abdalla:2006vb, Lin:2021fth, Vaidya, Capuano:2026tjy}. Characterizing this dependence is essential if BH spectroscopy is to be used as a quantitative probe of the matter distribution around astrophysical BHs.\par

The simplest analytically tractable model of a dynamical BH is the Vaidya spacetime~\cite{VaidyaOriginal}, an exact spherically symmetric solution sourced by ingoing or outgoing null dust. Linear perturbations of the Vaidya BH have been studied as both a toy model for dynamical ringdown and a benchmark for time-domain numerical schemes~\cite{Vaidya, VaidyaPL, Capuano:2026tjy}. While instructive, null dust is highly idealized. 
It provides no parametric handle on the equation of state (EoS) of the surrounding fluid. A more realistic description must allow for a generic EoS and for the source term based on fluid perturbations.
In this work, we analyze the ringdown waveform assuming that the accreting matter is a perfect fluid. This allows us to see how, even at the same accretion rate, the waveform exhibits different shifts depending on the fluid parameters.

In Ref.~\cite{BG}, a perturbative construction of the BG spacetime sourced by a steady, spherically symmetric perfect-fluid accretion onto a Schwarzschild BH was presented in Eddington-Finkelstein (EF) coordinates, working to first order in the accretion rate. In the present work we use this dilute-accretion BG to study, directly in the time domain, how the odd-parity ringdown waveform is modified by the surrounding fluid. Restricting to odd parity allows us to write the master equation in the Regge-Wheeler (RW) gauge in a particularly compact form~\cite{MasterGS, MasterGS2, RW}, and ensures that the metric and fluid perturbations decouple~\cite{GundlachMG}. 
Because the BG is itself dynamical, the master equation is integrated in double-null (DN) coordinates with $U$ and $V$ (and angular coordinates of $\theta$ and $\phi$) by means of the DN formalism (DNF) of Ref.~\cite{DNF1}, which provides second-order accuracy on a characteristic grid. To circumvent the well-known late-time instability of the DNF near the horizon~\cite{DNFproblem}, we adaptively redefine the $U$ coordinate at each step to keep the radial grid spacing uniform.\par

Our central observable is the dimensionless ratio $\Xi$, defined as the relative deviation of $\omI/\omR$ from its Schwarzschild value denoted by $\omI^{(\Sch)}/\omR^{(\Sch)}$, 
\eqn{\Xi\coloneq \left. \frac{\omI}{\omR}\middle/\frac{\omI^{(\Sch)}}{\omR^{(\Sch)}} \right.-1 \,,}
where $\omR$ and $\omI$ are the real and imaginary parts of the instantaneous frequency extracted from the waveform. By construction, $\Xi$ is insensitive to a uniform redshift and to the slow secular growth of the BH mass, isolating the genuine effect of the matter distribution. 
Complementarily, we introduce a time-domain estimator $\tilde{\scr{A}}$ for the accretion rate, 
defined from the secular drift of the frequency, and verify that it reproduces the input value 
 of the accretion rate $\scr{A}$ to first order in the dilute approximation. 
The parameters of the matter distribution are the accretion rate $\scr{A}$, the EoS parameter $w$ defined by the ratio of the pressure to the energy density $p/\rho$, and $\scr{F}$, which appears as an integration constant 
of the equation of motion (EoM). We will comprehensively investigate the dependence of $\Xi$ and $\tilde{\scr{A}}$ on these parameters, including the regime $\abs{w}>1$, in which the dominant energy condition is violated, since deviations from GR arising from modified gravity may behave as an effective energy-momentum tensor violating some energy condition
~\cite{MGtestI, Clifton}. 
Then, we examine what information about the BH environment can be extracted from the measurements of $\Xi$ and $\tilde{\scr{A}}$.\par

The paper is organized as follows. In Sec.~\ref{2.Master}, the master equation for odd-parity perturbations on a general spherically symmetric BG is reviewed in DN coordinates within the RW gauge. Section~\ref{3.BG} describes the dilute-accretion BG in EF coordinates and reduces it to the perfect-fluid hydrostatic problem governed by the EoS $p=w\rho$. Section~\ref{4.Source} derives the source term and confirms the decoupling of the metric and fluid perturbations. Section~\ref{5.Method} formulates the coupled differential equations applied to the DNF and defines the observables $\Xi$ and $\tilde{\scr{A}}$. The numerical results, including the dependence on $\scr{A}$, $w$, $\scr{F}$, $l$, and observer's location $r_\obs$ are presented in Sec.~\ref{6.Results}, and Sec.~\ref{7.Conclusion} summarizes our findings together with prospects for future work. Appendix~\ref{A.numerical} reviews the DNF and the treatment for the 
$U$ coordinate, and Appendix~\ref{NB.2param} collects numerical results not presented in the main text. Appendix~\ref{C.cutoff} attempts to introduce a cutoff into the matter distribution in order to make the BG spacetime asymptotically flat and obtain the waveform received by an observer at infinity (but we found that this involved practical difficulties). 
Throughout, we use geometric units with $G=c=1$ and the metric signature $(-,\,+,\,+,\,+)$.\par

\section{Master equation for general spherically symmetric background}\label{2.Master}
In this Section, we present the master equation for the odd-parity perturbations used in this work. 
In this paper, our main interest is in gravitational perturbations of a dynamical spacetime. 
Therefore, one approach that maintains generality is to use the gauge-invariant master equation derived in Refs.~\cite{MasterGS,MasterGS2}
[hereafter referred to as the Gerlach-Sengupta (GS) equation].
 However, since the zeroth-order BG will be set to the Schwarzschild spacetime in our setting, the standard RW formalism~\cite{RW} 
 would be more familiar.
We therefore derive the master equation based on the RW formalism and appropriately mention its consistency with the GS equation. 
 For performing dynamical simulations in the two-dimensional real space spanned by the time and radial directions, we adopt the DN coordinate system.\par

The general spherically symmetric line element in DN coordinates $(U,\,V,\,\theta,\,\phi)$ is given by
\eqs{2.1}{
 ds^2 &= g_{\mu\nu}^{\text{(BG)}} dx^{\mu} dx^{\nu} = -e^{\sigma (U,V)}dUdV +r(U,V)^2d\Omega^2 \,,\\
 d\Omega^2 &\coloneq d\theta^2 + \sin^2\theta d\phi^2 \,.
}
The BG spacetime is characterized by the two functions $r$ and $\sigma$. 
 This metric form is invariant under the coordinate transformation $U\rightarrow U'=U'(U)$ and $V\rightarrow V'=V'(V)$ with $\sigma\rightarrow \sigma'=\sigma'(U',V')$ defined by 
 \eq{2.2}{
 e^{\sigma'}=\deri{U}{U'}\deri{V}{V'}e^\sigma \,.
 }
Then, we consider the components $g_{\mu\nu}$ of a full metric $\mathbf{g}$ as
\eq{gh}{g_{\mu\nu} = g^{\text{(BG)}}_{\mu\nu}+ h_{\mu\nu} \,,}
where $h_{\mu\nu}$ is the components of a metric perturbation $\mathbf{h}$.
We expand $\mathbf{h}$ in terms of the odd-parity components of the tensor spherical harmonics $\tsh{i}$ as follows:
\eqa{2.3}{
\mathbf{h}=\sum_{l,m} \mathbf{h}^{\text{odd}}_{lm} 
&=\sum_{l,m}\Biggl[-\frac{\sqrt{2 l (l+1)}}{r(U,V)}H_{6 l m}(U,V)\tsh{6} \nn
&\qquad\quad +\frac{i\sqrt{2 l (l+1)}}{r(U,V)}H_{7 l m}(U,V)\tsh{7} \nn
&\qquad\quad +\frac{\sqrt{2l(l+1) (l-1) (l+2)}}{2 r(U,V)^2}H_{8 l m}(U,V)\tsh{8}\Biggr] \,.
}
The perturbation of the energy-momentum tensor $\bdelT$ is simply expanded as
\eqa{2.4}{
\bdelT&=\sum_{l,m} \bdelT^{\text{odd}}_{lm} =\sum_{l,m}\sum_{i=6,7,8}\delta T_{ilm}(U,V) \tsh{i} \,.
}
Here, the components of $\tsh{i}$ are defined in the DN coordinate basis as
\subeq{S2.1}{
\eq{32.2}{
\(\tsh{6}\)_{\mu\nu}\coloneq \frac{r(U,V)}{\sqrt{2l(l+1)}}\begin{pmatrix}
                                  0 & 0 & \dfrac{1}{\sin\theta}\dph\sh & -\sin\theta\,\dth\sh \\
                                  0 & 0 & 0 & 0 \\
                                  Sym & 0 & 0 & 0 \\
                                  Sym & 0 & 0 & 0 \end{pmatrix},
}
\eq{32.3}{
\left(\tsh{7}\right)_{\mu\nu}\coloneq \frac{ir(U,V)}{\sqrt{2l(l+1)}}\begin{pmatrix}
                                  0 & 0 & 0 & 0 \\
                                  0 & 0 & \dfrac{1}{\sin\theta}\dph\sh & -\sin\theta\,\dth\sh \\
                                  0 & Sym & 0 & 0 \\
                                  0 & Sym & 0 & 0 \end{pmatrix},
}
\eq{32.4}{
\(\tsh{8}\)_{\mu\nu}\coloneq -\frac{i r(U,V)^2}{{\sqrt{2 l (l+1) (l-1) (l+2)}}}\begin{pmatrix}
                                  0 & 0 & 0 & 0 \\
                                  0 & 0 & 0 & 0 \\
                                  0 & 0 & -\dfrac{1}{\sin\theta}X_{lm} & \sin\theta\, W_{lm} \\
                                  0 & 0 & Sym & \sin\theta\, X_{lm} \end{pmatrix},
}
\eq{22.13}{
X_{lm}\coloneq 2\dph\(\dth-\cot\theta\)\sh\,,
}
\eq{22.14}{
W_{lm}\coloneq \(\dth^2-\cot\theta\,\dth-\frac{1}{\sin^2\theta}\dph^2\)\sh \,,
}}
where $\sh$ denotes the standard scalar spherical harmonics and $Sym$ indicates symmetric components. 

Under a reparametrization of $U$ and $V$, $H_{i l m}$ transforms as
 \eqs{2.5}{
 H'_{6 l m}&=\deri{U}{U'}H_{6 l m} \nc
 H'_{7 l m}&=\deri{V}{V'}H_{7 l m} \nc
 H'_{8 l m}&=H_{8 l m} \,.
 }
The coefficients $\delta T_{ilm}$ transform in the same manner. 
For the angular coordinates, the odd-parity gauge transformation is given by
\eq{2.6}{
x^\mu\rightarrow x'^\mu=x^\mu-\frac{\sqrt{l(l+1)}}{r(U,V)} \Lambda_{lm}(U,V) \(\vsh{3}\)^{\mu} \,,
}
where $\Lambda_{lm}(U,V)$ is an arbitrary function and $\vsh{3}$ is the odd-parity component of the vector spherical harmonics 
defined by 
\eq{2.6.5}{
\(\vsh{3}\)_{\mu}\coloneq \frac{r(U,V)}{\sqrt{l(l+1)}} \tk{ 0,\,0,\,\frac{1}{\sin{\theta}} \partial_\phi Y_{lm},\,-\sin{\theta}\, \partial_\theta Y_{lm} } \,. 
}
Under this transformation, $H_{i l m}$ and $\delta T_{ilm}$ transform as
\eqs{2.7}{
H_{6lm} &\rightarrow H_{6lm} +2\frac{r\dU}{r}\Lambda_{lm} -\Lambda_{lm,U} \nc
H_{7lm} &\rightarrow H_{7lm} +2\frac{r\dV}{r}\Lambda_{lm} -\Lambda_{lm,V} \nc
H_{8lm} &\rightarrow H_{8lm} -2i\Lambda_{lm} \nc
}
\eqs{2.8}{
\delT_{6lm} &\rightarrow \delT_{6lm} -\frac{\sqrt{2l(l+1)}}{r^3} T^{(\text{BG})}_{22} \(2\frac{r\dU}{r}\Lambda_{lm} -\Lambda_{lm,U}\) \nc
\delT_{7lm} &\rightarrow \delT_{7lm} +i\frac{\sqrt{2l(l+1)}}{r^3} T^{(\text{BG})}_{22} \(2\frac{r\dV}{r}\Lambda_{lm} -\Lambda_{lm,V}\) \nc
\delT_{8lm} &\rightarrow \delT_{8lm} -i\frac{\sqrt{2l(l+1) (l-1) (l+2)}}{r^4} T^{(\text{BG})}_{22} \Lambda_{lm} \,.
}
Hereafter, by choosing
\eq{2.9}{
\Lambda_{lm}=-\frac{i}{2}H_{8lm} \,,
}
we adopt the RW gauge,
\eq{2.10}{
H_{8lm}=0 \,.
}
In this gauge, the Einstein equations linearized to first order in $\mathbf{h}$ take the following form (hereafter, we suppress the subscripts ``$l\,m$" and the superscripts ``$\odd$" when no confusion arises):
\subeq{S2.1.5}{
\eqa{2.11}{8\pi \,\delT_6=
& -\frac{\sqrt{2 l (l+1)}}{r}e^{-\sigma } \biggl[\frac{1}{2 r^2}\tk{(l-1) (l+2) e^{\sigma }-4r_{,U}r_{,V}-12rr_{,U V}+4rr_{,V}\sigma _{,U}-4r^2\sigma _{,U V}}H_6 \nn
&\qquad\qquad\qquad\qquad\; -\frac{2}{r^2}\(-r_{,U}^2-rr_{,U U}+rr_{,U}\sigma _{,U}\)H_7-\frac{1}{r}\(-2r_{,U}+r\sigma _{,U}\)H_{6,V} \nn
&\qquad\qquad\qquad\qquad\; -\frac{2r_{,V}}{r}H_{6,U}+\sigma _{,U}H_{7,U}+H_{6,U V}-H_{7,U U}\biggr] 
\,,
}\eqa{2.12}{8\pi \,\delT_7=
& i\frac{\sqrt{2 l (l+1)}}{r}e^{-\sigma } \biggl[\frac{1}{2 r^2}\tk{(l-1) (l+2) e^{\sigma }-4r_{,U}r_{,V}-12rr_{,U V}+4rr_{,U}\sigma _{,V}-4r^2\sigma _{,U V}}H_7 \nn
&\qquad\qquad\qquad\qquad\; -\frac{2}{r^2}\(-r_{,V}^2-rr_{,V V}+rr_{,V}\sigma _{,V}\)H_6-\frac{1}{r}\(-2r_{,V}+r\sigma _{,V}\)H_{7,U} \nn
&\qquad\qquad\qquad\qquad\; -\frac{2r_{,U}}{r}H_{7,V}+\sigma _{,V}H_{6,V}+H_{7,U V}-H_{6,V V}\biggr] 
\,,
}\eq{2.13}{8 \pi \,\delT_8 =
-i\frac{\sqrt{2 l(l+1) (l-1) (l+2)}}{r^2} e^{-\sigma }\(H_{6,V}+H_{7,U}\)
\,.
}}
\par
Next, we define the master variable $\psi_{lm}^{\odd}(U,V)$ by
\eq{2.14}{
\psi _{l m}^{\odd}(U,V) \coloneq-\frac{4 e^{-\sigma }}{(l-1) (l+2)}\tk{2\(r_{,V}H_6-r_{,U}H_7\)+r \(H_{7,U}-H_{6,V}\)} \,.
}
Using Eq.~\eqref{2.7}, one can verify that this quantity is gauge invariant. Furthermore, from Eqs.~\eqref{2.2} and \eqref{2.5}, it follows that $\psi$ is invariant under reparametrizations of the $U$ and $V$ coordinates. 
To derive the master equation, we first rewrite Eq.~\eqref{2.14} as
\eq{2.15}{
H_{7,U}=-\frac{(l-1)(l+2)}{4re^{-\sigma }}\psi-\frac{2}{r}\(r_{,V}H_6-r_{,U}H_7\)+H_{6,V} \,.
}
Substituting this into Eqs.~\eqref{2.11} and \eqref{2.12} and simplifying them, we obtain
\subeq{S2.2}{
\eq{2.16}{
H_6=-\frac{(l-1)(l+2) e^{\sigma }\(r_{,U}\psi +r\psi _{,U}\)}{2\zeta }-\frac{8 \sqrt{2}\pi r^3 e^{\sigma } \delT_{6}}{\sqrt{l(l+1)}\zeta} \,,
}\eq{2.17}{
H_7=\frac{(l-1)(l+2) e^{\sigma }\(r_{,V}\psi +r\psi _{,V}\)}{2\zeta }-i\frac{8 \sqrt{2}\pi r^3 e^{\sigma } \delT_{7}}{\sqrt{l(l+1)}\zeta} \,,
}}
where we have defined
\eq{2.18}{
\zeta _l(U,V) \coloneq (l-1) (l+2) e^{\sigma }-4r\(2r_{,U V}+r\sigma _{,U V}\) \,.
}
Once $\psi_{lm}^{\odd}$ is determined, $\mathbf{h}^{\text{odd}}_{lm}$ can be fully reconstructed from Eqs.~\eqref{2.16} and \eqref{2.17}. Substituting Eqs.~\eqref{2.16} and \eqref{2.17} back into Eq.~\eqref{2.14} and simplifying it, we arrive at the master equation
\eq{2.19}{
\[-4\frac{\partial ^2}{\partial U\partial V}+\gamma_{l,V} \pard{}{U}+\gamma_{l,U}\pard{}{V}-V_l^{\odd}\]\psi _{l m}^{\odd}=S_{l m}^{\odd} \,,
}
where
\eq{2.20}{
\gamma_l (U,V)\coloneq 2\(\ln{\zeta}-\sigma\) \,,
}
and the potential $V_l^{\odd}$ and the source term $S_{l m}^{\odd}$ are given by
\eq{2.21}{
V_l^{\odd}(U,V)\coloneq -\frac{r_{,U}\gamma_{,V}+r_{,V}\gamma_{,U}-4r_{,UV}}{r} +\frac{\zeta -8r_{,U}r_{,V}}{r^2} \,,
}\eq{2.215}{
S_{l m}^{\odd}(U,V)\coloneq \frac{16\sqrt{2}\pi r^2}{\sqrt{l(l+1)}(l-1)(l+2)} \tk{\(2\frac{r_{,V}}{r}-\gamma_{,V}\)\delT_{6}-i\(2\frac{r_{,U}}{r}-\gamma_{,U}\) \delT_{7}+2\delT_{6,V}-2i\delT_{7,U}} \,.
}
We note that 
in some cases, one of the BG Einstein equations,
\eq{2.22}{
-4r\(2r_{,U V}+r\sigma _{,U V}\)=16\pi e^\sigma T^{(\text{BG})}_{22}
}
can be substituted into Eq.~\eqref{2.18} to simplify the computation. 
We also note that for a Schwarzschild BG, Eq.~\eqref{2.19} reduces to the RW equation, but $V_l^{\odd}$ and $S_{lm}^{\odd}$ are not invariant under reparametrizations of the $U$ and $V$ coordinates.\par

Yet another note here is
that, although the GS equation (17) in Ref.~\cite{MasterGS}\footnote{Equation~(17) in Ref.~\cite{MasterGS} contains a sign error. For the correct sign, see Eq.~(6.5b') in Ref.~\cite{MasterGS2}.} explicitly contains $\mathbf{h}$ on its right-hand side, 
by eliminating it using Eqs.~\eqref{2.16} and \eqref{2.17}, one can verify consistency with Eq.~\eqref{2.19}. If the matter is a perfect fluid, this $\mathbf{h}$ automatically cancels out with the $\mathbf{h}$ contained in the energy-momentum tensor; 
while Eq.~\eqref{2.19} remains valid
even when this is not the case.\par

\section{Background with accretion}\label{3.BG}
\subsection{Setup}\label{3..order}
We assume that the accretion is sufficiently small, and denote its order by $\kappa$. We also denote the order of the nonspherical perturbation associated with the ringdown GWs 
by $\varepsilon$, and expand the metric $g_{\mu\nu}$, Einstein tensor $G_{\mu\nu}$, and energy-momentum tensor $T_{\mu\nu}$ as follows:
\eqs{3A.1}{
g_{\mu\nu}&=\(g_{\mu\nu}^{(0,0)}+\kappa g_{\mu\nu}^{(1,0)}+\kappa ^2g_{\mu\nu}^{(2,0)}+\cdots \)+\(\varepsilon g_{\mu\nu}^{(0,1)}+\kappa \varepsilon g_{\mu\nu}^{(1,1)}+\cdots \)+\cdots \,,\\
G_{\mu\nu}&=\(G_{\mu\nu}^{(0,0)}+\kappa G_{\mu\nu}^{(1,0)}+\kappa ^2G_{\mu\nu}^{(2,0)}+\cdots \)+\(\varepsilon G_{\mu\nu}^{(0,1)}+\kappa \varepsilon G_{\mu\nu}^{(1,1)}+\cdots \)+\cdots \,,\\
T_{\mu\nu}&=\(T_{\mu\nu}^{(0,0)}+\kappa T_{\mu\nu}^{(1,0)}+\kappa ^2T_{\mu\nu}^{(2,0)}+\cdots \)+\(\varepsilon T_{\mu\nu}^{(0,1)}+\kappa \varepsilon T_{\mu\nu}^{(1,1)}+\cdots \)+\cdots \,.
}
In this work, we consider the regime $1\gg \kappa \gg \varepsilon$ and solve the master equation up to order $\kappa\varepsilon$ (neglecting terms of order $\ep^2$ and higher). That is, we solve\footnote{In practice, we solve by combining them as $h_{\mu\nu}=\ep g_{\mu\nu}^{(0,1)}+ \kappa\ep g_{\mu\nu}^{(1,1)}$.}
\eq{3A.2}{
\varepsilon G_{\mu\nu}^{(0,1)}\[g_{\mu\nu}^{(0,0)}, g_{\mu\nu}^{(0,1)}\]+\kappa\varepsilon G_{\mu\nu}^{(1,1)}\[g_{\mu\nu}^{(0,0)}, g_{\mu\nu}^{(1,0)},g_{\mu\nu}^{(0,1)},g_{\mu\nu}^{(1,1)}\]=8\pi\kappa\varepsilon T_{\mu\nu}^{(1,1)} \,.
}
Therefore, it is sufficient to determine the BG spacetime to first order in the accretion parameter,\footnote{The $\kappa^2,\,\kappa^3,...$ terms do not contribute to GWs if spherical symmetry is assumed.} i.e., we solve
\eq{3A.3}{
\kappa G_{\mu\nu}^{(1,0)}\[g_{\mu\nu}^{(0,0)}, g_{\mu\nu}^{(1,0)}\]=8\pi\kappa T_{\mu\nu}^{(1,0)}\,.
}
According to the perturbation scheme of Ref.~\cite{BG}, the first-order energy-momentum tensor $T_{\mu\nu}^{(1,0)}$ is obtained by solving the EoM on the zeroth-order (Schwarzschild) metric $g_{\mu\nu}^{(0,0)}=g_{\mu\nu}^{(\Sch)}$, 
\eq{42.2}{
 \nabla_{\mathbf{g}^{(0,0)}}^\mu \(\kappa T_{\mu\nu}^{(1,0)}\)
=0\,,
}
where $\nabla_{\mathbf{g}^{(0,0)}}^\mu$ is the covariant derivative on the BG Schwarzschild spacetime.

\subsection{First-order metric}\label{3..metric}
In this work, we employ the BG spacetime with steady, spherically symmetric accretion onto a Schwarzschild BH, as presented in Ref.~\cite{BG}. The BG spacetime is constructed in EF coordinates $(V,\,r,\,\theta,\,\phi)$. The general spherically symmetric line element in EF coordinates is given by
\eq{43.1}{
ds^2=-\(1-\frac{2 M(V,r)}{r}\) e^{2 \lambda (V,r)}dV^2 +2 e^{\lambda (V,r)} dV dr +r^2 d\Omega^2\,.
}
At zeroth order, using 
$V=t+r^*$ with the tortoise coordinate $r^*\coloneq r+2M_0\ln{(r/2M_0-1)}$, we have
\eq{43.3}{
M(V,r)=M_0 \,,\quad\lambda(V,r)=0\,,
}
where $M_0$ denotes the mass of the Schwarzschild BH. Considering the coordinate transformation to the DN system, we find
\subeq{S3B.0}{
\eq{43.4A}{
r_{,V}=\frac{1}{2} e^{\lambda } \(1-\frac{2 M}{r}\) \,,
}\eq{43.4B}{
e^{\sigma }=-2e^{\lambda }r_{,U}\,.
}}
Writing out the nontrivial components of the Einstein equations in EF coordinates, we obtain
\subeq{S3B.1}{
\eq{44.1}{
M\dr=-4\pi r^2\T{0}{0} \,,
}\eq{44.2}{
M\dV=4\pi r^2\T{0}{1} \,,
}\eq{44.3}{
e^{-\lambda} \lambda\dr=4\pi r\T{1}{0} \,,
}\eq{44.4}{
(r-2 M) \lambda\dr- M\dr=4\pi r^2\T{1}{1} \,,
}\eq{44.5}{
\(-3 r M\dr+M+r\) \lambda\dr-r M_{,rr}+r (r-2 M) \lambda\dr^2+r (r-2 M) \lambda_{,rr}+r^2 e^{-\lambda } \lambda_{,Vr}=8\pi r^2\T{2}{2} =8\pi r^2\T{3}{3} \,,
}
}
Here, Eqs.~\eqref{44.4} 
and \eqref{44.5} can be expressed as combinations of Eqs.~\eqref{44.1}, \eqref{44.2}, and \eqref{44.3} using the EoM $\nabla_\nu \T{\mu}{\nu}=0$. Therefore, we need to consider only Eqs.~\eqref{44.1}, \eqref{44.2}, and \eqref{44.3}.\par
  Since the system is steady and spherically symmetric, $T_{\mu\nu}$ is a function of only $r$. We define the accretion rate $\scr{A}$ by
\eq{44.7}{
  \scr{A}\coloneq M\dV =4\pi r^2 \T{0}{1}\,.
}
Then from Eq.~\eqref{44.1},
\eq{44.8}{
  \scr{A}\dr=0\,,
}
and integrating Eqs.~\eqref{44.1} and \eqref{44.2}, we obtain
\eq{44.9}{
M(V,r)=M\(0,r_0\)+\scr{A}V-4\pi\integ{r_0}{r}{r'}r'^2\T{0}{0}(r')\,.
}
By an appropriate shift $V'= V+\text{const.}$, we can set $M(0,r_0) =M_0$. While $r_0$ can be chosen arbitrarily, we set $r_0=2M_0$.\par
  Since $\lambda$ is of order $\kappa$, writing Eq.~\eqref{44.3} to first order in $\kappa$ gives
\eq{44.10}{
\lambda\dr=4\pi r\T{1}{0}\,.
}
Integrating this, we obtain
\eq{44.11}{
\lambda(V,r)=4\pi\integ{r_1}{r}{r'}r'\T{1}{0}(r')+\lambda(V,r_1)\,.
}
By the reparametrization $dV'= e^{\lambda(V,r_1)}dV$, we can set $\lambda(V,r_1)=0$. Under this reparametrization, the accretion rate transforms as
\eq{3B.0}{
\scr{A}'=\deri{V}{V'}\scr{A}=\scr{A}+O\(\kappa^2\)\,. 
}
Therefore, the choice of $r_1$ does not affect the result as long as the dilute approximation is relevant, and we set $r_1=20M_0$ in this work.\footnote{In practice, $r_1$ cannot be taken to infinity because the dilute condition \eqref{3B.1} must be satisfied (see \tablref{fluidinfinity} and Subsec.~\ref{6..Nocut}).} Summarizing the above, at first order in $\kappa$ we have
\subeq{S3B.2}{
\eq{44.12}{
\delta M(V,r) \coloneq M-M_0 = \scr{A}V-4\pi\integ{2M_0}{r}{r'}r'^2\T{0}{0}(r') \,,
}\eq{44.13}{
\lambda(r)=4\pi\integ{20M_0}{r}{r'}r'\T{1}{0}(r')\,.
}}
Since $\kappa\ll 1$, $\delta M$ and $\lambda$ must satisfy the dilute condition
\eq{3B.1}{
\abs{\delta M}\ll M_0\,,\quad \abs{\lambda}\ll 1\,.
}

\subsection{
Perfect fluid cases
}\label{3..EnergyTensor}
In this work, we assume matter to be a perfect fluid,
\eq{45.1}{
T_{\mu\nu}=(\rho+p)u_\mu u_\nu+pg_{\mu\nu}\,.
}
The problem of spherically symmetric steady accretion was formulated for Newtonian gravity in Ref.~\cite{Bondi} and extended to Schwarzschild spacetime in Ref.~\cite{Michel}. The formulation in this subsection follows Refs.~\cite{Koga:2016jjq,Koga:2018ybs,fluid}.
Here, $\rho(r)$ and $p(r)$ are the energy density and pressure in the fluid rest frame, respectively, and $u^\mu(r)$ is the four velocity of the fluid. Since $\rho$ and $p$ are of order $\kappa$, we may set $g_{\mu\nu}=g^{(0,0)}_{\mu\nu}$ in Eq.~\eqref{45.1}. Defining $u(r)\coloneq  \abs{u^r(r)}$, the four velocity $u^\mu$ can be written as
\eq{45.2}{
u^\mu=\tk{\frac{1}{\sqrt{f_0+u^2}+u},\,-u,\,0,\,0}\,,
}
where $f_0\coloneq1-2M_0/r$. Using these expressions, we find
\eq{45.3}{
\T{0}{0}=\frac{p u-\rho  \sqrt{f_0+u^2}}{\sqrt{f_0+u^2}+u},\qquad \T{1}{0}=\frac{\rho+p}{\(\sqrt{f_0+u^2}+u\)^2}\,.
}
\par
  The quantities $\rho$, $p$, and $u$ are determined by the EoS and the EoM. In this work, we assume the EoS,
\eq{45.5}{
p=w\rho\,,
} where $w$ is a constant. The nontrivial components of the EoM 
\eqref{42.2}
are
\subeq{S3C.1}{
\eq{45.6}{
(1+w) \frac{r^2 \(f_0+2u^2\)u\dr+u \(2r-3 M_0+2ru^2\)}{r^2 \sqrt{f_0+u^2}}\rho+(1+w)  u \sqrt{f_0+u^2} \rho\dr=0 \,,
}\eq{45.7}{
(1+w) \frac{\(u-\sqrt{f_0+u^2}\)r^2 u\dr -2ru \sqrt{f_0+u^2}+M_0}{r^2\tk{f_0+u\(\sqrt{f_0+u^2}+u\)}}\rho+\frac{w \sqrt{f_0+u^2}-u}{\sqrt{f_0+u^2}+u}\rho\dr=0\,.
}}
Eliminating $u\dr$ from Eqs.~\eqref{45.6} and \eqref{45.7} and simplifying it, we have
\eq{45.8}{
\rho\dr=\frac{(1+w) \(2ru^2-M_0\)}{r^2 \(wf_0+(w-1)u^2\)}\rho\,.
}
Similarly, eliminating $\rho\dr$ from Eqs.~\eqref{45.6} and \eqref{45.7} and simplifying it, we obtain
\eq{45.9}{
u\dr=\frac{u \(-2wrf_0-2wru^2+M_0\)}{r^2\(wf_0+(w-1)u^2\)}\,.
}
From Eqs.~\eqref{45.8} and \eqref{45.9}, it follows that
\eq{45.10}{
\rec{1+w}\frac{\rho\dr}{\rho}+\frac{u\dr}{u}=-\frac{2}{r}\,.
}
Integrating both sides of Eq.~\eqref{45.10} with respect to $r$ and rearranging it, we derive
\eq{3C.1}{
\rho\(r^2u\)^{1+w}=\frac{\scr{A}\scr{F}M_0^{2w}}{4\pi (1+w)}\,,
}
where $\scr{F}$ is an arbitrary constant, and we have defined the integration constant on the right-hand side for later convenience. On the other hand, from Eq.~\eqref{44.7},
\eq{45.12}{
4\pi (1+w)\rho r^2 u \sqrt{f_0+u^2}=\mathcal{A}\,,
}
and dividing Eq.~\eqref{3C.1} by Eq.~\eqref{45.12}, we obtain
\eq{45.13}{
\frac{\(r^2 u\)^{w}}{\sqrt{f_0+u^2}} =\scr{F}M_0^{2w}\,.
}
Thus, the BG spacetime is fully determined by specifying the parameters $w$, $\scr{A}$, and $\scr{F}$ and solving Eqs.~\eqref{45.13} and \eqref{3C.1} 
 for $u$ and $\rho$ as functions of $r$.
Assuming $\rho>0$,
 from Eq.~\eqref{45.12}, 
 we find 
\eq{3C.5}{
\sgn(\scr{A})=\sgn (1+w)\,,
}
except static solution\footnote{The static
solution is obtained by substituting $u=0$ into Eq.~\eqref{45.8} and solving it.} $u=0$.
For $w=-1$,
\eq{3C.2}{
T_{\mu\nu}=-\rho g_{\mu\nu}\,,
} and the energy-momentum tensor does not depend on $u$. 
From Eqs.~\eqref{45.12} and \eqref{45.7}, 
 we obtain
\eq{3C.3}{
\scr{A}=0\,,\quad\rho=\frac{\Lambda}{8\pi}=\const 
} 
Since we are interested in nonstatic solutions, we consider only solutions with $\scr{A}\ne 0$.\par

 We numerically solve Eq.~\eqref{45.13} since it cannot be analytically solved in general. 
Qualitative behavior of the solutions for each value of $w$ can be understood by plotting the contours of $\mathcal F$ as a function of $r$ and $u$ as in Fig.~\ref{FluidContour}. 
\fig{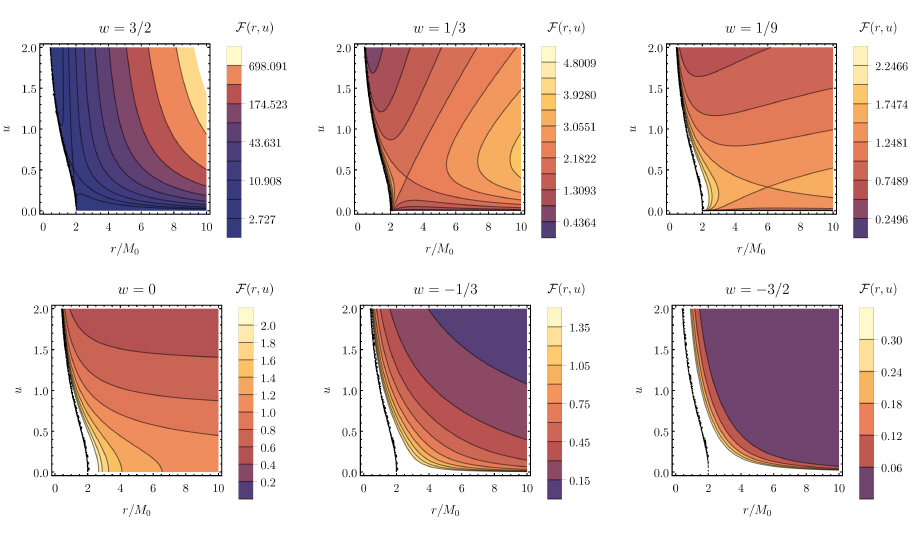}{\linewidth}{
Contour plots of $\scr{F}=\scr{F}(r,u)$ for $w=3/2$, $1/3$, $1/9$, $0$, $-1/3$, and $-3/2$. Each contour line represents a solution of $u(r)$. According to the properties of $\scr{F}(r,u)$, the behavior can be classified into four cases: (i) $w>1$, (ii) $0<w\le1$, (iii) $w=0$, and (iv) $w<-1,\,-1<w<0$. The thick black lines indicate the boundaries where $\scr{F}(r,u)$ takes real values. 
}
First, for the solution to be physically reasonable, we impose the following condition: $u(r)$ and $\rho(r)$ must take finite values in the entire region $r>0$\footnote{For $w>-1$ (positive accretion rate), it is sufficient for them to be finite at $r\ge 2M_0$, but 
 the possible parameter region is not extended by this relaxed condition. 
Note that solutions with $u(r\to 2M_0)=0$ are excluded because $\rho$ diverges.}. Then, for $0<w\le1$, the only allowed solution is the monotonically decreasing one passing through the saddle point. In this case, $\scr{F}$ at the saddle point is determined as
\eq{3C.6}{
\scr{F}_{\text{saddle}}=4^{-w} w^{-\frac{3}{2}w} (1+3w)^{\frac{1}{2}(1+3w)}\,.
}
For $w>1$, the saddle point lies in $r<2M_0$, and the value of $u$ vanishes at $r=2M_0$ for $\scr{F}<\scr{F}_\text{saddle}$. 
On the other hand, $\scr{F}\geq\scr{F}_\text{saddle}$ can be 
well defined on the entire region of $r>0$, so $\scr{F}\ge\scr{F}_{\text{saddle}}$ is permitted. 
For $w\le0$, no saddle point exists, but for $w=0$, only solutions with $\scr{F}\le1$ take finite values at infinity.\footnote{As shown in Table~\ref{fluidinfinity}, imposing $u(r\to\infty)=0$ yields $\scr{F}=1$.} For $w<0$, any $\scr{F}>0$ is allowed. In summary, the allowed values of $\scr{F}$ are restricted to
\eq{3C.7}{
\begin{cases}
    \scr{F}\ge \scr{F}_{\text{saddle}} & (w>1) \\
    \scr{F}=\scr{F}_{\text{saddle}} & (0<w\le1) \\
    0<\scr{F}\le1                   & (w=0) \\
    \scr{F}>0                       & (w<-1,\,-1<w<0)
\end{cases} \,.
}
\par
Finally, we discuss the asymptotic behavior of the solution at large $r$. By successively using Eqs.~\eqref{45.13}, \eqref{3C.1}, \eqref{45.3}, \eqref{44.12}, and \eqref{44.13}, we obtain Table~\ref{fluidinfinity}. As can be seen, the dilute condition \eqref{3B.1} is violated at sufficiently large $r$, so the solution cannot be used in that region. In practice, it would be natural to assume that accretion exists only within a certain region. We attempted to introduce a cutoff in the spacetime to obtain a more realistic asymptotically flat BG, but found it 
 technically
difficult to eliminate the nontrivial effects of the cutoff because the frequency is sensitive to 
 relatively steep potential deformations within a technically manageable range 
(see Appendix~\ref{C.cutoff}). Therefore, we do not introduce such a modification in the subsequent calculations, and note that our result is valid only in the region where the dilute approximation is valid. 
\begin{table}[htbp]
\caption{Asymptotic behavior of the fluid and the BG metric at large $r$.}
\tabcolsep = 10pt
\begin{ruledtabular}
\begin{tabular}{rccc}
                & $w\ne 0,-1$   & $w=0,\,\scr{F}<1$ & $w=0,\,\scr{F}=1$ \\
\colrule
  $u\simeq$     & $\scr{F}^{\rec{w}}\(\dfrac{r}{M_0}\)^{-2}$ & $\sqrt{\scr{F}^{-2}-1}$ & $\sqrt{\dfrac{2M_0}{r}}$ \\
  $\rho M_0^2\simeq$  & $\dfrac{\scr{A}}{4\pi(1+w)\scr{F}^{\rec{w}}}$ & $\dfrac{\scr{A}\scr{F}}{4\pi\sqrt{\scr{F}^{-2}-1}} \(\dfrac{r}{M_0}\)^{-2}$ & $\dfrac{\scr{A}}{4\sqrt{2}\pi} \(\dfrac{r}{M_0}\)^{-\frac{3}{2}}$ \\
  $\delta M/M_0\simeq$& $\dfrac{\scr{A}}{3(1+w)\scr{F}^{\rec{w}}}\(\dfrac{r}{M_0}\)^3$ & $\dfrac{\scr{A}}{\scr{F}^{-2}-1+\scr{F}^{-1}\sqrt{\scr{F}^{-2}-1}}\dfrac{r}{M_0}$ & $\dfrac{\sqrt{2}\scr{A}}{3}\(\dfrac{r}{M_0}\)^{\frac{3}{2}}$ \\
  $\lambda\simeq$ & $\dfrac{\scr{A}}{2\scr{F}^{\rec{w}}}\(\dfrac{r}{M_0}\)^2$ & $\dfrac{\scr{A}\scr{F}}{\sqrt{\scr{F}^{-2}-1}\(\scr{F}^{-1}+\sqrt{\scr{F}^{-2}-1}\)^2}\ln\({\dfrac{r}{M_0}}\)$ & $\sqrt{2}\scr{A}\sqrt{\dfrac{r}{M_0}}$
\end{tabular}
\end{ruledtabular}
\label{fluidinfinity}
\end{table}

\section{The source term}\label{4.Source}
\subsection{Form of the source term}\label{4..FormS}
The source term $S_{lm}^{\odd}(U,V)$ is determined from the energy-momentum tensor at order $\kappa\varepsilon$, $T_{\mu\nu}^{(1,1)}$. To include terms up to order $\kappa\varepsilon$ in the energy-momentum tensor, we make the replacements in Eq.~\eqref{45.1},
\eqs{51.1}{
\rho&\rightarrow \rho+\sum_{l,m}\delta\rho_{lm}Y_{lm} \,,\\
u_\mu&\rightarrow u_\mu+ \sum_{i,l,m}\delta u_{ilm}\(\vsh{i}\)_\mu \,,\\
g_{\mu\nu}&= g^{(\Sch)}_{\mu\nu}+\ep g^{(0,1)}_{\mu\nu}\,.
}
Here, $\delta\rho_{lm}$ and $\delta u_{ilm}$ represent the fluid perturbations, where $\delta\rho_{lm}$ is of order $\kappa\ep$ and $\delta u_{ilm}$ is of order $\ep$. Then, for the odd parity mode, we have
\eqs{51.3}{
\(\delta\B{T}_{lm}^{\odd}\)_{\mu\nu}=
(1+w)\rho \tk{u_\mu\delta u_{3lm}\(\vsh{3}\)_\nu+u_\nu\delta u_{3lm}\(\vsh{3}\)_\mu}+w\rho \(\B{h}^{\odd}_{lm}\)_{\mu\nu}\,,
}
and by comparing both sides in DN coordinates, we obtain\footnote{Note that $r\dV=f_0/2$ from Eq.~\eqref{51.5} has already been used when transforming $u_\mu$ to the DN system.}
\eqs{51.4}{
\delT_{6}&=(1+w) \frac{\sqrt{2}r\dU}{\sqrt{f_0+u^2}+u}\rho\,\delta u_{3} -w\frac{\sqrt{2l(l+1)}}{r} \rho H_{6} \,,\\
\delT_{7}&=i(1+w) \frac{\sqrt{f_0+u^2}+u}{\sqrt{2}}\rho\,\delta u_{3}+ iw\frac{\sqrt{2l(l+1)}}{r}\rho H_{7} \,,\\
\delT_{8}&=0\,.
}
Using Eq.~\eqref{S2.2}, $H_i$ can be replaced by expressions involving the master variable. Since $H_i$ need only 
be included up to order $\ep$, we can omit the contribution from $\delT_6$ and $\delT_7$ in Eq.~\eqref{S2.2}. 
Using 
Eqs.~\eqref{2.18}, \eqref{2.22}, and \eqref{S3B.0} evaluated at zeroth order, 
\eqs{51.5}{
r\dV&=\rec{2}f_0 \,,\\
e^\sigma&=-2r\dU \,,\\
\zeta&=-2 (l-1)(l+2) r\dU \,,
}
we can write
\subeq{4A.1}{
\eq{51.6}{
H_6=-\frac{1}{2} \(r\dU \psi+r\psi\dU\) \,,
}\eq{51.7}{
H_7=\frac{1}{2}\(\frac{f_0}{2} \psi+r\psi\dV\)\,.
}}
Substituting Eqs.~\eqref{51.4}, \eqref{51.5}, and \eqref{4A.1} into the definition of the source term \eqref{2.215}, we derive
\eqa{51.8}{
S_{l m}^{\odd}=&\frac{32\pi(1+w)r^2}{\sqrt{l(l+1)}(l-1)(l+2)} \Biggl[\( \sqrt{f_0+u^2}+u\)\rho\,\delta u_{3,U} +\frac{2 r\dU}{\sqrt{f_0+u^2}+u}\rho\,\delta u_{3,V} \nn
&\qquad\qquad\qquad\qquad\qquad+r\dU\tk{\frac{1+f_0+2u\(u+ru\dr\)}{r \sqrt{f_0+u^2}}\rho+ 2\sqrt{f_0+u^2}\rho\dr} \delta u_3 \Biggr] \nn
&+\frac{32\pi wr}{(l-1)(l+2)} \tk{2r\rho\psi\dUV+ \frac{f_0}{2}\(2\rho +r \rho\dr\)\psi\dU +r\dU \(2 \rho +r \rho\dr\)\psi\dV+r\dU \(\frac{2M_0}{r^2} \rho +f_0 \rho\dr \)\psi }\,.
}
Furthermore, using the master equation at order $\ep$,
\eq{51.9}{
-4\psi\dUV=-2r\dU \tk{\frac{l(l+1)}{r^2}-\frac{6 M_0}{r^3}}\psi\,,
}
we can eliminate $\psi\dUV$ to obtain
\eqa{51.10}{
S_{l m}^{\odd}=&\frac{32\pi(1+w)r^2}{\sqrt{l(l+1)}(l-1)(l+2)} \Biggl[\( \sqrt{f_0+u^2}+u\)\rho\,\delta u_{3,U} +\frac{2 r\dU}{\sqrt{f_0+u^2}+u}\rho\,\delta u_{3,V} \nn
&\qquad\qquad\qquad\qquad\qquad+r\dU\tk{\frac{1+f_0+2u\(u+ru\dr\)}{r \sqrt{f_0+u^2}}\rho+ 2\sqrt{f_0+u^2}\rho\dr } \delta u_3 \Biggr] \nn
&+\frac{32\pi wr}{(l-1)(l+2)} \[\frac{f_0}{2}\(2\rho +r \rho\dr\)\psi\dU +r\dU \(2 \rho +r \rho\dr\)\psi\dV+r\dU\tk{\frac{(l-1)(l+2)+2f_0}{r}\rho +f_0 \rho\dr}\psi \]\,.
}
Here, it is sufficient  
for $\psi$ appearing in $S_{l m}^{\odd}$ to include terms only up to order $\ep$, 
 but we also include the order $\kappa \ep$ terms to avoid the effort of handling the order $\ep$ and $\kappa\ep$ terms separately.\footnote{
This is possible because 
the EoM \eqref{52.2} does not depend on $\psi$.}

\subsection{Equation for the fluid perturbation}\label{4..FPeq}
Similarly to Eq.~\eqref{42.2}, $T_{\mu\nu}^{(1,1)}$ satisfies the EoM at the $\kappa \varepsilon$ order, 
\eq{50.1}{
\nabla_{\mathbf{g}^{(0,0)}+\mathbf{g}^{(0,1)}}^\mu \(\kappa\ep T_{\mu\nu}^{(1,1)} +\kappa T_{\mu\nu}^{(1,0)}\)=0\,, 
}
where $\nabla_{\mathbf{g}^{(0,0)}+\mathbf{g}^{(0,1)}}^{\mu}$ is the covariant derivative up through the $\kappa\epsilon$ order, and the terms at order $\kappa$ disappear due to Eq.~\eqref{42.2}. 
Writing Eq.~\eqref{50.1} in the DN system, we obtain
\eq{52.1}{
(1+w)\Biggl[-\frac{\sqrt{f_0+u^2}+u}{2r\dU}\rho\,\delta u_{3,U} +\frac{1}{\sqrt{f_0+u^2}+u}\rho\,\delta u_{3,V} -\tk{\(u\dr+3\frac{u}{r}\)\rho +u\rho\dr }\delta u_3 \Biggr] \vsh{3} =0\,,
}
which gives
\eq{52.2}{
-\frac{\sqrt{f_0+u^2}+u}{2r\dU}\rho\,\delta u_{3,U} +\frac{1}{\sqrt{f_0+u^2}+u}\rho\,\delta u_{3,V} -\tk{\(u\dr+3\frac{u}{r}\)\rho +u\rho\dr }\delta u_3 =0\,.
}
As Eq.~\eqref{52.2} shows, the EoM does not depend on the metric perturbation $H_i$, indicating that the odd-parity perturbations of the perfect fluid decouple from odd-parity GWs. 
This is consistent with the result obtained in Ref.~\cite{GundlachMG}. This also implies that if $\delta u_3=0$ on some initial surface, then $\delta u_3=0$ at all times; hereafter, we assume $\delta u_3=0$.

\section{
Equations and extraction of physical quantities
}\label{5.Method}
\subsection{Equations to be solved}\label{5..Solvedeq}
Since the BG used in this work is dynamical, we derive the waveform by directly solving Eq.~\eqref{2.19} in the time domain. In general, an equation of the form
\eq{61.0}{
x\dUV=F(x,x\dU,x\dV)
}
can be solved numerically using the DNF~\cite{DNF1}, which is described in Appendix~\ref{A.numerical}.\par
In practice, to obtain the solution for $\psi$ via the DNF given the BG metric components $M$ and $\lambda$, we need to solve the following coupled differential equations for $r$ and $\psi$. First, expanding Eq.~\eqref{43.4A} to order $\kappa$, we obtain
\eq{62.1}{
r\dV=\frac{f_0}{2}(1+\lambda)-\frac{\delta M}{r}\,,
}
and then differentiating both sides with respect to $U$, we have
\eq{5C.1}{
r\dUV=r\dU\(\frac{M_0+\delta M+M_0\lambda}{r^2}-\frac{\delta M\dr}{r}+\frac{f_0}{2}\lambda\dr\) \,.
}
The other differential equation is the master equation
\eq{5C.2}{
\[-4\frac{\partial ^2}{\partial U\partial V}+\gamma_{l,V} \pard{}{U}+\gamma_{l,U}\pard{}{V}-V_l^{\odd}\]\psi_{lm}^{\odd}=S_{l m}^{\odd}\,,
}
where the potential term
\eq{5C.3}{
V_l^{\odd}= -\frac{r_{,U}\gamma_{,V}+r_{,V}\gamma_{,U}-4r_{,UV}}{r} +\frac{\zeta -8r_{,U}r_{,V}}{r^2}
}
follows from the definition, and the source term
\eq{5C.4}{
S_{l m}^{\odd}=\frac{32\pi wr}{(l-1)(l+2)} \[\frac{f_0}{2}\(2\rho +r \rho\dr\)\pard{}{U} +r\dU \(2 \rho +r \rho\dr\)\pard{}{V} +r\dU\tk{\frac{(l-1)(l+2)+2f_0}{r}\rho +f_0 \rho\dr}\] \psi\,,
}
is obtained by substituting $\delta u_3=0$ into Eq.~\eqref{51.10}. In addition, 
substituting Eq.~\eqref{43.4B} into Eq.~\eqref{2.18} and expanding to order $\kappa$, we obtain
\eq{5C.5}{
\zeta=-2(l-1) (l+2) (1+\lambda)r\dU -4r\(2r_{,U V}+r\sigma _{,U V}\)\,.
}
To obtain $\sigma\dUV$, we differentiate the logarithm of Eq.~\eqref{43.4B} with respect to $V$,
\eq{B.1}{
\sigma\dV=r\dV \lambda\dr+\frac{r\dUV}{r\dU}\,.
}
Substituting Eqs.~\eqref{62.1} and \eqref{5C.1} and expanding to order $\kappa$, we obtain
\eq{B.2}{
\sigma\dV=\frac{M_0+\delta M+M_0\lambda}{r^2}-\frac{\delta M\dr}{r}+f_0\lambda\dr\,,
}
and differentiating both sides with respect to $U$, we derive
\eq{5C.6}{
\sigma\dUV=r\dU\(-2\frac{M_0+\delta M+M_0\lambda}{r^3}+\frac{2\delta M\dr+3M_0\lambda\dr}{r^2}-\frac{\delta M_{,rr}}{r}+f_0\lambda_{,rr}\)\,.
}
Substituting Eqs.~\eqref{43.4B}, \eqref{5C.5}, \eqref{5C.1}, and \eqref{5C.6} into Eq.~\eqref{2.20} cancels $r\dU$, and $\gamma$ is determined as
\eq{B.3}{
\gamma=\frac{4}{(l-1)(l+2)}\tk{(r+M_0)\lambda\dr -r\delta M_{,rr}+r^2f_0 \lambda_{,rr}} +2\ln\tk{(l-1)(l+2)}\,,
}
and differentiating with respect to $U$ and $V$, respectively, we obtain
\subeq{S5C.0}{
\eq{5C.7}{
\gamma_{,U}=\frac{4r_{,U}}{(l-1)(l+2)}\tk{\lambda_{,r}-\delta M_{,rr}+\(3r-M_0\)\lambda_{,rr} -r\delta M_{,rrr}+r^2f_0\lambda_{,rrr}} \,,
}
\eq{5C.8}{
\gamma_{,V}=\frac{4r_{,V}}{(l-1)(l+2)}\tk{\lambda_{,r}-\delta M_{,rr}+\(3r-M_0\)\lambda_{,rr} -r\delta M_{,rrr}+r^2f_0\lambda_{,rrr}}\,.
}}
In the above
 equations, 
$r$ and its derivatives can be obtained from Eq.~\eqref{S5A.1}. \par
However, as discussed in Sec.~\ref{2.Master}, using
\eq{5C.9}{
-4r(2r\dUV+r\sigma\dUV)=16\pi e^\sigma T_{22}^{(\text{BG})}=16\pi e^\sigma w\rho r^2\,,
}
 Eq.~\eqref{5C.2} can be simplified as 
follows.
First, substituting Eq.~\eqref{5C.9} into Eq.~\eqref{2.18}, $\zeta$ becomes
\eq{B.8}{
\zeta = (l-1)(l+2)e^{\sigma }+16\pi e^\sigma w\rho r^2\,.
}
Substituting into Eq.~\eqref{2.20} and expanding to order $\kappa$, we obtain
\eq{B.9}{
\gamma=\frac{32 \pi w\rho r^2}{(l-1)(l+2)}+2 \ln\tk{(l-1)(l+2)}\,,
}
and then differentiating with respect to $U$ and $V$, respectively, we have
\subeq{SB.2}{
\eq{B.10}{
\gamma_{,U}=\frac{32\pi wr}{(l-1)(l+2)} r\dU\(2\rho +r\rho\dr\) \,,
}
\eq{B.11}{
\gamma_{,V}=\frac{32 \pi wr}{(l-1)(l+2)} r\dV\(2\rho +r\rho\dr\)\,.
}}
Substituting Eqs.~\eqref{B.8}, \eqref{SB.2}, and \eqref{43.4B} into Eq.~\eqref{5C.3}, we obtain
\eq{B.12}{
V_l^{\odd}= \frac{-2(l-1)(l+2)(1+\lambda)r\dU -8r\dU r\dV+4rr\dUV}{r^2} -\frac{32\pi wr\dU}{(l-1)(l+2)} \tk{(l-1)(l+2)\rho +2r\dV \(2\rho +r\rho\dr\)}\,.
}
Furthermore, substituting these into Eq.~\eqref{5C.2} and moving $S_{l m}^{\odd}$ to the left-hand side, we derive another master equation
\eq{5C.10}{
\[-4\frac{\partial ^2}{\partial U\partial V}-\frac{-2(l-1)(l+2)(1+\lambda)r\dU-8r\dU r\dV +4rr\dUV}{r^2}\]\psi_{lm}^{\odd}=0\,.
}
Equation~\eqref{5C.10} is equivalent 
to the GS equation with $(\text{rhs})=0$:
\eq{5C.11}{
\[-4\frac{\partial ^2}{\partial U\partial V}-\frac{(l-1)(l+2)e^\sigma-8r\dU r\dV +4rr\dUV}{r^2}\]\psi_{lm}^{\odd}=0\,.
}
In the end, to obtain the time evolution using the DNF, it suffices to solve Eqs.~\eqref{5C.1} and \eqref{5C.10} simultaneously.\footnote{The other master equations, Eqs.~\eqref{5C.2}--\eqref{S5C.0}, are used in Appendix~\ref{C.cutoff}.}
\par

In this paper, we set the initial conditions as follows:
\eqs{64.2}{
\psi(U,V_0)&=A\exp\tk{-\frac{(r(U,V_0)-r_c)^2}{2s^2}}\,,\\
\psi(U_0,V)&=0\,,
}
where $A$, $r_c$, and $s$ are constants. The $U$ dependence of $r$, including $r(U,V_0)$, is provided at each step by Eq.~\eqref{63.10}, and $r(U_0,V)$ is obtained by solving Eq.~\eqref{62.1} using a second-order Runge-Kutta method. Unless otherwise stated, the parameters are set to
\eqs{5C.13}{
(A,\,r_c,\,s)&=(0.1,\,2.5 M_0,\,0.1 M_0)\,,\\
\(r\minn,\,r\maxx(V_0)\) &=(1.5M_0,\,3.5M_0)\,,\\
(U_0,\,U\maxx)=(V_0,\,V\maxx)&=(0,\,200 M_0)\,,\\
(N_U,\,N_V) \coloneq\(\frac{U\maxx}{\Delta U},\,\frac{V\maxx}{\Delta V}\) &=(163840,\,163840) \,.\\
}
Here, $r\minn$, $r\maxx$, $U_0$, $U\maxx$, $V_0$, and $V\maxx$ specify the computational domain and $\Delta U$ and $\Delta V$ denote the grid spacing, whose definitions follow Appendix~\ref{A.numerical}.\par

Finally, the BG components entering these equations are determined by specifying $u(r)$ as described in Sec.~\ref{3.BG}. The function $u(r)$ is obtained by solving Eq.~\eqref{45.13} using the Newton-Raphson method while incrementally varying $r$; however, for $w>0$, two solutions pass through the saddle point, so care must be taken regarding the amount of change and convergence of $u$ to avoid selecting the wrong branch.

\subsection{
Extraction of useful physical quantities
}\label{5..Potential}

To read off the waveform of $\psi$, we fix the observer at 
$r=r_\obs$ and examine the $V$ dependence. 
 In an asymptotically flat situation, unlike our case, 
to obtain the waveform at $r_\obs\rightarrow\infty$, one usually needs to compute a correction term arising from the tail of the potential. Such a correction term can be obtained as a time integral of $\psi|_{r=r_\obs}$ 
 in an asymptotically flat BG~\cite{Lousto:2010qx,InfCorrection},
but this is not possible for our current BG. Therefore, we also need to pay attention to the $r_\obs$-dependence of the observables. This will be discussed in Subsec.~\ref{6..TimeAverage}.

The frequency $\omega$ can be determined by extracting the extrema from the waveform of $\psi$. Assuming that the timescale of the variation of $\omega$ is sufficiently longer than the oscillation period, the real part $\omR$ and imaginary part $\omI$ of $\omega$ can be obtained using the peak intervals $\delta V_\text{peak}$ and $\delta\ln{|\psi_\text{peak}|}$ as
\subeq{S5D.1}{
\eq{5D.1}{
\omR \simeq\frac{\pi}{\delta V_\text{peak}} \,,
}\eq{5D.2}{
\omI \simeq\frac{\delta\ln{|\psi_\text{peak}|}}{\delta V_\text{peak}}\,.
}}
In the Schwarzschild BG, $\omR$ and $\omI$ are inversely proportional to the BH mass $M_0$, but their ratio $\omI^{(\Sch)}/\omR^{(\Sch)}$ is a constant specific to each mode. 
Since the effects of red/blue shift are expected to cancel 
by taking the ratio $\omI/\omR$, 
the observable
\eq{5D.2.5}{
\Xi\coloneq \left. \frac{\omI}{\omR}\middle/\frac{\omI^{(\Sch)}}{\omR^{(\Sch)}} \right.-1
}
is of particular importance as a measure of the deviation from Schwarzschild spacetime.\par
To investigate what information about the matter distribution can be extracted from the ringdown gravitational waveform, we consider a method for estimating the accretion rate $\scr{A}$ from the time dependence of $\omega$. We assume that $\omega$ can be written in the following form
with (nearly) constant $\tilde{\scr{A}}$ and $\scr{M}$:
\eq{5D.3}{
\omega=\frac{M_0\omega^{(\Sch)}}{\tilde{\scr{A}}\,V+\scr{M}}\,.
}
This assumption is based on the following considerations. Let us generally take $\scr{A}$ to be a function of $V$ and $r$. If the deviation of the BG from Schwarzschild spacetime is sufficiently small, the frequency of an oscillation generated at a point $r=r_\gen$ is expected to satisfy
\eq{5D.4}{
\omega(V,r_\gen)\simeq \frac{M_0\omega^{(\Sch)}}{M(V,r_\gen)}= \frac{M_0\omega^{(\Sch)}}{\integ{0}{V}{V'}\scr{A}(V',r_\gen)+M(0,r_\gen)}\,.
}
Observing this wave at $r=r_\obs$, we have
\eq{5D.5}{
\omega(V,r_\obs) \simeq \frac{M_0\omega^{(\Sch)}}{M(V_\gen(V),r_\gen)}\scr{R}(V) = \frac{M_0\omega^{(\Sch)}}{\integ{0}{V_\gen(V)}{V'}\scr{A}(V',r_\gen)+M(0,r_\gen)}\scr{R}(V)\,,
}
where $\scr{R}(V)$ is the redshift factor arising from the nonstationarity of the BG, and $V_\gen(V)$ is the time at which the wave observed at $(V,r_\obs)$ was generated. 
 The redshift factor can be estimated as 
\eq{5D.7}{
\scr{R}(V)=\deri{V_\gen(V)}{V}\,. 
}
If $\scr{A}(V,r_\gen)$ can be regarded as constant for $V_\gen(V_1)\le V\le V_\gen(V_2)$, then
\eq{5D.6}{
\omega(V_2,r_\obs) \simeq \frac{M_0\omega^{(\Sch)}}{\scr{A}\, \(V_\gen(V_2)-V_\gen(V_1)\)+M(V_\gen(V_1),r_\gen)}\scr{R}(V_2)\,.
}
Using the relation \eqref{5D.7}, 
we find that, when $\scr{R}$ can be regarded as constant for $V_1\le V\le V_2$,
\eq{5D.8}{
\omega(V_2,r_\obs) \simeq \frac{M_0\omega^{(\Sch)}}{\scr{A}\, \(V_2-V_1\) +\scr{R}^{-1}M(V_\gen(V_1),r_\gen)}\,.
}
Therefore, the $V$ dependence of $\omega$ can be expressed in the form of Eq.~\eqref{5D.3} independently of the redshift effect,\footnote{In our current approximation, $\scr{A}\,\scr{R}\simeq\scr{A}$, but this argument holds regardless of that.} and we expect $\tilde{\scr{A}}\simeq\scr{A}$. However, when $\Xi\ne 0$, $\omega/\omega^{(\Sch)}$ is not real. Therefore, we allow $\tilde{\scr{A}}$ and $\scr{M}$ to take different values for $\omR$ and $\omI$, respectively. Eliminating $\scr{M}$ from the frequencies at two different times, $\omega_1\coloneq \omega(V_1,r_\obs)$ and $\omega_2\coloneq \omega(V_2,r_\obs)$, we obtain\footnote{Note that $\tilAR$ and $\tilAI$ do not denote $\text{Re}(\tilde{\scr{A}})$ and $\text{Im}(\tilde{\scr{A}})$, respectively.}
\eqs{5D.9}{
\tilAR&\coloneq \frac{M_0\omega^{(\Sch)}_\text{R}}{V_2-V_1}\(\rec{\omega_{2\text{R}}}-\rec{\omega_{1\text{R}}}\) \,,\\
\tilAI&\coloneq \frac{M_0\omega^{(\Sch)}_\text{I}}{V_2-V_1}\(\rec{\omega_{2\text{I}}}-\rec{\omega_{1\text{I}}}\)\,.
}
However, note that the expectation $\tilde{\scr{A}}\simeq\scr{A}$ is valid only when
\eq{5D.10}{
\abs{\scr{A}\dV\,(V_2-V_1)},\,\abs{\scr{M}\dV}\ll \scr{A}
}
is satisfied. For steady accretion with $\scr{A}\dV=0$, we have $\scr{M}\dV\simeq\scr{R}\dV M_0=O(\scr{A}^2)$, so Eq.~\eqref{5D.10} should always be satisfied.\par
All quantities are computed with appropriate interpolation so that the final error achieves second-order accuracy.\par

\section{Results}\label{6.Results}
\subsection{Plots in the time domain}\label{6..Nocut}
We first present the numerical results for several representative cases. Figure~\ref{damp_l3} shows the waveform of $\psi$ received by an observer fixed at $r=20M_0$. After 
 a certain period of 
time, the fundamental mode becomes dominant, and its frequency differs slightly depending on the BG. 
For comparison, we also compute the waveform in an ingoing linear Vaidya BG obtained by setting $\T{0}{0}=\T{1}{0}=0$ in Eq.~\eqref{S3B.2}, which corresponds to a special case of exact spherically symmetric BH solutions with null dust accretion. 

Figure~\ref{BG} plots the numerical solutions of the BG functions for these cases 
 (except for
the vacuum and Vaidya cases). The BG functions are computed at $r$ intervals sufficiently fine relative to the DNF grid spacing and then appropriately interpolated for use in the computation. Note that in the DNF, $r\maxx(V)$ increases as the computation progresses, and regions where the dilute approximation breaks down may arise. However, if the effect of the potential term is neglected, oscillations only propagate outward, so the influence of such regions is expected to be 
 highly suppressed. 
Indeed, the approximation is shown to be 
 valid 
for $\abs{\scr{A}}=3\times 10^{-5}$ in Subsec.~\ref{6..TimeAverage}. 
Therefore, hereafter we consider 
that our calculation is valid even if 
the condition \eqref{3B.1} is satisfied
 only for $r\lesssim r_\obs$.
\fig{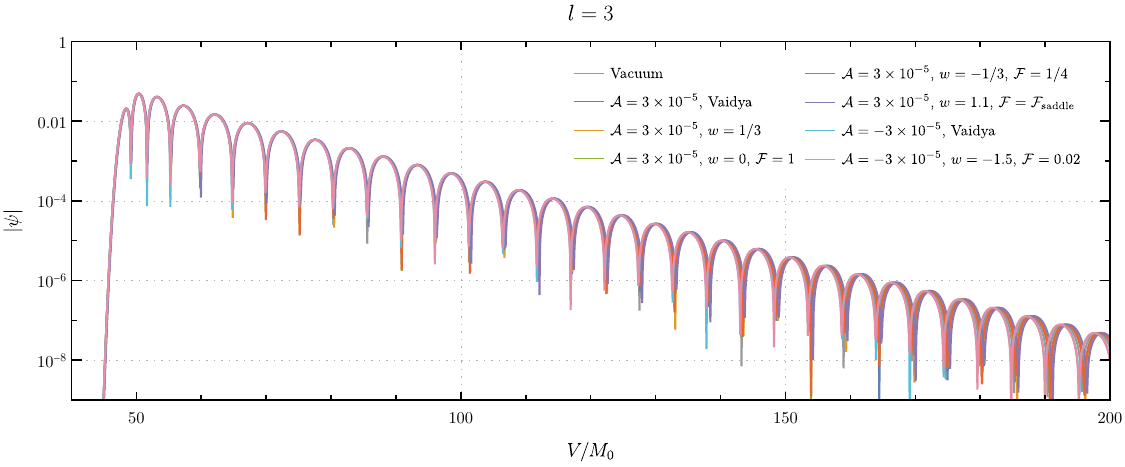}{0.938\linewidth}{
$V$ dependence of $\abs{\psi}$ at $r_\obs=20M_0$ for $l=3$. The cases of the vacuum BG, the ingoing linear Vaidya BG with positive/negative accretion rate, and several representative perfect fluid solutions are plotted simultaneously.
}
\fig{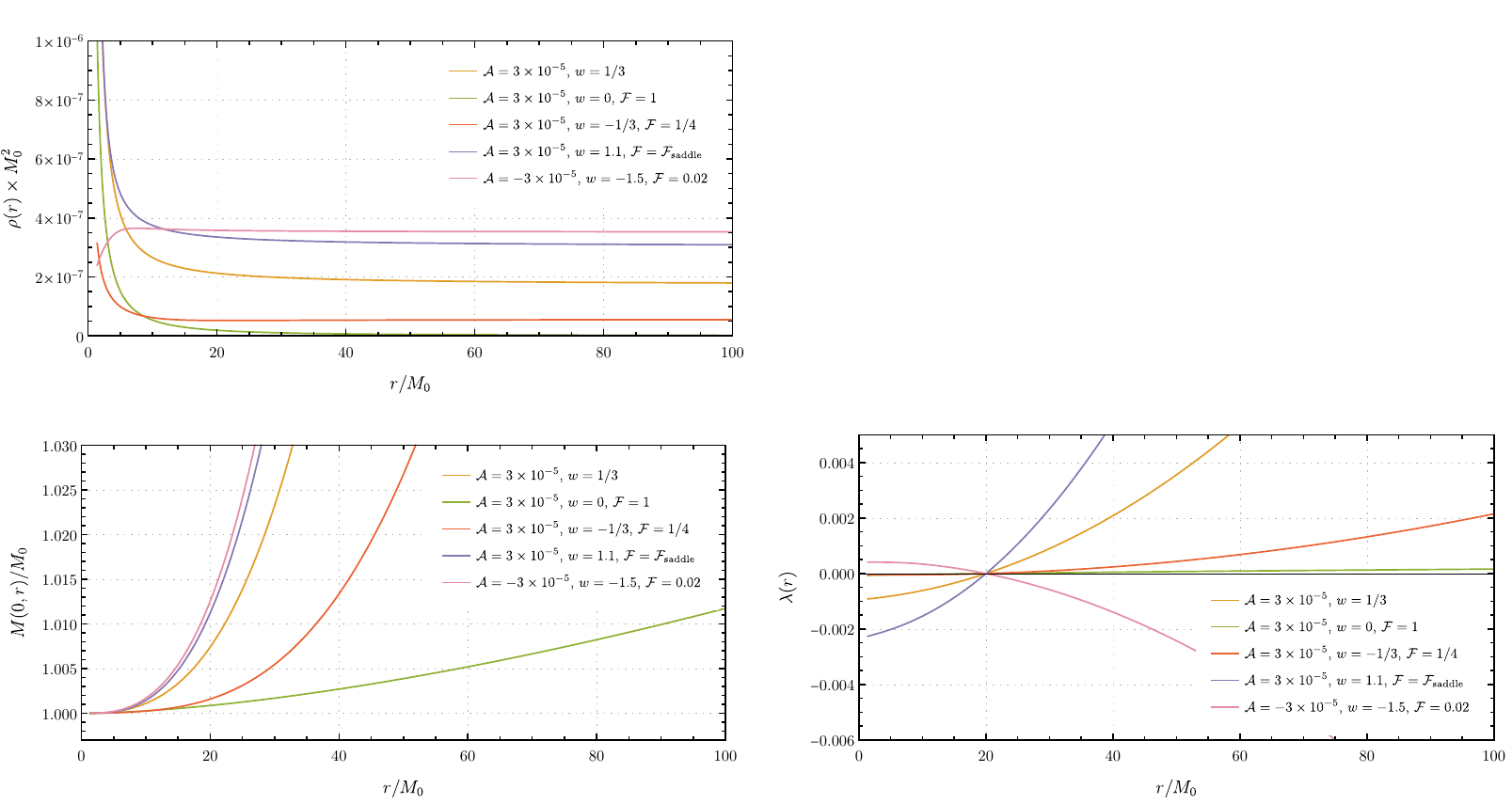}{\linewidth}{
BG functions $\rho$ (top), $M$ (bottom-left), and $\lambda$ (bottom-right) for $\abs{\scr{A}}=3\times 10^{-5}$ with respect to $r$. The asymptotic behavior at large $r$ follows TABLE~\ref{fluidinfinity}.
}
\par
Next, we examine the behavior of the frequency extracted from the waveform. Figure~\ref{omR_omI} plots the time evolution of the frequency for the same set of BGs as in Fig.~\ref{damp_l3}. One can observe how the frequency changes as the mass evolves.
\fig{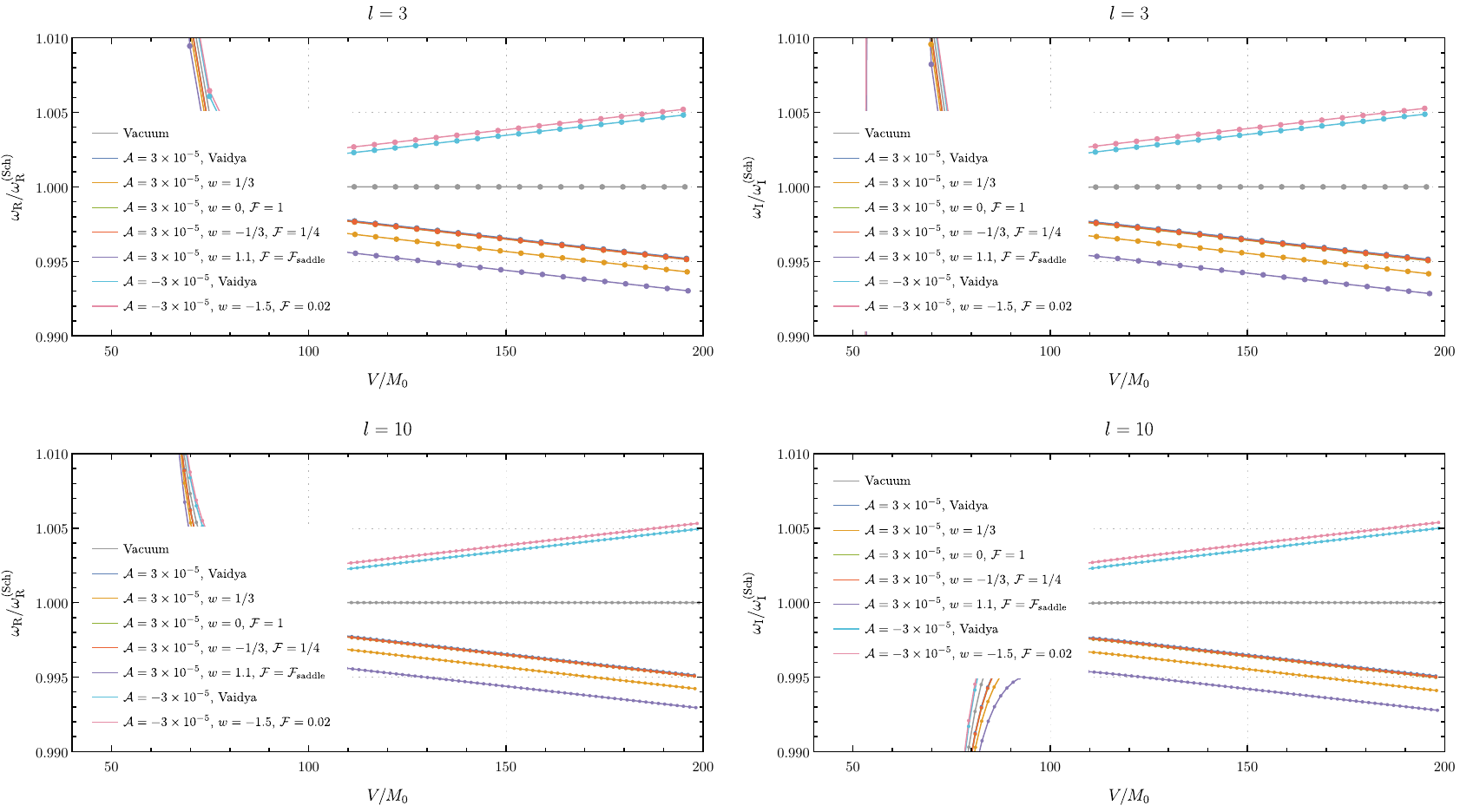}{\linewidth}{
Frequencies extracted from the waveform at $r_\obs=20M_0$ for $l=3$ (top) and $10$ (bottom). The real part (left) and imaginary part (right) are each divided by the fundamental mode value of the Schwarzschild QNM. Each point is computed from two adjacent peaks of $\abs{\psi}$, and the horizontal axis uses the midpoint time between them.
}
Figure~\ref{Xiplot} plots the time dependence of $\Xi$ based on these ratios. By taking the ratio, the mass dependence is canceled, and $\Xi$ becomes constant in the region where the fundamental mode is dominant. However, 
for a relatively small value of $l$, the results are contaminated by the 
influence of the tail part~\cite{Pricelaw,Capuano:2026tjy},
and we cannot extract a constant value of $\Xi$. 
In particular, since it is difficult to measure the mean value of $\Xi$ for $l=2$ with the accuracy we require, we exclude 
 this case 
from subsequent calculations.
\fig{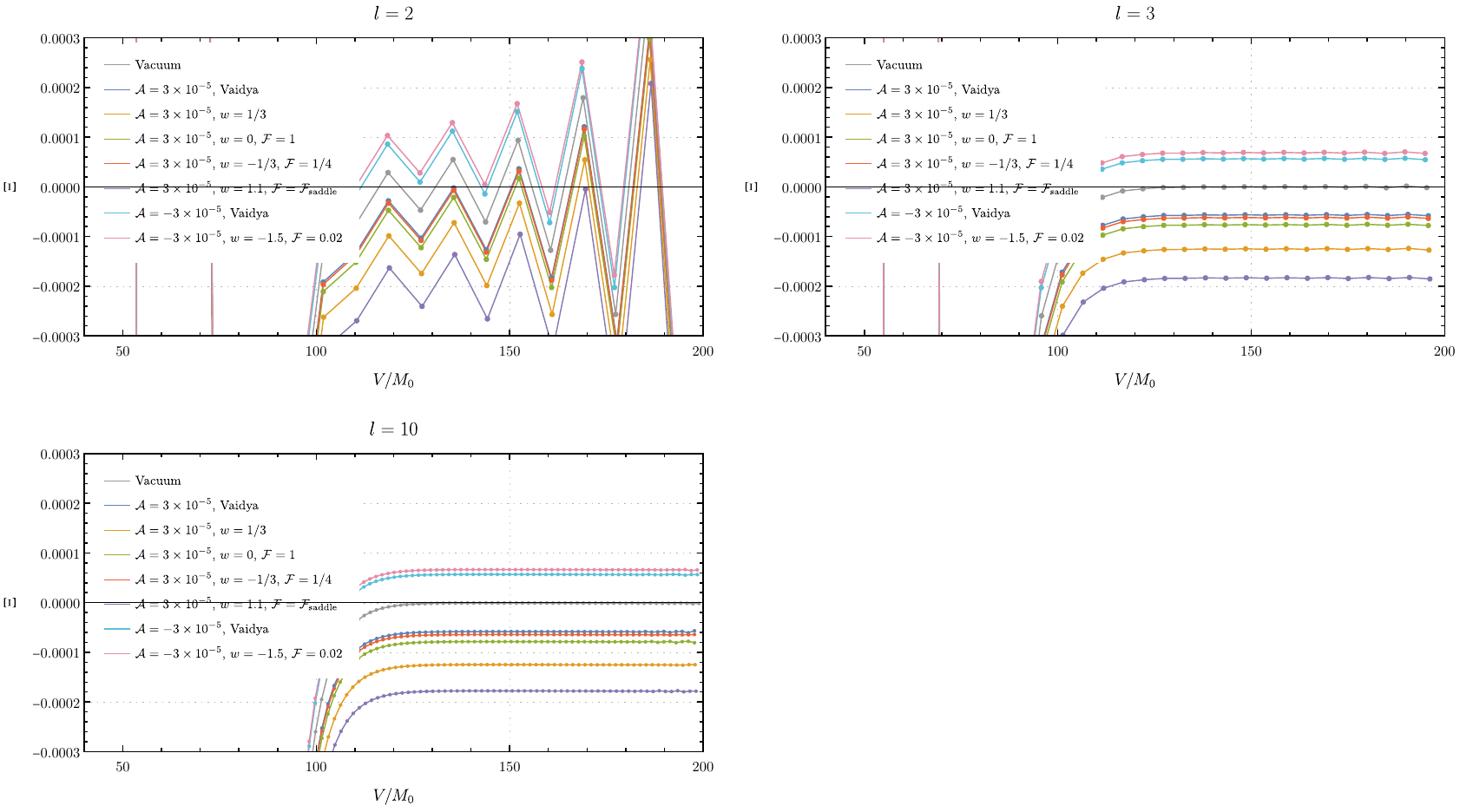}{\linewidth}{
Time dependence of $\Xi$ at $r_\obs=20M_0$ for $l=2$ (top-left), $3$ (top-right), and $10$ (bottom). The horizontal axis is defined in the same way as in Fig.~\ref{omR_omI}, and the fundamental mode values are used for $\omR^{(\Sch)}$ and $\omI^{(\Sch)}$. $\Xi$ converges to a constant of $O(\scr{A})$ over time. For $l=2$, the influence of the 
tail part
is significantly larger.
}
\par

Finally, Fig.~\ref{Atilplot} shows the behavior of $\tilde{\scr{A}}$ computed from these values of $\omega$. Apart from the repeated oscillation due
to the tail part, no significant difference is observed between the behavior of $\tilAR$ and $\tilAI$; both take constant mean values in the region where the fundamental mode is dominant, confirming that $\tilde{\scr{A}}\approx\scr{A}$.
\fig{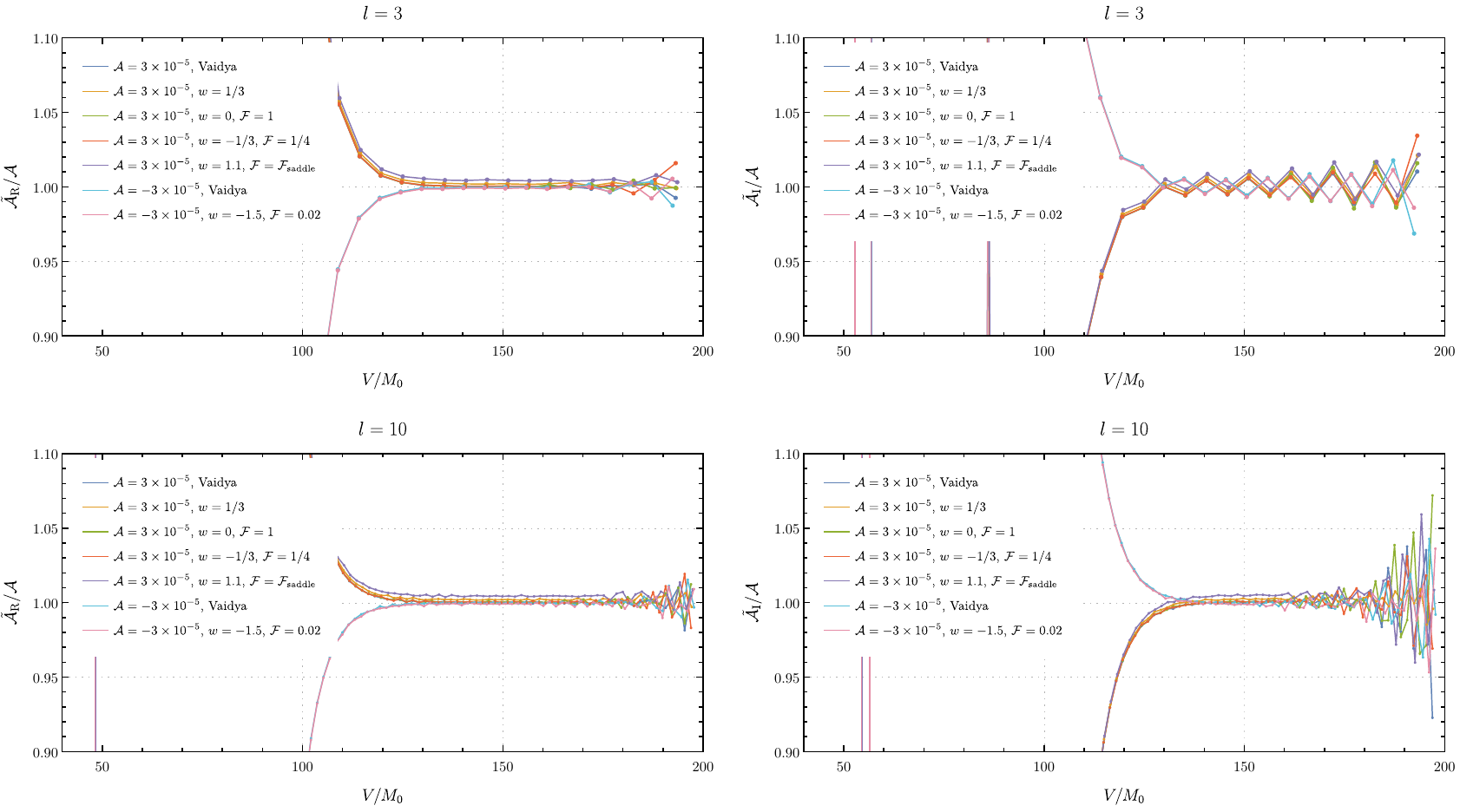}{\linewidth}{
Time dependence of $\tilAR/\scr{A}$ (left) and $\tilAI/\scr{A}$ (right) at $r_\obs=20M_0$ for $l=3$ (top) and $10$ (bottom). Each point is computed from two adjacent values of $\omega$, and the horizontal axis uses the midpoint time between them.
}

\subsection{Time average of $\Xi$ and $\tilde{\scr{A}}$}\label{6..TimeAverage}
We have found that the mean values of $\Xi$ and $\tilde{\scr{A}}$ 
 take approximately constant after a sufficiently long time. 
 To make our analysis more concrete, hereafter, we use their 
time-averaged values. 
Specifically, we compute $\Xi$ and $\tilde{\scr{A}}$ using peaks satisfying $V>V_{\text{peak}0}+100M_0$, where $V_{\text{peak}0}$ is the time of the first peak of the waveform, and define $\Xi$ and $\tilde{\scr{A}}$ as the average of the eight earliest values among these
 to eliminate the effects of the random numerical errors that grow with time. 
 For $\Xi$, in addition to the random numerical error, we found a systematic shift error that drifts in the negative direction over time. 
Figure~\ref{ShiftVacuum} plots the behavior of this shift in the vacuum BG. 
In particular, the shift error becomes dominant for large $l$. 
Such a shift can also be confirmed to appear similarly in nonvacuum cases. 
 Since this systematic error converges to zero at the limit of infinite resolution ($N_U=N_V=N\rightarrow \infty$), we prepare a run with $N$ halved and take the $N\rightarrow\infty$ limit assuming the second-order convergence, that is, its magnitude is inversely proportional to $N^2$. 
 Then, we estimate the extent of the numerical error by just calculating the standard deviation of the values taken at eight points. 
Note that for $l=3$, the standard deviation is
 remarkably larger due to the influence of the tail part. 

\fig{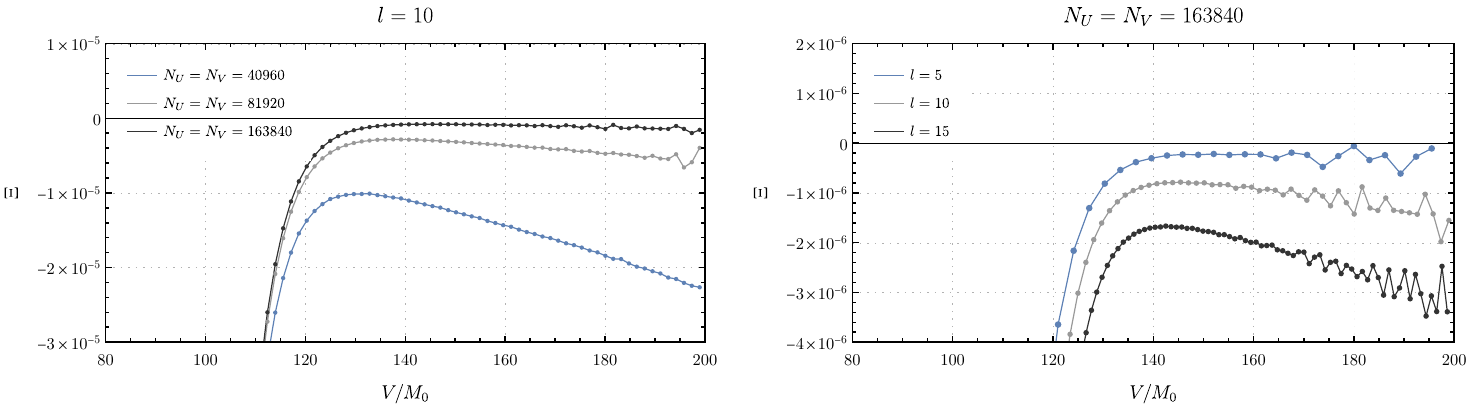}{\linewidth}{
Behavior of the shift in the vacuum case: varying the number of grid points at $l=10$ (left) and varying $l$ at the same number of grid points (right). The shift width is approximately inversely proportional to $N^2$, confirming that it originates from numerical error. The shift width also increases with larger $l$.
}
\par

We first examine the $\scr{A}$ dependence of $\Xi$ and $\tilde{\scr{A}}$. Since $\Xi=0$ in the vacuum BG, under the dilute approximation we expect
\eq{6B.1}{
\frac{\Xi}{\abs{\scr{A}}}=\frac{\const\times \scr{A}+\mathcal O(\kappa^2)}{\abs{\scr{A}}}=\const+O(\kappa)\,, 
}
where ``$\const$'' here means independent of the magnitude of $\scr{A}$.
Similarly, we also expect
\eq{6B.2}{
\frac{\tilde{\scr{A}}}{\scr{A}}=1+O(\kappa)\,.
}
In the present case, since $\delta M/\scr{A}\sim10^2$ and $\lambda/\scr{A}\sim10^2$, we can regard $\kappa\sim10^2\scr{A}$. The numerical results for representative cases are shown in Tables~\ref{DepofAl3} and \ref{DepofAl10}. For the vacuum BG, all values are consistent with zero, in agreement with the known Schwarzschild QNM. In the presence of accretion, 
 the numerical error in $\Xi/\abs{\scr{A}}$ 
 at $\abs{\scr{A}}=3\times 10^{-5}$ is at the few-percent level for $l=3$ and at the subpercent level for $l=10$, becoming about an order of magnitude larger at $\abs{\scr{A}}=3\times 10^{-6}$.
Comparing the values for $\abs{\scr{A}}=3\times 10^{-5}$ and $\abs{\scr{A}}=3\times 10^{-6}$, we find the value of $\Xi/\abs{\scr{A}}$ is sufficiently convergent at $\abs{\scr{A}}=3\times 10^{-5}$ as expected. 
The ratio $\tilde{\scr{A}}/\scr{A}$ also converges to $1$ in the limit $\scr{A}\rightarrow 0$, showing that the estimation using $\tilde{\scr{A}}$ is correct to first order in the accretion. 
These results suggest that for a BH with dilute accretion, 
from the measurement of 
 $\tilde{\scr{A}}$, in principle, one can extract the information of the accretion rate $\scr{A}$.  
\tabl{lcccccc}{DepofAl3}{
Time average of $\Xi$ and $\tilde{\scr{A}}$ and their $\scr{A}$ dependence at $l=3$ and $r_\obs=20M_0$. For the vacuum BG, $\scr{A}=0$, but the tabulated values are divided by $\abs{\scr{A}}$ of each column.
The extent of the numerical error is derived by calculating the standard deviation (see the text for details).
}{
        &\multicolumn{2}{c}{$\Xi/\abs{\scr{A}}$}    &\multicolumn{2}{c}{$\tilAR/\scr{A}$}   &\multicolumn{2}{c}{$\tilAI/\scr{A}$}\\
BG      &$\abs{\scr{A}}=3\times 10^{-5}$    &$\abs{\scr{A}}=3\times 10^{-6}$    &$\abs{\scr{A}}=3\times 10^{-5}$    &$\abs{\scr{A}}=3\times 10^{-6}$    &$\abs{\scr{A}}=3\times 10^{-5}$    &$\abs{\scr{A}}=3\times 10^{-6}$\\
\colrule
$\text{Vacuum}$ & $0.00\pm0.05$ & $0.0\pm0.5$ & $0.000\pm0.006$ & $0.00\pm0.06$ & $0.000\pm0.010$ & $0.00\pm0.10$ \\
$\text{Vaidya},\,\scr{A}=+\abs{\scr{A}}$ & $-1.88\pm0.04$ & $-1.9\pm0.4$ & $0.9995\pm0.0027$ & $1.00\pm0.07$ & $1.000\pm0.010$ & $1.01\pm0.19$ \\
$w = 1/3$ & $-4.17\pm0.04$ & $-4.2\pm0.4$ & $1.0017\pm0.0012$ & $0.988\pm0.033$ & $1.003\pm0.011$ & $1.00\pm0.12$ \\
$w = 0,\, \scr{F} = 1$ & $-2.54\pm0.04$ & $-2.6\pm0.4$ & $1.0003\pm0.0018$ & $1.00\pm0.06$ & $1.001\pm0.013$ & $1.01\pm0.08$ \\
$w = - 1/3,\, \scr{F} = 1/4$ & $-2.06\pm0.04$ & $-2.1\pm0.4$ & $1.002\pm0.006$ & $0.99\pm0.04$ & $1.003\pm0.014$ & $1.00\pm0.10$ \\
$w = 1.1,\, \scr{F} = \scr{F}_\text{saddle}$ & $-6.11\pm0.04$ & $-6.1\pm0.4$ & $1.0044\pm0.0015$ & $0.994\pm0.016$ & $1.006\pm0.011$ & $1.00\pm0.13$ \\
$\text{Vaidya},\,\scr{A}=-\abs{\scr{A}}$ & $1.89\pm0.04$ & $1.9\pm0.4$ & $0.999\pm0.005$ & $1.01\pm0.06$ & $0.998\pm0.015$ & $1.01\pm0.09$ \\
$w = - 1.5,\, \scr{F} = 0.02$ & $2.30\pm0.04$ & $2.3\pm0.4$ & $0.999\pm0.004$ & $0.98\pm0.06$ & $0.998\pm0.010$ & $0.97\pm0.18$
}
\tabl{lcccccc}{DepofAl10}{
Time average of $\Xi$ and $\tilde{\scr{A}}$ and their $\scr{A}$ dependence at $l=10$ and $r_\obs=20M_0$. For the vacuum BG, the same convention as in \tablref{DepofAl3} applies.
}{
        &\multicolumn{2}{c}{$\Xi/\abs{\scr{A}}$}    &\multicolumn{2}{c}{$\tilAR/\scr{A}$}   &\multicolumn{2}{c}{$\tilAI/\scr{A}$}\\
BG      &$\abs{\scr{A}}=3\times 10^{-5}$    &$\abs{\scr{A}}=3\times 10^{-6}$    &$\abs{\scr{A}}=3\times 10^{-5}$    &$\abs{\scr{A}}=3\times 10^{-6}$    &$\abs{\scr{A}}=3\times 10^{-5}$    &$\abs{\scr{A}}=3\times 10^{-6}$\\
\colrule
$\text{Vacuum}$ & $-0.0004\pm0.0024$ & $-0.004\pm0.024$ & $0.00001\pm0.00027$ & $0.0001\pm0.0027$ & $0.0004\pm0.0008$ & $0.004\pm0.008$ \\
$\text{Vaidya},\,\scr{A}=+\abs{\scr{A}}$ & $-1.9180\pm0.0025$ & $-1.924\pm0.022$ & $1.0001\pm0.0004$ & $0.999\pm0.004$ & $1.0005\pm0.0007$ & $1.004\pm0.010$ \\
$w = 1/3$ & $-4.1364\pm0.0027$ & $-4.131\pm0.026$ & $1.0020\pm0.0004$ & $1.000\pm0.006$ & $1.0024\pm0.0004$ & $1.002\pm0.007$ \\
$w = 0,\, \scr{F} = 1$ & $-2.5924\pm0.0024$ & $-2.598\pm0.026$ & $1.00032\pm0.00027$ & $1.000\pm0.007$ & $1.0007\pm0.0008$ & $1.0038\pm0.0033$ \\
$w = - 1/3,\, \scr{F} = 1/4$ & $-2.1170\pm0.0022$ & $-2.124\pm0.024$ & $1.0002\pm0.0004$ & $0.999\pm0.006$ & $1.00071\pm0.00030$ & $1.002\pm0.009$ \\
$w = 1.1,\, \scr{F} = \scr{F}_\text{saddle}$ & $-5.8978\pm0.0028$ & $-5.871\pm0.024$ & $1.00435\pm0.00030$ & $1.001\pm0.007$ & $1.0051\pm0.0006$ & $1.004\pm0.013$ \\
$\text{Vaidya},\,\scr{A}=-\abs{\scr{A}}$ & $1.9223\pm0.0028$ & $1.915\pm0.025$ & $0.99978\pm0.00032$ & $0.999\pm0.004$ & $0.9995\pm0.0011$ & $0.9964\pm0.0030$ \\
$w = - 1.5,\, \scr{F} = 0.02$ & $2.2420\pm0.0024$ & $2.239\pm0.026$ & $0.9991\pm0.0005$ & $1.000\pm0.005$ & $0.9985\pm0.0018$ & $0.996\pm0.009$
}
\par

Next, we investigate the $r_\obs$ dependence. The $r_\obs$ dependence of $\Xi$ and $\tilde{\scr{A}}$ is shown in Tables~\ref{DepofroXil3}, \ref{DepofroXil10}, \ref{DepofroAtill3}, and \ref{DepofroAtill10}. We find that $\tilde{\scr{A}}$ is independent of $r_\obs$, while $\Xi$ depends on it significantly. 
The origin of this dependence should be in the time-dependent BG geometry. 
This is because, 
when the frequency is time independent, 
the correction term for $r_\obs\rightarrow\infty$ calculated\footnote{In practice, this correction term cannot be calculated unless the potential term is of $O(r^{-2})$, but at least under the dilute approximation, it is expected to behave qualitatively in the same way.} as a time integral of $\psi\sim e^{-i\omega V}$ [see, e.g., Eq.~(5) in Ref.~\cite{InfCorrection}] only corrects 
the amplitude and phase of $\psi$, 
which does not affect the value of $\Xi$.
 Therefore, the time-dependent 
term $\scr{A}V$ is responsible for the $r_{\rm obs}$ dependence of $\Xi$. 
Since this term is independent of the EoS parameter, the $r_{\rm obs}$ dependence is expected to be EoS independent. Indeed, we find
that 
the $r_\obs$ dependence of
$\Xi/\abs{\scr{A}}$ depends only on $l$ and $\sgn{(\scr{A})}$ under our approximation and is independent of the fluid parameters. 
 The $r_\obs$ dependence exhibits remarkably interesting behavior, which may suggest that the influence of geometrical dynamics in regions far from the source on the observable quantity is rather nontrivial. Investigating these details would be important for comparison with observations, but due to the technical difficulties associated with the dilute approximation used in this study, we will leave this issue as a future problem. 
\tabl{lccccc}{DepofroXil3}{
$r_\obs$ dependence of the time average of $\Xi$ at $l=3$ and $\abs{\scr{A}}=3\times 10^{-5}$. For the vacuum BG, the same convention as in \tablref{DepofAl3} applies.
}{
        &\multicolumn{3}{c}{$\Xi/\abs{\scr{A}}$}    &\multicolumn{2}{c}{Difference of $\Xi/\abs{\scr{A}}$}\\
BG      &$r_\obs=10M_0$    &$r_\obs=20M_0$    &$r_\obs=30M_0$    &$(20M_0)-(10M_0)$    &$(30M_0)-(20M_0)$\\
\colrule
$\text{Vacuum}$ & $-0.002\pm0.005$ & $0.00\pm0.05$ & $0.00\pm0.12$ & $0.00\pm0.05$ & $0.00\pm0.13$ \\
$\text{Vaidya},\,\scr{A}=+\abs{\scr{A}}$ & $-1.060\pm0.007$ & $-1.88\pm0.04$ & $-2.23\pm0.13$ & $-0.82\pm0.04$ & $-0.35\pm0.14$ \\
$w = 1/3$ & $-3.344\pm0.006$ & $-4.17\pm0.04$ & $-4.52\pm0.13$ & $-0.82\pm0.04$ & $-0.36\pm0.13$ \\
$w = 0,\, \scr{F} = 1$ & $-1.718\pm0.008$ & $-2.54\pm0.04$ & $-2.90\pm0.13$ & $-0.82\pm0.04$ & $-0.36\pm0.13$ \\
$w = - 1/3,\, \scr{F} = 1/4$ & $-1.242\pm0.006$ & $-2.06\pm0.04$ & $-2.42\pm0.13$ & $-0.82\pm0.04$ & $-0.35\pm0.14$ \\
$w = 1.1,\, \scr{F} = \scr{F}_\text{saddle}$ & $-5.286\pm0.004$ & $-6.11\pm0.04$ & $-6.47\pm0.14$ & $-0.83\pm0.04$ & $-0.35\pm0.15$ \\
$\text{Vaidya},\,\scr{A}=-\abs{\scr{A}}$ & $1.064\pm0.005$ & $1.89\pm0.04$ & $2.25\pm0.12$ & $0.82\pm0.04$ & $0.36\pm0.13$ \\
$w = - 1.5,\, \scr{F} = 0.02$ & $1.482\pm0.009$ & $2.30\pm0.04$ & $2.66\pm0.12$ & $0.82\pm0.04$ & $0.36\pm0.13$
}
\tabl{lccccc}{DepofroXil10}{
$r_\obs$ dependence of the time average of $\Xi$ at $l=10$ and $\abs{\scr{A}}=3\times 10^{-5}$. For the vacuum BG, the same convention as in \tablref{DepofAl3} applies.
}{
        &\multicolumn{3}{c}{$\Xi/\abs{\scr{A}}$}    &\multicolumn{2}{c}{Difference of $\Xi/\abs{\scr{A}}$}\\
BG      &$r_\obs=10M_0$    &$r_\obs=20M_0$    &$r_\obs=30M_0$    &$(20M_0)-(10M_0)$    &$(30M_0)-(20M_0)$\\
\colrule
$\text{Vacuum}$ & $-0.0005\pm0.0024$ & $-0.0004\pm0.0024$ & $-0.0007\pm0.0022$ & $0.0000\pm0.0034$ & $-0.0003\pm0.0032$ \\
$\text{Vaidya},\,\scr{A}=+\abs{\scr{A}}$ & $-1.1863\pm0.0023$ & $-1.9180\pm0.0025$ & $-2.2306\pm0.0018$ & $-0.7317\pm0.0034$ & $-0.3126\pm0.0031$ \\
$w = 1/3$ & $-3.4027\pm0.0025$ & $-4.1364\pm0.0027$ & $-4.4502\pm0.0020$ & $-0.734\pm0.004$ & $-0.3138\pm0.0034$ \\
$w = 0,\, \scr{F} = 1$ & $-1.8607\pm0.0023$ & $-2.5924\pm0.0024$ & $-2.9051\pm0.0016$ & $-0.7317\pm0.0033$ & $-0.3128\pm0.0028$ \\
$w = - 1/3,\, \scr{F} = 1/4$ & $-1.3854\pm0.0023$ & $-2.1170\pm0.0022$ & $-2.4300\pm0.0023$ & $-0.7317\pm0.0032$ & $-0.3129\pm0.0032$ \\
$w = 1.1,\, \scr{F} = \scr{F}_\text{saddle}$ & $-5.1612\pm0.0031$ & $-5.8978\pm0.0028$ & $-6.2129\pm0.0021$ & $-0.737\pm0.004$ & $-0.315\pm0.004$ \\
$\text{Vaidya},\,\scr{A}=-\abs{\scr{A}}$ & $1.1933\pm0.0024$ & $1.9223\pm0.0028$ & $2.2338\pm0.0020$ & $0.729\pm0.004$ & $0.3115\pm0.0035$ \\
$w = - 1.5,\, \scr{F} = 0.02$ & $1.5137\pm0.0024$ & $2.2420\pm0.0024$ & $2.5527\pm0.0021$ & $0.7283\pm0.0034$ & $0.3107\pm0.0032$
}
\tabl{lcccccc}{DepofroAtill3}{
$r_\obs$ dependence of the time average of $\tilde{\scr{A}}$ at $l=3$ and $\abs{\scr{A}}=3\times 10^{-5}$. For the vacuum BG, the same convention as in \tablref{DepofAl3} applies.
}{
        &\multicolumn{3}{c}{$\tilAR/\scr{A}$}    &\multicolumn{3}{c}{$\tilAI/\scr{A}$}\\
BG      &$r_\obs=10M_0$    &$r_\obs=20M_0$    &$r_\obs=30M_0$    &$r_\obs=10M_0$    &$r_\obs=20M_0$    &$r_\obs=30M_0$\\
\colrule
$\text{Vacuum}$ & $-0.002\pm0.005$ & $0.000\pm0.006$ & $0.002\pm0.009$ & $-0.002\pm0.004$ & $0.000\pm0.010$ & $0.00\pm0.04$ \\
$\text{Vaidya},\,\scr{A}=+\abs{\scr{A}}$ & $0.998\pm0.008$ & $0.9995\pm0.0027$ & $1.0006\pm0.0024$ & $0.999\pm0.006$ & $1.000\pm0.010$ & $1.00\pm0.04$ \\
$w = 1/3$ & $1.0010\pm0.0032$ & $1.0017\pm0.0012$ & $1.002\pm0.005$ & $1.0012\pm0.0023$ & $1.003\pm0.011$ & $1.01\pm0.04$ \\
$w = 0,\, \scr{F} = 1$ & $1.0009\pm0.0021$ & $1.0003\pm0.0018$ & $1.0006\pm0.0015$ & $1.001\pm0.004$ & $1.001\pm0.013$ & $1.00\pm0.04$ \\
$w = - 1/3,\, \scr{F} = 1/4$ & $1.002\pm0.006$ & $1.002\pm0.006$ & $1.0002\pm0.0028$ & $1.002\pm0.006$ & $1.003\pm0.014$ & $1.00\pm0.04$ \\
$w = 1.1,\, \scr{F} = \scr{F}_\text{saddle}$ & $1.0044\pm0.0024$ & $1.0044\pm0.0015$ & $1.003\pm0.006$ & $1.0045\pm0.0026$ & $1.006\pm0.011$ & $1.01\pm0.04$ \\
$\text{Vaidya},\,\scr{A}=-\abs{\scr{A}}$ & $1.002\pm0.008$ & $0.999\pm0.005$ & $0.9993\pm0.0035$ & $1.001\pm0.007$ & $0.998\pm0.015$ & $1.00\pm0.04$ \\
$w = - 1.5,\, \scr{F} = 0.02$ & $0.9993\pm0.0020$ & $0.999\pm0.004$ & $0.9988\pm0.0011$ & $0.999\pm0.004$ & $0.998\pm0.010$ & $1.00\pm0.04$
}
\tabl[\footnotesize]{lcccccc}{DepofroAtill10}{
$r_\obs$ dependence of the time average of $\tilde{\scr{A}}$ at $l=10$ and $\abs{\scr{A}}=3\times 10^{-5}$. For the vacuum BG, the same convention as in \tablref{DepofAl3} applies.
}{
        &\multicolumn{3}{c}{$\tilAR/\scr{A}$}    &\multicolumn{3}{c}{$\tilAI/\scr{A}$}\\
BG      &$r_\obs=10M_0$    &$r_\obs=20M_0$    &$r_\obs=30M_0$    &$r_\obs=10M_0$    &$r_\obs=20M_0$    &$r_\obs=30M_0$\\
\colrule
$\text{Vacuum}$ & $0.00003\pm0.00021$ & $0.00001\pm0.00027$ & $-0.00002\pm0.00025$ & $0.00029\pm0.00029$ & $0.0004\pm0.0008$ & $0.0000\pm0.0014$ \\
$\text{Vaidya},\,\scr{A}=+\abs{\scr{A}}$ & $1.0001\pm0.0008$ & $1.0001\pm0.0004$ & $1.00006\pm0.00033$ & $1.0005\pm0.0007$ & $1.0005\pm0.0007$ & $1.0003\pm0.0007$ \\
$w = 1/3$ & $1.0020\pm0.0005$ & $1.0020\pm0.0004$ & $1.00192\pm0.00029$ & $1.0024\pm0.0006$ & $1.0024\pm0.0004$ & $1.0023\pm0.0008$ \\
$w = 0,\, \scr{F} = 1$ & $1.0003\pm0.0006$ & $1.00032\pm0.00027$ & $1.0003\pm0.0006$ & $1.0007\pm0.0007$ & $1.0007\pm0.0008$ & $1.0007\pm0.0007$ \\
$w = - 1/3,\, \scr{F} = 1/4$ & $1.0002\pm0.0004$ & $1.0002\pm0.0004$ & $1.0003\pm0.0006$ & $1.0006\pm0.0004$ & $1.00071\pm0.00030$ & $1.0004\pm0.0015$ \\
$w = 1.1,\, \scr{F} = \scr{F}_\text{saddle}$ & $1.00442\pm0.00019$ & $1.00435\pm0.00030$ & $1.0044\pm0.0004$ & $1.0051\pm0.0004$ & $1.0051\pm0.0006$ & $1.0052\pm0.0013$ \\
$\text{Vaidya},\,\scr{A}=-\abs{\scr{A}}$ & $0.9997\pm0.0006$ & $0.99978\pm0.00032$ & $0.9999\pm0.0005$ & $0.9994\pm0.0005$ & $0.9995\pm0.0011$ & $0.9996\pm0.0013$ \\
$w = - 1.5,\, \scr{F} = 0.02$ & $0.99914\pm0.00030$ & $0.9991\pm0.0005$ & $0.9992\pm0.0004$ & $0.9988\pm0.0004$ & $0.9985\pm0.0018$ & $0.9986\pm0.0016$
}

\subsection{Dependence on fluid parameters}\label{6..FluParam}
Finally, we investigate the relationship between the fluid parameters $w$, $\scr{F}$ and $\Xi/\abs{\scr{A}}$. Figure~\ref{Xiw0to1p5} plots $\Xi/\abs{\scr{A}}$ with $\scr{F}=\scr{F}_{\text{saddle}}$ held fixed. We find that $\Xi/\abs{\scr{A}}$ is monotonically decreasing in $w$. In particular, for $0<w\le1$, $\Xi/\abs{\scr{A}}$ is determined solely by $w$, and they are in one-to-one correspondence. Regarding the $l$ dependence, $\Xi/\abs{\scr{A}}$ looks to converge to a certain curve in the geometric optics limit $l\rightarrow\infty$, while for small $l$ there is a finite deviation from this curve. 
 The existence of the $l$ dependence suggests that, in principle, by observing the $l$ dependence, we may obtain further information. 
\fig{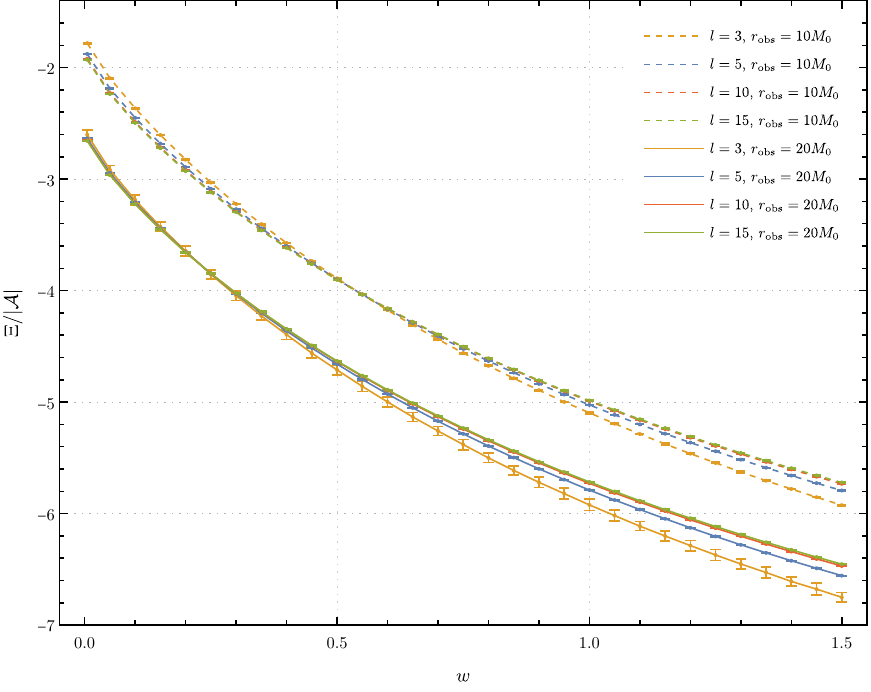}{0.7\linewidth}{
$w$ dependence of $\Xi$ for $w>0$, $\scr{F}=\scr{F}_{\text{saddle}}$, and $\scr{A}=3\times 10^{-5}$. 
We present the results for $l=3$, $5$, $10$, and $15$ at $r_\obs=10M_0$ and $20M_0$.
The curves asymptote to a certain curve as $l\rightarrow\infty$.
The intersection points of lines for different $l$ depend on $r_\obs$ and have no physical significance.
The error bar denotes the extent of the numerical error.
}
\par

For $w<0$, 
to maintain the dilute approximation, we fix $\rho_\infty/\abs{\scr{A}}\coloneq \rho(r\rightarrow\infty)/\abs{\scr{A}} \propto(1+w)^{-1}\scr{F}^{-1/w}$ instead of fixing $\scr{F}$, and vary $w$. 
 Then $\scr{F}$ depends on $w$ as explicitly shown in Fig.~\ref{Fbyrhoinf}.
\fig{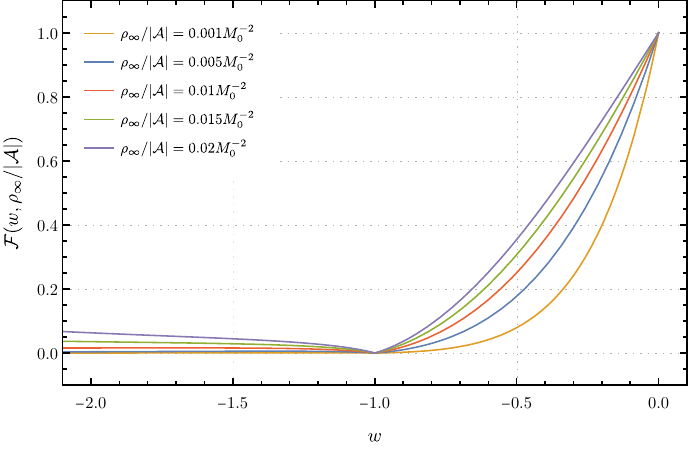}{0.55\linewidth}{
$w$ dependence of $\scr{F}=\scr{F}(w,\rho_\infty/\abs{\scr{A}})$ with fixed $\rho_\infty/\abs{\scr{A}}$.
}
 Here, it should be noted that, in our prescription, the sign of $\scr{A}$ flips at $w=-1$
, so $\Xi/\abs{\scr{A}}$ is discontinuous at $w=-1$. 
 This expected behavior can be confirmed in Fig.~\ref{Xiw-2to0}, in which the value of $\Xi/\abs{\scr{A}}$ is plotted as a function of $w$ for each value of $\rho_\infty/\abs{\scr{A}}$. 
 From Fig.~\ref{Xiw-2to0}, we find that $\Xi/\abs{\scr{A}}$ approaches the Vaidya BG values as $\rho_\infty/\abs{\scr{A}}$ decreases. 
 This is because the BG metric approaches the Vaidya metric as $\rho_\infty/\abs{\scr{A}}$ decreases [see Eq.~\eqref{44.12}]. 
In the limits $w\rightarrow-1\pm0$ and $w\rightarrow-0$, $\scr{F}$ takes a constant value independent of $\rho_\infty/\abs{\scr{A}}$. 
In the limit $w\rightarrow-1\pm0$, $\Xi/\abs{\scr{A}}$ asymptotes to the Vaidya--de Sitter (VdS) BG values with 
 $\Lambda=8\pi \rho_\infty$ and the accretion rate $\scr{A}=\pm\abs{\scr{A}}$. 
Indeed, from Eqs.~\eqref{3C.1}, \eqref{45.13}, \eqref{44.13}, and \eqref{45.3},
\eqs{6C.1}{
\rho &\rightarrow \rho_\infty=\const\,,\\
u &\rightarrow \infty\,,\\
\lambda &\rightarrow 0\,,
} which is precisely the VdS spacetime. The deviation due to differences in $\rho_\infty/\abs{\scr{A}}$ at $w\rightarrow-1\pm0$ becomes particularly pronounced for small $l$. 
Conversely, in the limit of $l\rightarrow\infty$, the VdS BG and the Vaidya BG become indistinguishable.
On the other hand, in the limit $w\rightarrow-0$, 
 the value of $\scr{F}$ approaches 1 as shown in Fig.~\ref{Fbyrhoinf}, and consistently, 
$\Xi/\abs{\scr{A}}$ always approaches the value of the dust solution with $w=0$ and $\scr{F}=1$.
\fig{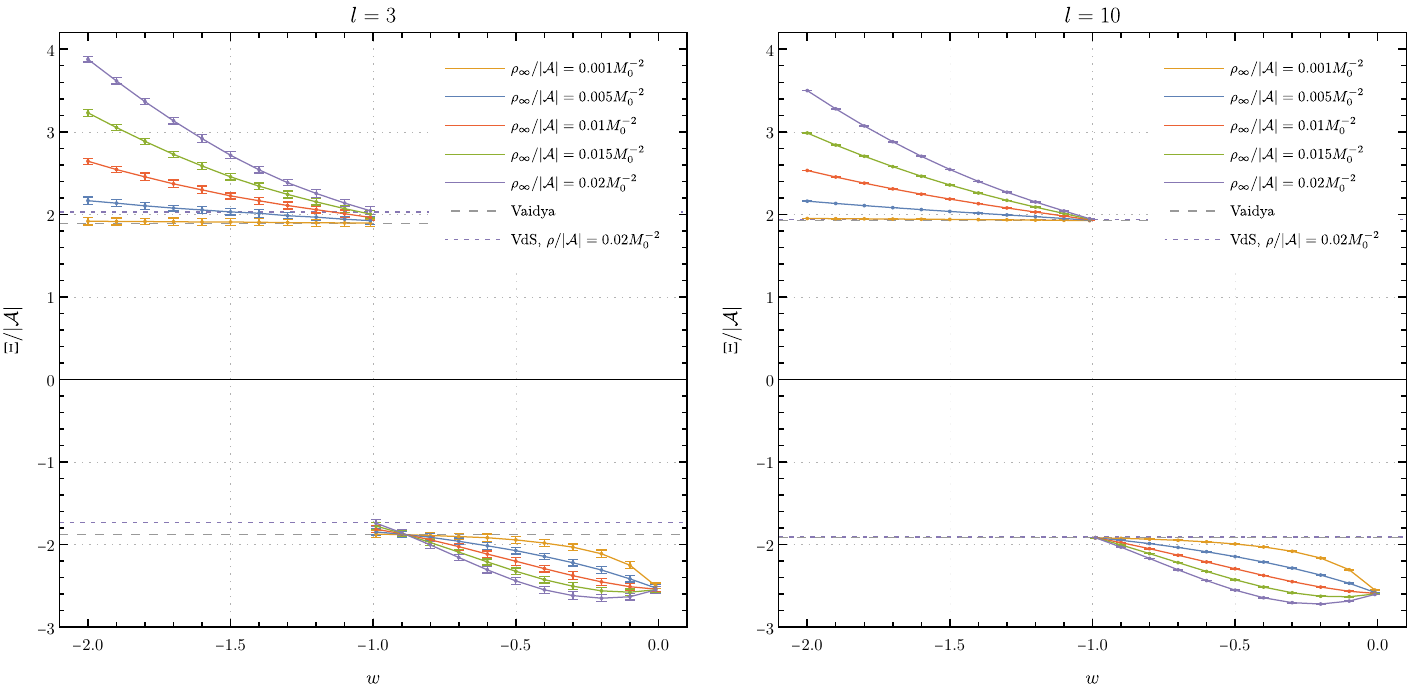}{\linewidth}{
$w$- and $\rho_\infty/\abs{\scr{A}}$ dependence of $\Xi$ for $-2\le w<0$, $r_\obs=20M_0$, and $\abs{\scr{A}}=3\times 10^{-5}$ in the $l=3$ (left) and $10$ (right) cases. For comparison, the values for the ingoing linear Vaidya BG with $\scr{A}=\pm3\times 10^{-5}$ and the ingoing linear VdS BG with $\scr{A}=\pm3\times 10^{-5}$ and $\rho/\abs{\scr{A}}=0.02M_0^{-2}$ are also plotted. For small $l$, the deviation due to $\rho_\infty/\abs{\scr{A}}$ at $w\rightarrow -1\pm0$ becomes pronounced.
}
\par
An important fact revealed by the present computation is that the absolute value $\abs{\Xi}/\abs{\scr{A}}$ is larger than its value in the Vaidya BG regardless of the fluid parameters, apart from a few exceptions. The same holds for all the other solutions presented in Appendix~\ref{NB.2param}.
In particular, focusing on the large-$l$ limit, this relation holds without exception. That is, under our assumptions, for any BG satisfying the weak energy condition ($w>-1$), the inequality
\eq{6C.3}{
\frac{\Xi}{\abs{\scr{A}}}\le -\abs{\frac{\Xi}{\scr{A}}}^{\(\text{Vaidya}\)}
}
always holds, and any $\Xi/\abs{\scr{A}}$ satisfying
\eq{6C.4}{
-\abs{\frac{\Xi}{\scr{A}}}^{\(\text{Vaidya}\)}< \frac{\Xi}{\abs{\scr{A}}}< \abs{\frac{\Xi}{\scr{A}}}^{\(\text{Vaidya}\)}
}
cannot be explained within our model.
These relations can also be interpreted by comparison with 
the analytical result in static BGs~\cite{Ariadna2026}. 
Let us assume that the value of $\Xi$ can be divided into the contribution $\Xi^{(\text{Vaidya})}$ from the dynamical term $\scr{A}V$ and the contribution $\Xi^{(\text{static})}$ from the remaining static terms in the metric form \eqref{S3B.2} as follows:
\eq{6C.5}{
\Xi=\Xi^{\(\text{Vaidya}\)}+\Xi^{\(\text{static}\)}+O(\kappa^2)\,.
}
Equation~(2.44) of Ref.~\cite{Ariadna2026} suggests that, for a static spherically symmetric BH with a perfect fluid satisfying $p=w\rho$ and $\rho>0$, in the limit of $l\rightarrow\infty$,
\eq{6C.6}{
\sgn\( \Xi^{\(\text{static}\)} \)=-\sgn(1+w)
}
holds, which is consistent with our results.\footnote{Note that, since $u\ne 0$ in our BG, Eq.~(2.44) of Ref.~\cite{Ariadna2026} does not hold as it stands, but Eq.~\eqref{6C.6} holds in this case.}
\par 

Finally, Fig.~\ref{XiDiffw-2to0} shows the differences in $\Xi/\abs{\scr{A}}$ between $l=4$ modes and $l=10$ modes for $w<0$. These differences also depend on $w$ and $\scr{F}$, and their measurement provides an additional constraint that differs from those obtained from the measurement of $\Xi/\abs{\scr{A}}$ itself.\par
\fig{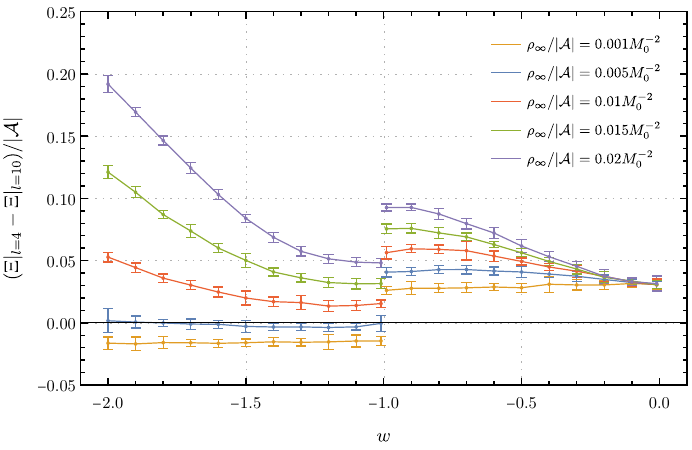}{0.575\linewidth}{
$w$- and $\rho_\infty/\abs{\scr{A}}$ dependence of $\Xi|_{l=4}-\Xi|_{l=10}$ for $-2\le w<0$, $r_\obs=20M_0$, and $\abs{\scr{A}}=3\times 10^{-5}$.
}
The basic properties of the other combinations of $w$ and $\scr{F}$, which are not discussed in this Subsection, are essentially the same, and they are presented in Appendix~\ref{NB.2param} to avoid redundancy.

\section{Conclusion and prospects}\label{7.Conclusion}
In this work, we 
 considered spherically symmetric BH solutions with dilute perfect fluid accretion, and analyzed the odd-parity master equation in DN coordinates. Then, we reconfirmed that the metric perturbation and the matter perturbation decouple for odd-parity perturbations. We used the DNF to derive the ringdown gravitational waveform in the time domain, and showed that in practice it is sufficient to solve the coupled differential equations for $r$ and $\psi$.\par

We showed that $\Xi$, defined by the deviation of the ratio of the real and imaginary parts of the frequency from that of the Schwarzschild spacetime,
is independent of time (in the region where the fundamental mode is dominant) and proportional to the accretion rate. On the other hand, we showed that $\tilde{\scr{A}}$, defined by the time dependence of the frequency, always agrees with the accretion rate $\scr{A}$ to first order of the dilute approximation of steady accretion. 
This makes it possible to obtain information about the matter distribution around the BH from the measurement of $\Xi/\abs{\scr{A}}$. In particular, when a perfect fluid with $0<w\le 1$ is assumed, the value of $w$ can be completely determined. It also 
turned out that even in a multiparameter case such as $w>1$ or $w\le 0$, an additional constraint can be obtained from the difference between $\Xi/\abs{\scr{A}}$ at small $l$ and at large $l$, but the details of that constraint are left for future work. 
In addition, the frequencies of the overtone modes, which were not considered in this work, may contain further information and constitute an interesting topic (see, e.g., Ref.~\cite{Berti:2025hly} and references therein). 
We also found that the value of $\Xi/\abs{\scr{A}}$ exhibits a jump across $w=-1$, and in the large-$l$ limit, $\abs{\Xi/\scr{A}}$ is always larger than $\abs{\Xi/\scr{A}}^{\(\text{Vaidya}\)}$. 

These results, however, assume globally 
steady accretion, which implies that the spacetime is not asymptotically flat. 
Therefore, all of our results concerning $\Xi$ are limited to describing its qualitative behavior, 
and the more realistic observed values at future null infinity may be shifted by a constant that differs for each $l$. 
In addition, for general nonsteady accretion, $\tilde{\scr{A}}\simeq\scr{A}$ does not hold, as shown in Appendix~\ref{C.cutoff}. 
Conversely, this result indicates that the possible differences between $\tilde{\scr{A}}$ and $\scr{A}$ may contain information about the matter distribution at intermediate distances that cannot be obtained from the measurement of $\Xi$ alone. 
 Therefore, it would be worthwhile to investigate methods for extracting such information from observations.
To address these problems,
 we need to construct more realistic asymptotically flat BG
 solutions that satisfy the Einstein equation everywhere and investigate ringdown waves in those spacetimes. 

We assumed that the accretion order $\kappa$ and the ringdown GW order $\ep$ satisfy $1\gg\kappa\gg\ep$. However, the effect of order $\kappa\ep$ in this case is very small, and it is more practical to perform actual measurements in the regime satisfying $1\gg\kappa\sim\ep$. In that case, the effect of order $\ep^2$ needs to be taken into account (see, e.g., Refs.~\cite{Ioka:2007ak,NakanoIoka}). 
We also note that, especially for small $l$, the influence of the 
tail part becomes non-negligible at a relatively early time. 
Conversely, how matter accretion affects the behavior of the 
tail part is an interesting open problem (see, e.g., \cite{Capuano:2026tjy}). 
 Needless to say, further research extending this approach to even-parity modes should be conducted in the future.

\begin{acknowledgments}
This work was financially supported by JST SPRING, Grant No. JPMJSP2125. R.O. would like to take this opportunity to thank the ``THERS Make New Standards Program for the Next Generation Researchers.'' 
This work was supported by JSPS KAKENHI Grants No. JP25K07281 (C.Y.), No. JP24K07027 (C.Y.), No. JP26K07074 (H.N.), No. JP25K17396 (K.O.), and No. JP23KK0048 (Y.K.).
H.N. also would like to acknowledge the valuable support of the Research Centre for Relational Studies of Ryukoku University.

\end{acknowledgments}

\appendix
\section{Numerical methods}\label{A.numerical}
\subsection{Double null formalism}\label{5..DNF}
The numerical method used in this work, the DNF, is known as a scheme that can solve the master equation on DN coordinates with second-order accuracy~\cite{DNF1}. 
In general, when solving an equation with the form of Eq.~\eqref{61.0},
it is sufficient to have the initial values of $x$ on $U=U_0,\;V_0\le V\le V_{\max}$ and $V=V_0,\;U_0\le U\le U\maxx$, together with a method to obtain $\xv{4}$ from the values $x\rvert_1,\,\xv{2}$, and $\xv{3}$ at grid points 1, 2, and 3 shown in Fig.~\ref{kousitenn}. \par
First, we write the various quantities at point 0 as
\subeq{S5A.1}{
\eq{61.1}{
x\dUV\rvert_0=\frac{\xv{4}-\xv{3}-\xv{2}+\xv{1}}{\Delta U\Delta V}+O(\Delta^2) \,,
}\eq{61.2}{
x\dU\rvert_0=\frac{\xv{3}-\xv{1}+\xv{4}-\xv{2}}{2\Delta U}+O(\Delta^2) \,,\qquad
x\dV\rvert_0=\frac{\xv{2}-\xv{1}+\xv{4}-\xv{3}}{2\Delta V}+O(\Delta^2) \,,
}\eq{61.3}{
\xv{0}=\frac{\xv{2}+\xv{3}}{2}+O(\Delta^2)\,.
}}
Here, the grid points are placed at intervals of $\Delta U$ and $\Delta V$ in the $U$ and $V$ directions, respectively, and $\Delta U$ and $\Delta V$ are of $O(\Delta)$.
Next, substituting these into Eq.~\eqref{61.0} yields an equation for $\xv{4}$. However, to avoid a nonlinear algebraic equation, we replace Eq.~\eqref{61.2} with
\eq{61.4}{
x\dU\rvert_0=\frac{\xv{3}-\xv{1}}{\Delta U}+O(\Delta) \,,\qquad
x\dV\rvert_0=\frac{\xv{2}-\xv{1}}{\Delta V}+O(\Delta)\,,
}
and then substituting into Eq.~\eqref{61.0}, which gives
\eq{61.5}{
\tilde{x}\rvert_4=\xv{3}+\xv{2}-\xv{1}+ F(\xv{0},x\dU\rvert_0,x\dV\rvert_0)\Delta U\Delta V+O(\Delta^3)\,.
}
Replacing $\xv{4}$ in Eq.~\eqref{61.2} with $\tilde{x}\rvert_4$ and substituting back into Eq.~\eqref{61.0}, we obtain
\eq{61.6}{
x\rvert_4=\xv{3}+\xv{2}-\xv{1}+ F(\xv{0},x\dU\rvert_0,x\dV\rvert_0)\Delta U\Delta V+O(\Delta^4)\,.
}
Since the number of required steps is inversely proportional to $\Delta U\Delta V$, the final error is $O(\Delta^2)$.
\fig{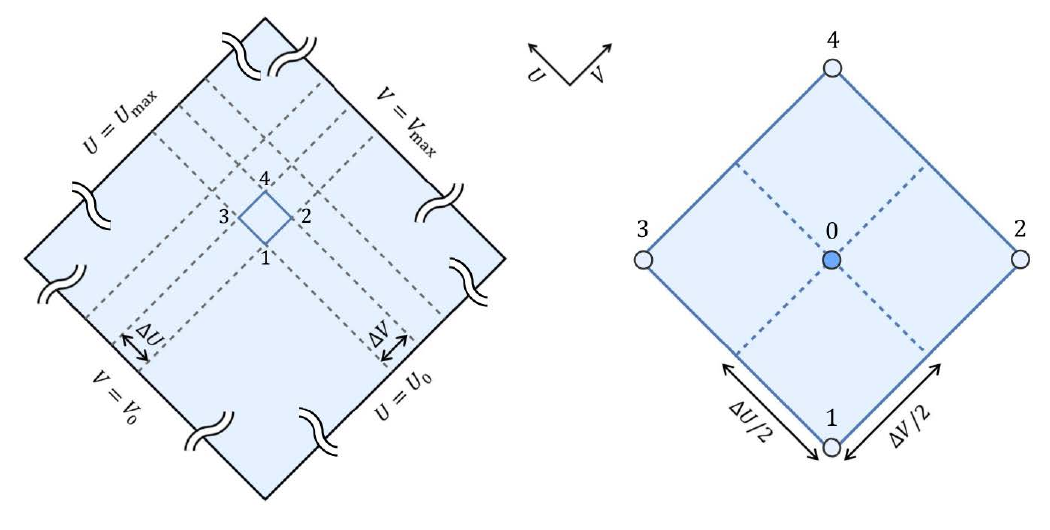}{14.5cm}{
Layout of grid points. The blue-shaded region in the left panel represents the computational domain. Grid points are placed at intervals of $\Delta U$ and $\Delta V$ in the $U$ and $V$ directions, respectively. The blue quadrilateral connects four adjacent grid points, whose bottom, right, left, and top vertices are labeled $1,\,2,\,3$, and $4$, respectively. The right panel is an enlarged view of this quadrilateral, and point $0$ is an auxiliary point taken at its center.
}

\subsection{
Treatment for the 
$U$ coordinate}\label{5..Ucoord}
It is known that the DNF suffers from a problem in which numerical errors become uncontrollable in finite time when the computation is started near the horizon~\cite{DNFproblem}. For simplicity, let us consider this problem in the Schwarzschild BG. Near the horizon $r\simeq 2M_0$,
\eq{63.1}{
r^*\simeq 2M_0\ln\(\frac{r-2M_0}{2M_0}\)\,,
}
and solving for $r$ gives
\eq{63.2}{
r\simeq 2M_0\tk{1+\exp\(\frac{r^*}{2M_0}\)}\,.
}
Using the fact that one can write\footnote{The freedom of the $V$ coordinate has already been fixed in Eq.~\eqref{S3B.2}.} $U=\alpha(t-r^*)$ and $V=t+r^*$ in Schwarzschild spacetime, we have
\eq{63.3}{
r^*=\rec{2}\(V-\alpha^{-1}(U)\)\,.
}
Substituting the above into Eq.~\eqref{63.2}, we obtain
\eq{63.4}{
r\simeq 2M_0\tk{1+\exp\(\frac{V-\alpha^{-1}(U)}{4M_0}\)}\,.
}
Examining the $V$ dependence of the difference $\delta r(V)$ in $r$ between adjacent grid points on $V=\const$, we find
\eqs{63.5}{
\delta r(V)&\simeq \delta r(V_0)\exp\(\frac{V-V_0}{4M_0}\) \,,\\
\delta r(V_0)&= 2M_0\exp\(\frac{V_0}{4M_0}\)\abs{\exp\(-\frac{\alpha^{-1}(U)}{4M_0}\)-\exp\(-\frac{\alpha^{-1}(U+\Delta U)}{4M_0}\)}\,.
}
Equation~\eqref{63.5} shows that the grid spacing grows exponentially with increasing $V$ near the horizon.\par
To avoid this problem, we adopt a method in which the $U$ coordinate is redefined at every step, as illustrated in Fig.~\ref{Utorinaosi}. Specifically, after obtaining $\,x(U,V_{n})\,$ from $x(U,V_{n-1})$ and $x(U_0,V_{n})$ for the $n$th step $V=V_n$, we redefine the $U$ coordinate by
\eq{63.10}{
r(U,V_{n})=r\maxx(V_n)-\(r\maxx(V_n)-r\minn\)\frac{U-U_0}{U\maxx-U_0}\,.
}
Here, $r\maxx$ is taken on $U=U_0$ as in Fig.~\ref{kousitenn}, while 
$r\minn$ is fixed at a point sufficiently inside $r=2M_0$. 
The initial $U$ coordinate is obtained by substituting $n=0$ into Eq.~\eqref{63.10}. The value of $x$ at the new grid point $3'$ is interpolated from three nearby points as
\eq{63.12}{
\xv{3'}=\frac{\(\rv{3'}-\rv{4}\)\(\rv{3'}-\rv{6}\)}{\(\rv{2}-\rv{4}\)\(\rv{2}-\rv{6}\)}\xv{2}+\frac{\(\rv{3'}-\rv{6}\)\(\rv{3'}-\rv{2}\)}{\(\rv{4}-\rv{6}\)\(\rv{4}-\rv{2}\)}\xv{4}+\frac{\(\rv{3'}-\rv{2}\)\(\rv{3'}-\rv{4}\)}{\(\rv{6}-\rv{2}\)\(\rv{6}-\rv{4}\)}\xv{6}+O(\Delta^3)\,.
}
With this method, we obtain
\eq{63.14}{
\delta r(V_n)=\(r\maxx(V_n)-r\minn\)\frac{\Delta U}{U\maxx-U_0}\,
} without the exponential growth. 
\fig{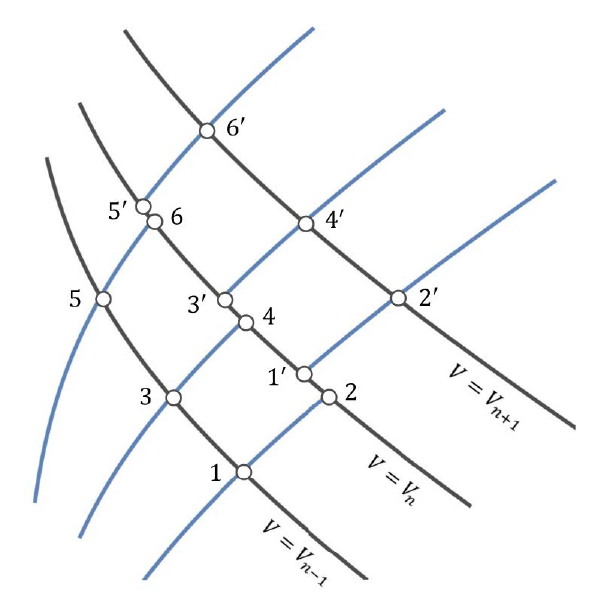}{8cm}{
Schematic of the $U$-coordinate redefinition. After computing $x(U,V_{n})$, the $U$ coordinate is redefined so that the grid spacing in $r$ becomes uniform. Black and blue lines indicate $V=\const$ and $U=\const$, respectively, and primed points represent grid points in the new $U$ coordinate.
}

\section{Numerical results for all other cases}\label{NB.2param}
In this Appendix, we present the parameter-dependence results for the combinations of $w$ and $\scr{F}$ not presented in Subsec.~\ref{6..FluParam}.\par
Figure~\ref{Xiw1to2} plots $\Xi/\abs{\scr{A}}$ for $w>1$. In this case, $\rho_\infty$ is monotonically decreasing in $\scr{F}$, and the dilute condition is always satisfied for $\scr{F}\ge\scr{F}_\text{saddle}$ (see Fig.~\ref{Fsaddle}). We therefore plot the behavior when varying $w$ with $\scr{F}/\scr{F}_\text{saddle}$ held fixed. As in the $w<0$ case, $\Xi/\abs{\scr{A}}$ approaches the Vaidya BG value as $\rho_\infty/\abs{\scr{A}}$ decreases. Figure~\ref{Xiw1to2pv2} is a magnified view of that region. We find that for certain values of $w$ and $\scr{F}$, the value can slightly exceed that of the Vaidya BG. This deviation becomes smaller in the large-$l$ limit.
\fig{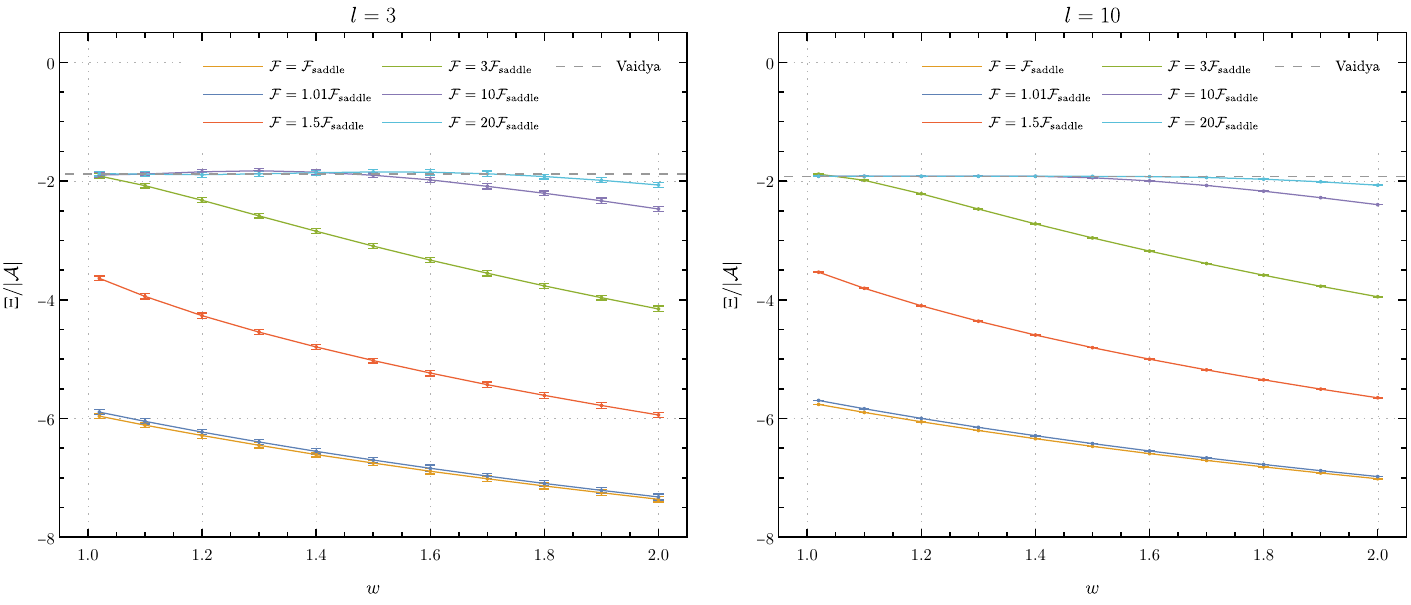}{\linewidth}{
$w$- and $\scr{F}$ dependence of $\Xi$ in the case of  $1<w\le2$, $r_\obs=20M_0$, and $\scr{A}=3\times 10^{-5}$ for $l=3$ (left) and $10$ (right). In the limit of $\scr{F}\rightarrow\infty$, the values asymptote to those of the Vaidya BG.
}
\fig{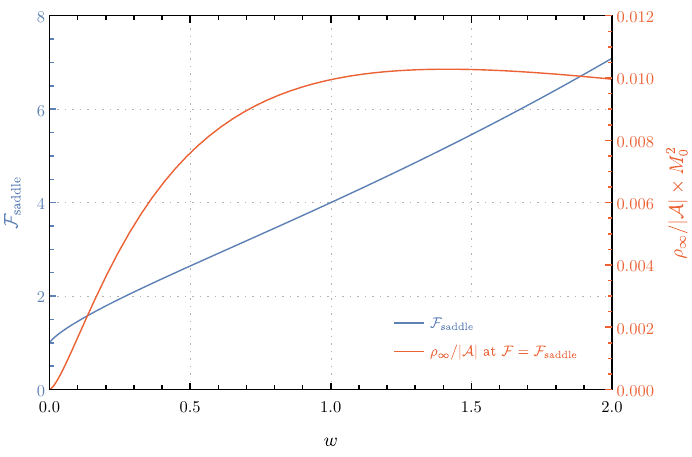}{0.575\linewidth}{
$w$ dependence of $\scr{F}_\text{saddle}$ (blue line) and $\rho_\infty/\abs{\scr{A}}$ at $\scr{F}=\scr{F}_\text{saddle}$ (red line). Since $\scr{F}\ge\scr{F}_\text{saddle}$ for $w>1$, the red line represents the maximum value of $\rho_\infty/\abs{\scr{A}}$.
}
\fig{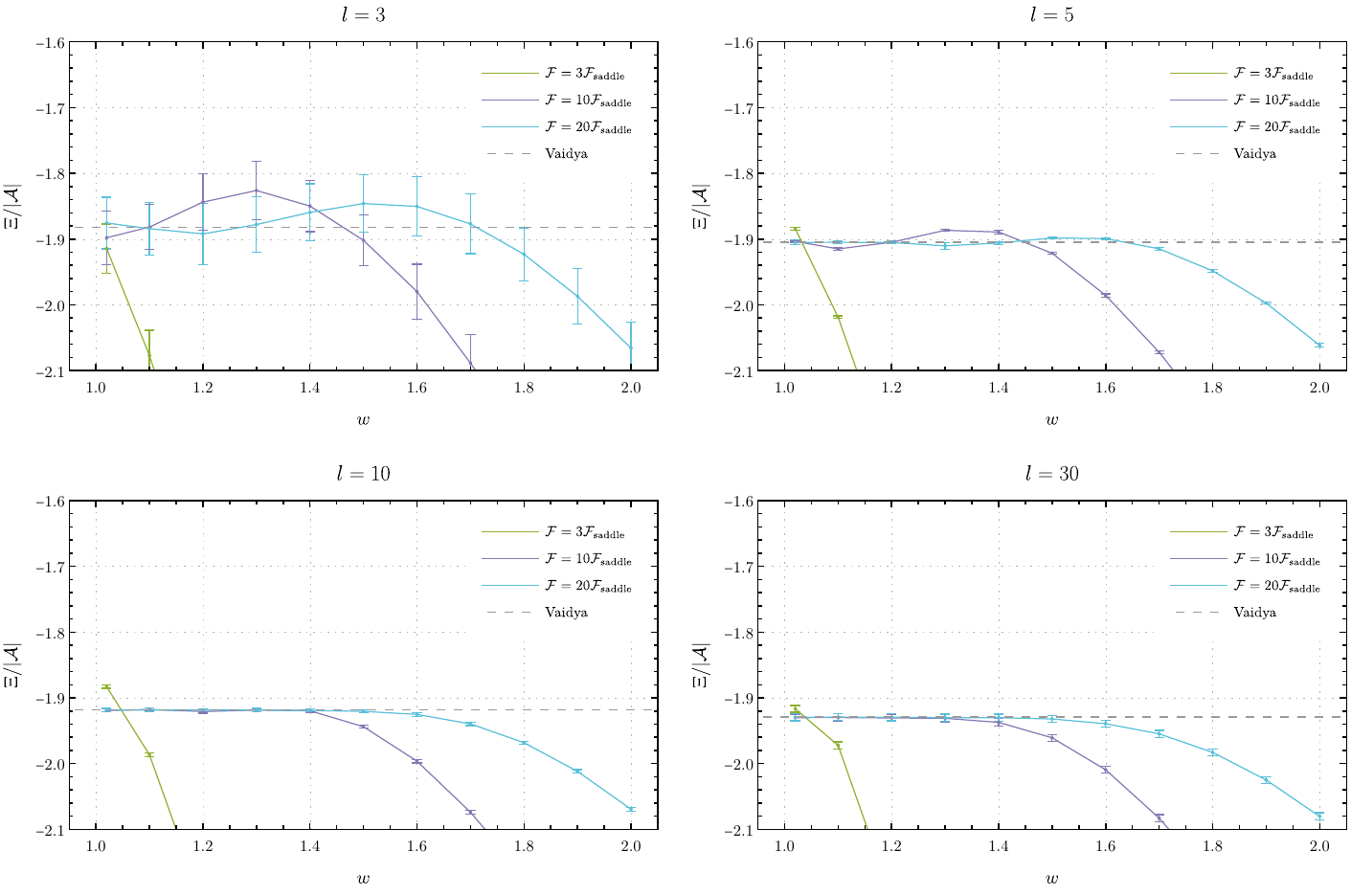}{\linewidth}{
Magnified view of a portion of Fig.~\ref{Xiw1to2} for $l=3$ (top-left) and $10$ (bottom-left), with the addition of the $l=5$ (top-right) and $30$ (bottom-right) cases in the same region. For small $l$, regions where the values exceed those of the Vaidya BG exist at $\scr{F}=10\scr{F}_\text{saddle}$ and $20\scr{F}_\text{saddle}$; for large $l$, such regions exist at $\scr{F}=3\scr{F}_\text{saddle}$.
In the limit of $l\rightarrow\infty$, however, this excess converges to $0$.
}
\par

Next, Fig.~\ref{Xiw0} plots the $\scr{F}$ dependence of $\Xi/\abs{\scr{A}}$ for $w=0$. From Eqs.~\eqref{45.13} and \eqref{3C.1}, we have
\eq{6C.3x}{
\rho M_0^2=\frac{\scr{A}\scr{F}}{4\pi\sqrt{\scr{F}^{-2}-1+\frac{2M_0}{r}}} \(\frac{r}{M_0}\)^{-2}\,,
}
so $\rho$ is monotonically increasing in $\scr{F}$, with $\rho\rightarrow0$ as $\scr{F}\rightarrow+0$. Indeed, we find that $\Xi/\abs{\scr{A}}$ asymptotes to the Vaidya BG value as $\scr{F}\rightarrow+0$.
\fig{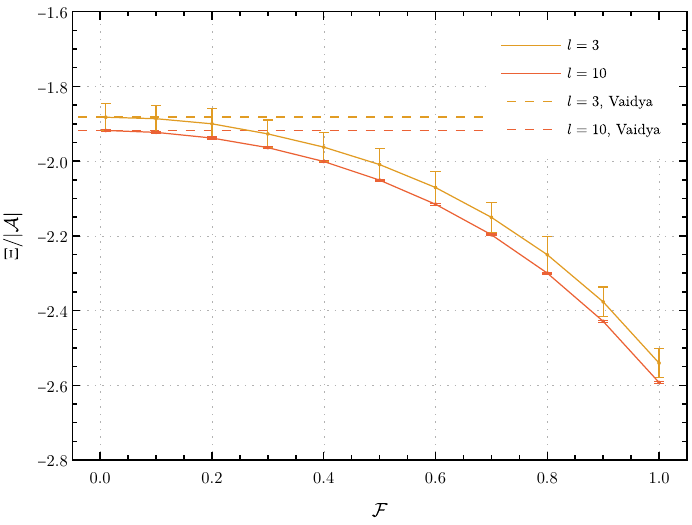}{0.575\linewidth}{
$\scr{F}$ dependence of $\Xi$ for $w=0$, $r_\obs=20M_0$, and $\scr{A}=3\times 10^{-5}$.
}
\par

Finally, the differences in $\Xi/\abs{\scr{A}}$ between $l=4$ modes and $l=10$ modes for the $w>1$ and $w=0$ cases are shown in Figs.~\ref{XiDiffw1to2} and \ref{XiDiffw0}, respectively. As in the case of $w<0$, we can 
obtain an additional constraint 
from the measurement of these differences.
\fig{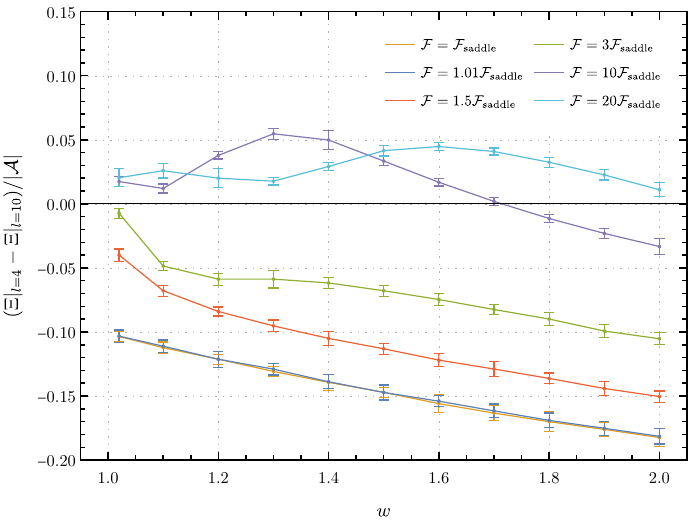}{0.575\linewidth}{
$w$- and $\scr{F}$ dependence of $\Xi|_{l=4}-\Xi|_{l=10}$ for $1<w\le2$, $r_\obs=20M_0$, and $\scr{A}=3\times 10^{-5}$
}
\fig{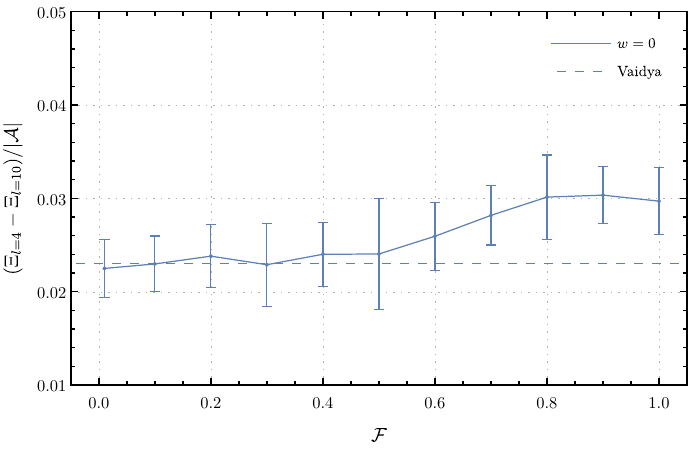}{0.575\linewidth}{
$\scr{F}$ dependence of $\Xi|_{l=4}-\Xi|_{l=10}$ for $w=0$, $r_\obs=20M_0$, and $\scr{A}=3\times 10^{-5}$.
}

\section{Introduction of cutoff}\label{C.cutoff}
\subsection{Background with cutoff}\label{C..BGcutoff}
A gas cloud around a realistic BH would have a finite mass and be localized within a finite region.
We attempt to introduce a cutoff into the BG spacetime in order to confine the accretion region to a finite radius and make it possible to compute the waveform at infinity. It is then natural to require that the Misner-Sharp mass $M(V,r)$ be constant as seen by a distant observer. We also choose the coordinate $V$ to coincide with the proper time of an observer at rest at future null infinity. That is, we require
\subeq{SC.1}{
\eq{C1.1}{
M|_{\scrI^+}=M_\text{fin}=\const\,,
}\eq{C1.2}{
\lambda|_{\scrI^+}=0
}}
to be satisfied.\par
We introduce a cutoff function $f_\text{cut}(x,\tau)$,
\eq{3D.2}{
f_\text{cut}(x,\tau)\coloneq
\begin{cases}
    1 & (x\le-\tau) \\
    \rec{2}\[1-\tanh{\tk{\tan{\(\frac{\pi}{2}\frac{x}{\tau}\)}}}\] & (-\tau<x<\tau) \\
    0 & (\tau\le x)
\end{cases} \,.
}
Here, $\tau$ is the cutoff width in $x$, and $f_\text{cut}$ is defined to be infinitely differentiable at $x=\pm \tau$. The new $M$ is defined by
\eq{3D.3}{
M^{(\text{new})}(M^{(\text{old})})\coloneq\int _{M_0}^{M^{(\text{old})}} dM^{(\text{old})'}\, f_{\text{cut}} \(M^{(\text{old})'}-M_{\text{fin}},\tau_M\)+M_0\,,
}
where $\tau_M$ denotes the cutoff width in $M$.
From Eqs.~\eqref{44.7} and \eqref{44.1}, we then have
\subeq{S3D.1}{
\eq{3D.4}{
\scr{A}^{(\text{new})}(V,r)=\scr{A}^{(\text{old})} f_{\text{cut}}\(M^{(\text{old})}(V,r)-M_{\text{fin}},\tau_M\) \,,
}
\eq{3D.5}{
\T{0}{0(\text{new})}(V,r)=\T{0}{0(\text{old})}(r) f_{\text{cut}}\(M^{(\text{old})}(V,r)-M_{\text{fin}},\tau_M\)\,.
}}
Since $\T{0}{0}$ and $\T{1}{0}$ are proportional to $\rho$, we similarly define
\subeq{S3D.2}{
\eq{3D.6}{
\rho^{(\text{new})}(V,r)\coloneq \rho^{(\text{old})}(r) f_{\text{cut}}\(M^{(\text{old})}(V,r)-M_{\text{fin}},\tau_M\) \,,
}
\eq{3D.7}{
\T{1}{0(\text{new})}(V,r)\coloneq \T{1}{0(\text{old})}(r) f_{\text{cut}}\(M^{(\text{old})}(V,r)-M_{\text{fin}},\tau_M\)\,.
}} 
From Eq.~\eqref{44.13}, $\lambda$ is modified as
\eq{3D.8}{
\lambda ^{(\text{new})}(V,r)=4\pi \int_{\infty}^r dr'\,r'\T{1}{0(\text{old})} (r')f_{\text{cut}}\(M^{(\text{old})}(V,r')-M_{\text{fin}},\tau_M\)\,,
}
where the lower limit of integration is set to $\infty$ in order to satisfy Eq.~\eqref{C1.2}. The way the cutoff enters is illustrated in Figs.~\ref{fcut} and \ref{Afcut}. The spacetime is divided into three regions: the accretion region, the cutoff region, and the Schwarzschild region. By taking $M_\text{fin}-M_0$ sufficiently small, the dilute condition~\eqref{3B.1} is satisfied in the entire region $V>0$ when $\scr{A}^{(\old)}>0$.

\fig{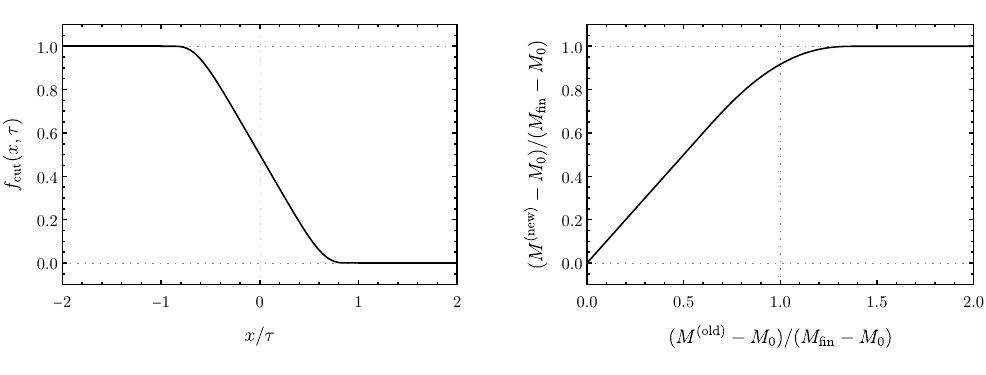}{0.847\linewidth}{
$f_\text{cut}(x,\tau)$ (left) and $M^{(\text{new})}(M^{(\text{old})})$ (right), with $\tau_M/(M_\text{fin}-M_0)=0.5$. $M^{(\text{new})}$ is defined so that $M^{(\text{new})}=M^{(\text{old})}$ for $M^{(\text{old})}\le M_\text{fin}-\tau_M$ and $M^{(\text{new})}=M_\text{fin}$ for $M^{(\text{old})}\ge M_\text{fin}+\tau_M$.
}
\fig{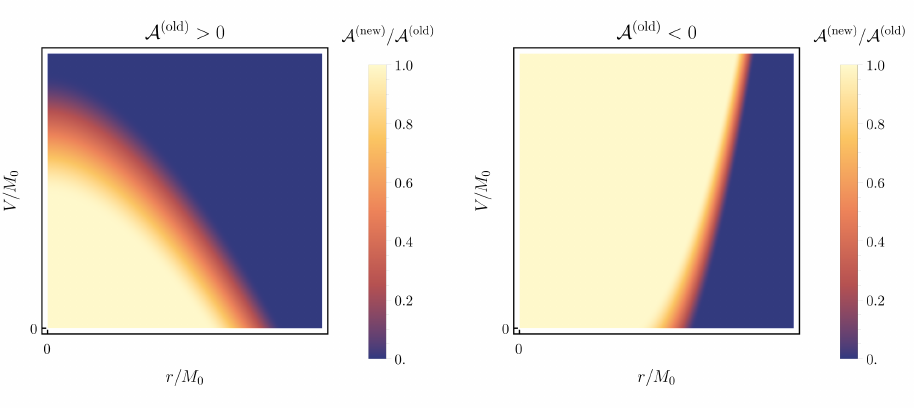}{0.895\linewidth}{
Schematic showing how the cutoff is implemented. The accretion region with $\scr{A}^{(\text{new})}=\scr{A}^{(\text{old})}$ and the Schwarzschild region with $M=M_\text{fin}$ are smoothly connected through the cutoff region. When $\scr{A}^{(\text{old})}>0$ (left), the mass increases with time, so the cutoff radius is generally monotonically decreasing in $V$; when $\scr{A}^{(\text{old})}<0$ (right), it is monotonically increasing. In either case (excluding Vaidya spacetime with $\scr{A}^{(\text{old})}<0$), $\scr{A}^{(\text{new})}=0$ always holds at $\scrI^+$. 
}
\par

Note that since all these physical quantities depend on $V$, the master equation is partially modified. Computing with $\lambda=\lambda(V,r)$ and $\rho=\rho(V,r)$, Eqs.~\eqref{5C.2}--\eqref{S5C.0} are modified to
\subeq{SB.1}{
\eq{5C.2'}{
\[-4\frac{\partial ^2}{\partial U\partial V}+\gamma_{l,V} \pard{}{U}+\gamma_{l,U}\pard{}{V}-V_l^{\odd}\]\psi_{lm}^{\odd}=S_{l m}^{\odd}\,,
}
\eq{5C.3'}{
V_l^{\odd}= -\frac{r_{,U}\gamma_{,V}+r_{,V}\gamma_{,U}-4r_{,UV}}{r} +\frac{\zeta -8r_{,U}r_{,V}}{r^2} \,,
}
\eqa{B.4}{
S_{l m}^{\odd}=\frac{32\pi wr}{(l-1)(l+2)} &\biggl[\tk{\frac{f_0}{2}\(2\rho +r \rho\dr\)+r\rho\dV}\pard{}{U} +r\dU \(2 \rho +r \rho\dr\)\pard{}{V} \nn
&+r\dU\tk{\frac{(l-1)(l+2)+2f_0}{r}\rho +f_0 \rho\dr+\rho\dV}\biggr] \psi \,,
}
\eq{5C.5'}{
\zeta=-2(l-1) (l+2) e^{\lambda}r\dU -4r\(2r_{,U V}+r\sigma _{,U V}\) \,,
}
\eq{B.5}{
\sigma\dUV=r\dU\(-2\frac{M_0+\delta M+M_0\lambda}{r^3}+\frac{2\delta M\dr+3M_0\lambda\dr}{r^2}-\frac{\delta M_{,rr}}{r}+f_0\lambda_{,rr}+\lambda_{,Vr}\) \,,
}
\eq{B.6}{
\gamma_{,U}=\frac{4r_{,U}}{(l-1)(l+2)}\tk{\lambda_{,r}-\delta M_{,rr}+\(3r-M_0\)\lambda_{,rr} +r\(2\lambda_{,Vr}-\delta M_{,rrr}\)+r^2\(f_0\lambda_{,rrr}+\lambda_{,Vrr}\)} \,,
}
\eqa{B.7}{
\gamma_{,V}=&\frac{4r_{,V}}{(l-1)(l+2)}\tk{\lambda_{,r}-\delta M_{,rr}+\(3r-M_0\)\lambda_{,rr} +r\(2\lambda_{,Vr}-\delta M_{,rrr}\)+r^2\(f_0\lambda_{,rrr}+\lambda_{,Vrr}\)} \nn
&+\frac{4}{(l-1)(l+2)} \tk{\(r+M_0\)\lambda_{,Vr}-r\delta M_{,Vrr}+r^2\(f_0\lambda_{,Vrr}+\lambda_{,VVr}\)}\,.
}}
Equations~\eqref{62.1}, \eqref{5C.1}, and \eqref{5C.10} retain the same form even after introducing the cutoff.\par

However, the physical interpretation of the spacetime modified by hand in the above way is nontrivial. 
While such a modification naturally satisfies Eqs.~\eqref{44.1}--\eqref{44.4}, $T_{22}$ and $T_{33}$ newly defined by Eq.~\eqref{44.5} no longer satisfy the perfect fluid form Eq.~\eqref{45.1}. 
The question then arises of how to define the source term. We therefore continue to assume a perfect fluid. 
This means that we use Eq.~\eqref{51.10} as the source term as before, and assume $\delta u_3=0$.\par

Even with this prescription, the same equations as before are still satisfied [up to a constant shift $\lambda\rightarrow \lambda+\delta\lambda(V)$] in the accretion region. Therefore, if the cutoff is sufficiently far outside the photon sphere, the properties of the generated GWs should remain unchanged. On the other hand, the Einstein equation of BG no longer holds in the cutoff region. This region may have a nontrivial effect on the waveform, so care is needed. In particular, since Eq.~\eqref{5C.9} does not hold, the two master equations \eqref{SB.1} and \eqref{5C.10} do not agree in the cutoff region, and the results may differ depending on which one is used.
We expected that, by taking the cutoff region sufficiently wide and tapering off the matter sufficiently slowly, the influence of anything other than the redshift and the correction term for $r_\obs\rightarrow\infty$ could be neglected. However, the results presented below show that this prescription entails practical difficulties.\par

\subsection{Result}\label{C..resultcutoff}
As before, we solve the master equation numerically. The parameters are
\eqs{C2.1}{
\scr{A}_0\coloneq \scr{A}^{(\text{old})} &=\pm3\times 10^{-5} \,,\\
(M_\text{fin},\,\tau_M) &=(1.0075M_0,\,0.0015M_0 ) \,,\\
r_\obs &=100M_0 \,,\\
(U_0,\,U\maxx)=(V_0,\,V\maxx)&=(0,\,400 M_0)\,,
}
with the remaining parameters following Eq.~\eqref{5C.13}. First, the numerical solutions of the BG functions at the initial time are presented in Fig.~\ref{BGcut}.
\fig{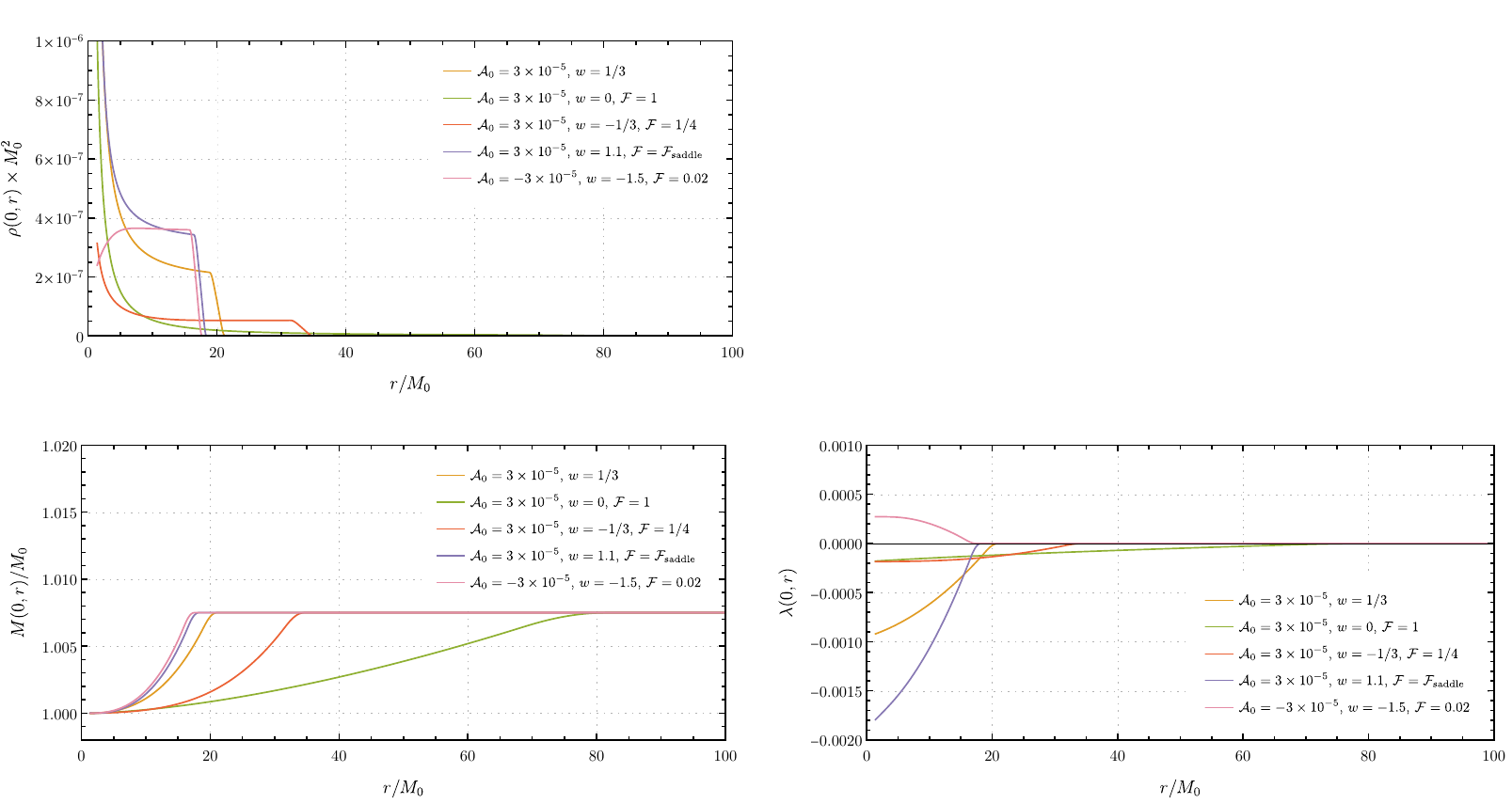}{\linewidth}{
BG functions $\rho$ (top), $M$ (bottom-left), and $\lambda$ (bottom-right) after the cutoff with respect to $r$. Note that $\rho=\rho(V,r)$ and $\lambda=\lambda(V,r)$ are now time-dependent, unlike the previous cases.
}
\par

Next, the numerical results for $\Xi/\abs{\scr{A}_0}$ using Eq.~\eqref{SB.1} are shown in Fig.~\ref{XicutnoGS}. Except in some cases, $\Xi$ does not settle to a constant. To examine the contribution from the cutoff region, we consider plotting the potential term. Since $V_l^{\odd}$ depends on the choice of $U$ coordinate, we define
\eq{C2.2}{
\tilde{V}_l^{\odd}\coloneq -\frac{f_0}{2r\dU}V_l^{\odd}\,,
}
to cancel this dependence. When using Eq.~\eqref{5C.10}, we employ
\eq{C2.3}{
V_l^{\odd}\coloneq \frac{-2(l-1)(l+2)e^\lambda r\dU-8r\dU r\dV +4rr\dUV}{r^2}\,.
}
At zeroth (vacuum) order, $\tilde{V}_l^{\odd}$ coincides with the RW potential,
\eq{C2.4}{
V_{l(\Sch)}^{\odd}=f_0\tk{\frac{l(l+1)}{r^2}-\frac{6M_0}{r^3}}\,.
}
Figure~\ref{potw-1p5cutnoGS} plots $\tilde{V}_l^{\odd}$ for a representative BG. Figure~\ref{potdifcutnoGS} shows the difference from $V_{l(\Sch)}^{\odd}$ to make the deformation of the potential more visible. New peaks appear in the potential in the cutoff region, suggesting that they may generate new modes. Such peaks can also be understood from the fact that $\gamma_{,U}$ and $\gamma_{,V}$ in Eq.~\eqref{SB.1} contain third-order derivatives of $\delta M$ and $\lambda$. This means that derivatives of $f_\text{cut}$ up to second order enter the potential, and the positive and negative peaks in Fig.~\ref{potdifcutnoGS} reproduce this behavior. Moreover, the effect on $\Xi$ becomes smaller for larger $l$ since the deformation of the potential becomes smaller for larger $l$ and the geometric optics approximation improves.
\fig{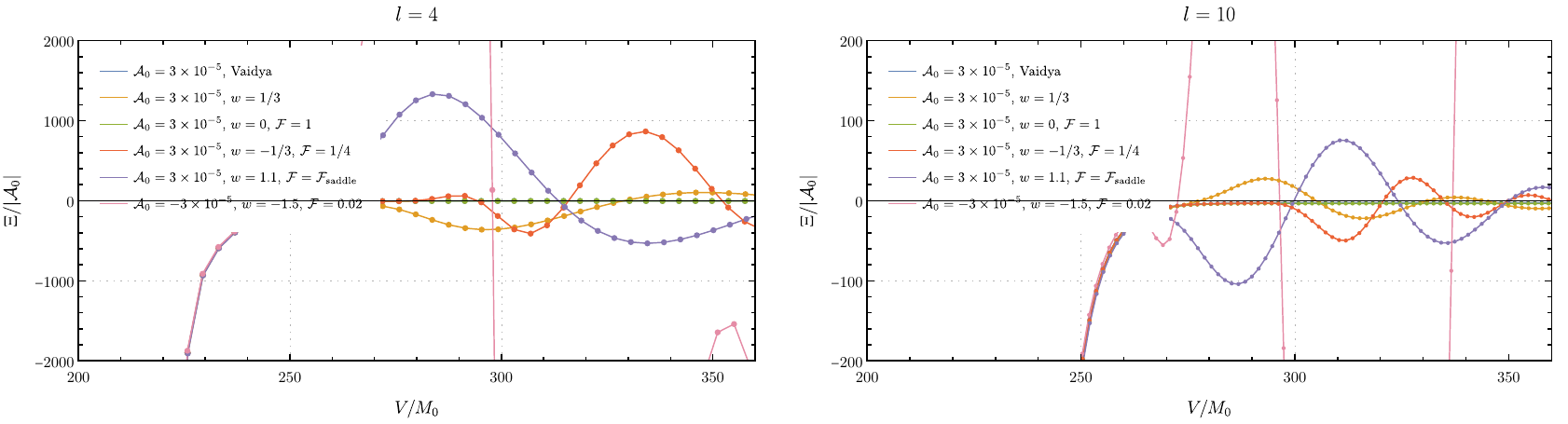}{\linewidth}{
Time dependence of $\Xi/\abs{\scr{A}_0}$ for $l=4$ (left) and $10$ (right). Since the influence of the tail part is large at $r_\obs=100M_0$, we use $l=4$ instead of $l=3$. Except in some cases, the contribution from the cutoff region is significant, making it difficult to compute the time average as before.
}
\fig{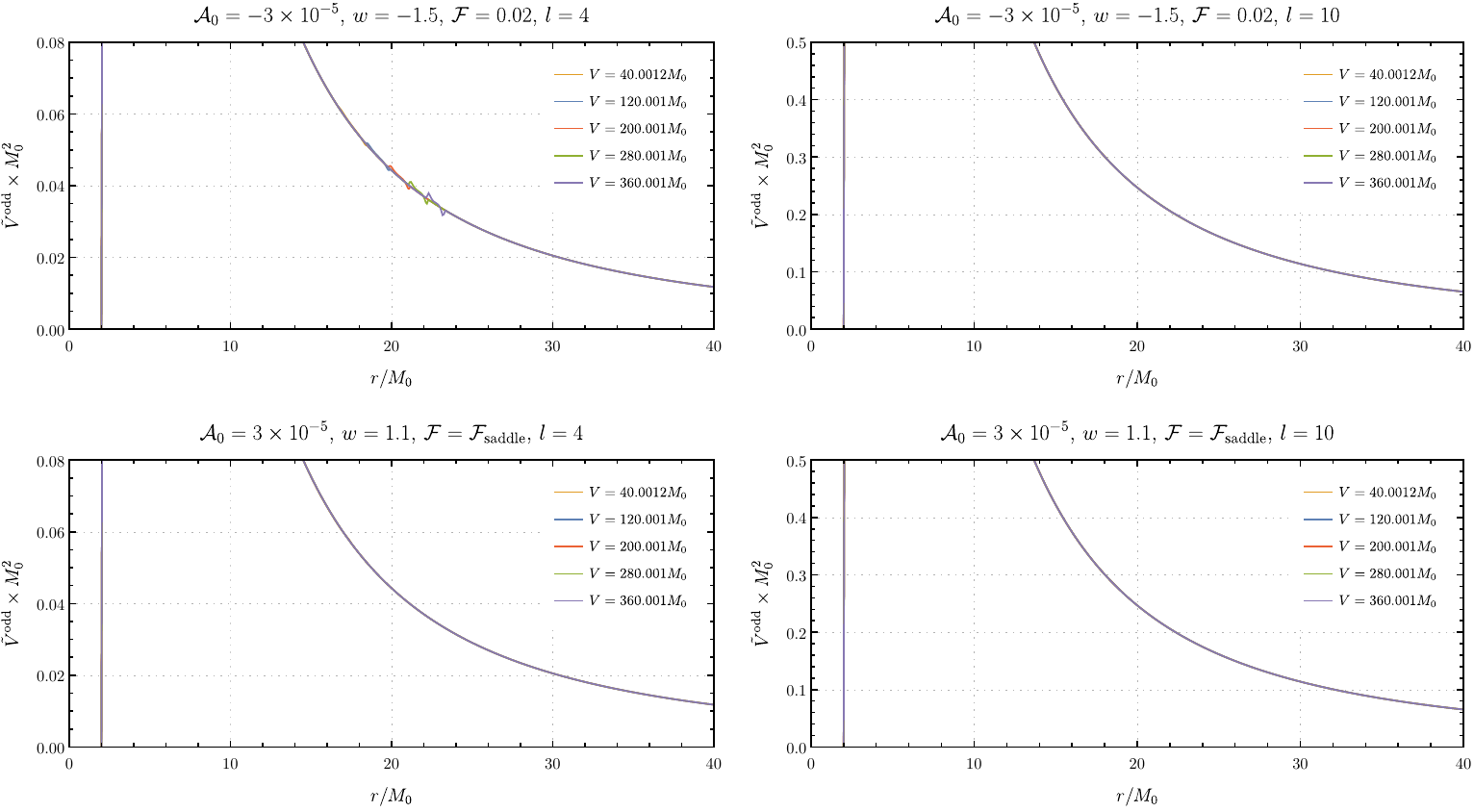}{\linewidth}{
Potential-like factor $\tilde{V}^{\odd}$ for $\scr{A}_0=-3\times10^{-5},\,w=-1.5,\,\scr{F}=0.02$ (top) and $\scr{A}_0=3\times10^{-5},\,w=1.1,\,\scr{F}=\scr{F}_\text{saddle}$ (bottom) with $l=4$ (left) and $10$ (right). The $r$ dependence is shown at fixed $V$. For $\scr{A}_0=-3\times10^{-5},\,w=-1.5,\,\scr{F}=0.02,\,l=4$, new peaks are visible in the potential.
}
\fig{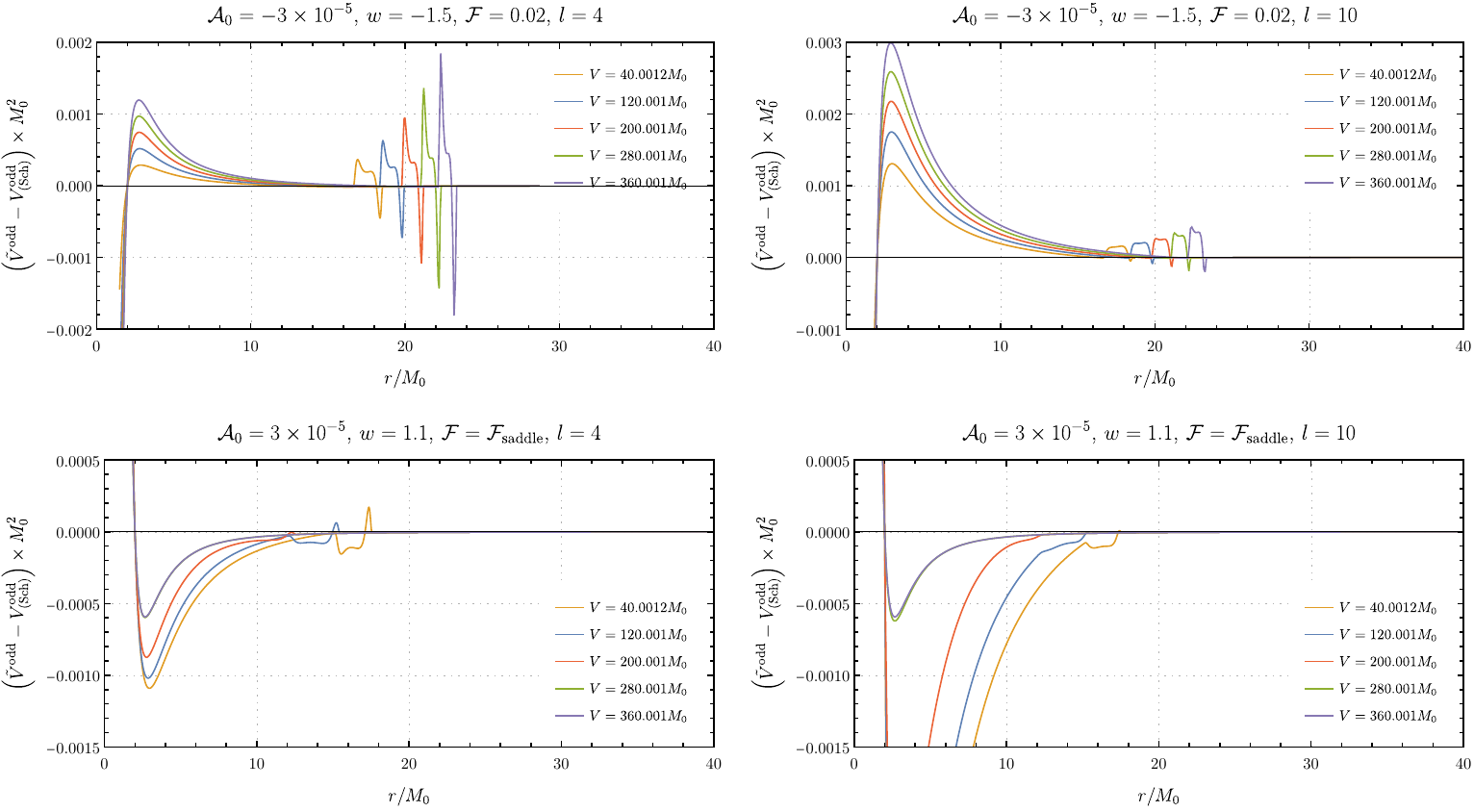}{\linewidth}{
Difference $\tilde{V}^{\odd}-V_{(\Sch)}^{\odd}$ for each $\tilde{V}^{\odd}$ in Fig.~\ref{potw-1p5cutnoGS}. Nontrivial deformations are observed in the cutoff region in all cases. Note that for $\scr{A}_0=3\times10^{-5},\,w=1.1,\,\scr{F}=\scr{F}_\text{saddle}$ at $V=360M_0$, the entire region $r>2M_0$ is already outside the cutoff, and the difference between $\tilde{V}^{\odd}$ and $V_{l(\Sch)}^{\odd}$ is solely due to the difference in Schwarzschild mass.
}
\par

Next, we consider the case using Eq.~\eqref{5C.10}. In this case, only first-order derivatives of $\delta M$ and $\lambda$ appear in the potential, so no derivatives of $f_\text{cut}$ exist. Indeed, no peaks like those in the case of Eq.~\eqref{SB.1} appear in Fig.~\ref{potdifcutGS}, which plots the potential deformation in the same manner as Fig.~\ref{potdifcutnoGS}. 
However, as seen in Fig.~\ref{XicutGS}, distortions remain in the plot of $\Xi/\abs{\scr{A}_0}$. While the effect can be neglected by taking $l$ large enough so that the wavelength is sufficiently short compared to the cutoff scale, this also amplifies numerical errors, resulting in a trade-off. In any case, defining a modified BG that allows computation for arbitrary $l$ remains an open problem.
\fig{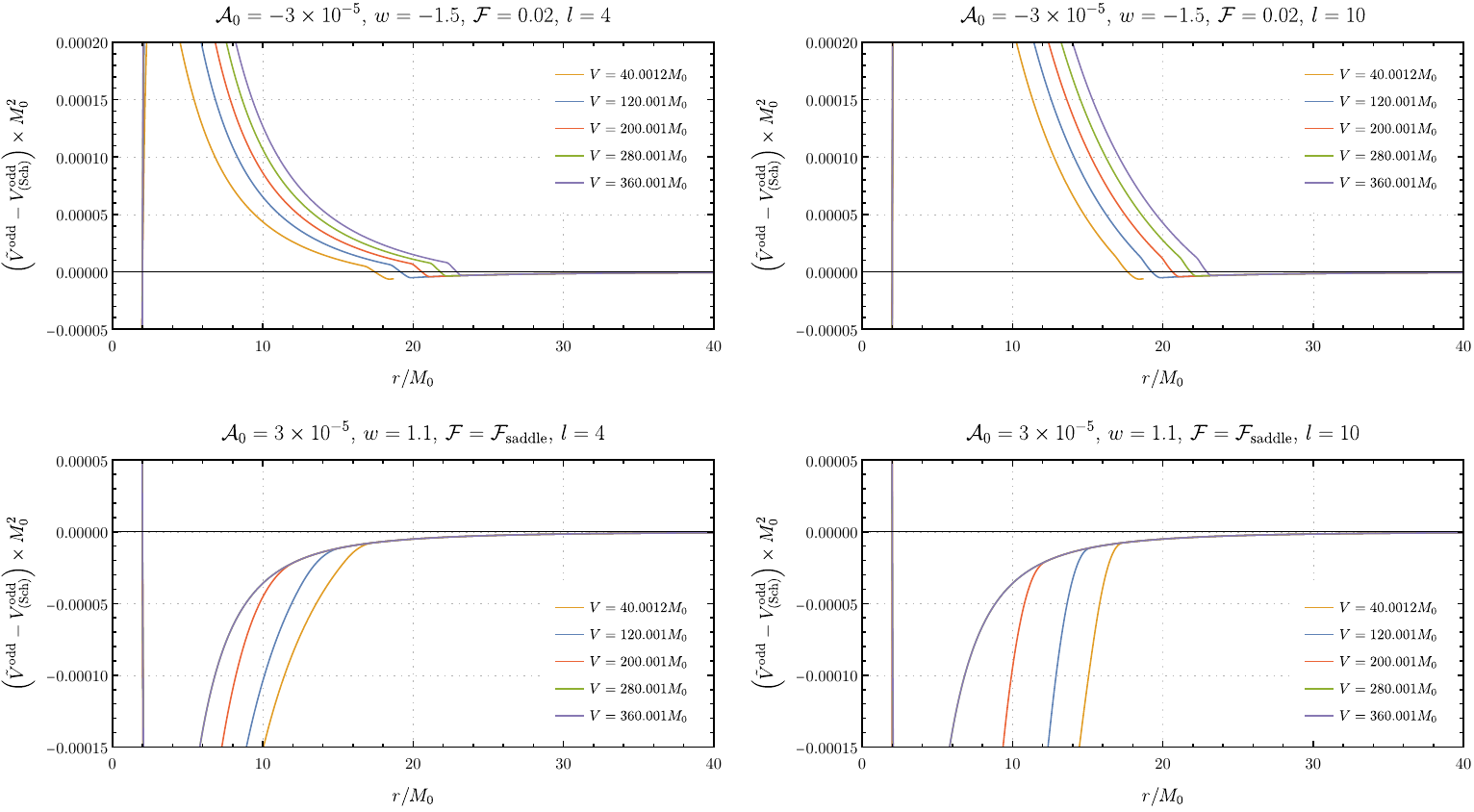}{\linewidth}{
Difference $\tilde{V}^{\odd}-V_{(\Sch)}^{\odd}$ when using Eq.~\eqref{5C.10}, analogous to Fig.~\ref{potdifcutnoGS}. The accretion region and the Schwarzschild region appear to be naturally connected through the monotonic function $f_\text{cut}$.
}
\fig{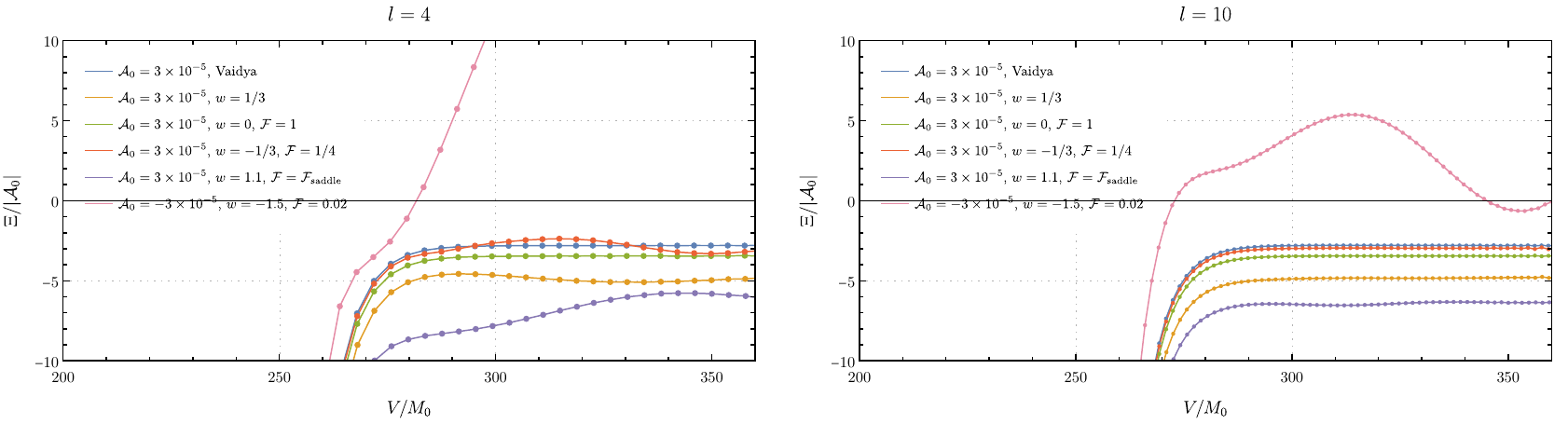}{\linewidth}{
Time dependence of $\Xi/\abs{\scr{A}_0}$ for $l=4$ (top) and $10$ (bottom) when using Eq.~\eqref{5C.10}. Compared to the case using Eq.~\eqref{SB.1}, the influence of the cutoff region is considerably reduced, but it cannot be completely removed.
}
\par

Finally, we discuss the effect of the cutoff on the estimation of the accretion rate. Figure~\ref{AtilcutGS} plots the time dependence of $\tilde{\scr{A}}$ when using Eq.~\eqref{5C.10}. We find that $\tilAR$ is less susceptible to the influence of the potential deformation than $\tilAI$, but $\tilde{\scr{A}}\simeq\scr{A}_0$ no longer holds. This is because the time dependence of the redshift factor $\scr{R}$ is no longer negligible, so the condition \eqref{5D.10} is no longer satisfied.\footnote{The condition $\abs{\scr{A}\dV\,(V_2-V_1)}\ll \scr{A}$ is satisfied near the horizon for $0\le V\le V_\gen(360M_0)\approx150M_0$, since the accretion is steady in that range.} This implies that, in actual observations, the estimation using $\tilde{\scr{A}}$ loses accuracy when $\scr{R}\dV$ cannot be neglected, i.e., when matter does not extend sufficiently far. On the other hand, this result means that $\tilde{\scr{A}}$ contains information about the matter distribution up to intermediate distances $r\sim100M_0$. Indeed, when $\scr{R}=1+O(\kappa)$ and $\scr{R}\dV=O(\kappa)$, taking $(V_2-V_1)$ sufficiently small gives
\eq{C2.5}{
\tilde{\scr{A}}\(V,r_\obs\)\simeq \scr{A}\(V_\gen(V),r_\gen\)-\scr{R}\dV M_0 +O(\kappa^2)\,,
}
suggesting that the measurement of $\scr{R}\dV$ may provide information about the matter distribution in the intermediate region.
\fig{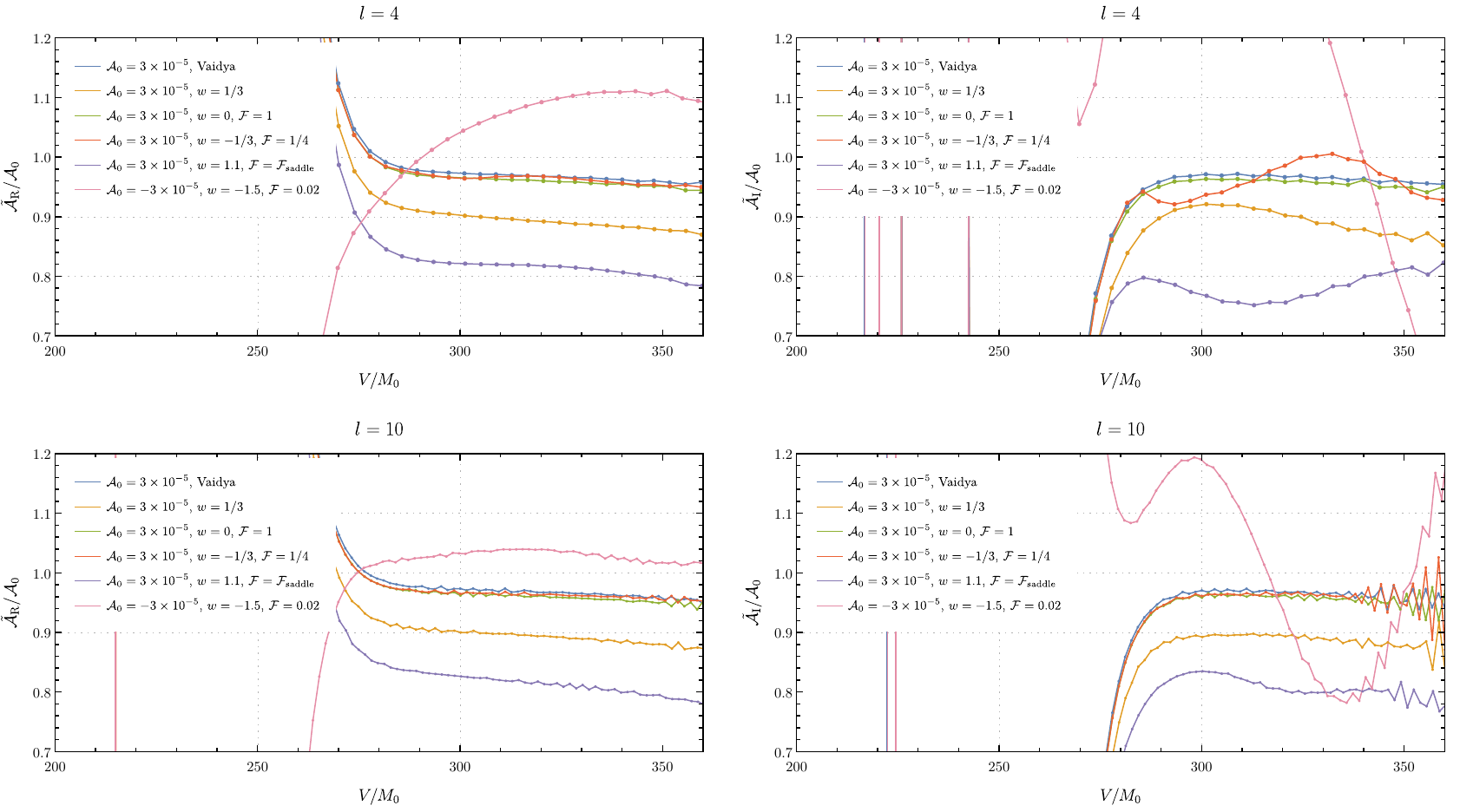}{\linewidth}{
Time dependence of $\tilde{\scr{A}}/\scr{A}_0$ when using Eq.~\eqref{5C.10}. We present $\tilAR/\scr{A}$ (left) and $\tilAI/\scr{A}$ (right) for $l=4$ (top) and $10$ (bottom).
The nontrivial influence of the cutoff region is less pronounced for $\tilAR$ than for $\tilAI$, and is also less pronounced for larger $l$. Unlike the case without a cutoff, a deviation of $O(\scr{A}_0)$ arises between $\tilde{\scr{A}}$ and $\scr{A}_0$.
}
\par


\bibliography{ref}

@ARTICLE{Vaidya,
   author       = "Jaime Redondo-Yuste and David Pereñiguez and Vitor Cardoso",
   title        = "Ringdown of a dynamical spacetime",
   year         = "2024",
   journal      = "Phys. Rev. D",
   volume       = "109",
   pages        = "044048",
   doi          = "10.1103/PhysRevD.109.044048"
}

@ARTICLE{MasterGS,
   author       = "U. H. Gerlach and U. K. Sengupta",
   title        = "Gauge-invariant perturbations on most general spherically symmetric space-times",
   year         = "1979",
   journal      = "Phys. Rev. D",
   volume       = "19",
   pages        = "2268",
   doi          = "10.1103/PhysRevD.19.2268"
}

@ARTICLE{MasterGS2,
   author       = "U. H. Gerlach and U. K. Sengupta",
   title        = "Gauge-invariant coupled gravitational, acoustical, and electromagnetic modes on most general spherical space-times",
   year         = "1980",
   journal      = "Phys. Rev. D",
   volume       = "22",
   pages        = "1300",
   doi          = "10.1103/PhysRevD.22.1300"
}

@ARTICLE{RW,
   author       = "T. Regge and J. A. Wheeler",
   title        = "Stability of a {S}chwarzschild Singularity",
   year         = "1957",
   journal      = "Phys. Rev.",
   volume       = "108",
   pages        = "1063",
   doi          = "10.1103/PhysRev.108.1063"
}

@ARTICLE{BG,
   author       = "E. Babichev and V. Dokuchaev and Yu. Eroshenko",
   title        = "Backreaction of accreting matter onto a black hole in the {E}ddington--{F}inkelstein coordinates",
   year         = "2012",
   journal      = "Classical Quantum Gravity",
   volume       = "29",
   number       = "11",
   pages        = "115002",
   doi          = "10.1088/0264-9381/29/11/115002"
}

@ARTICLE{Koga:2016jjq,
    author = "Koga, Yasutaka and Harada, Tomohiro",
    title = "Correspondence between sonic points of ideal photon gas accretion and photon spheres",
    reportNumber = "RUP-16-5",
    doi = "10.1103/PhysRevD.94.044053",
    journal = "Phys. Rev. D",
    volume = "94",
    number = "4",
    pages = "044053",
    year = "2016"
}

@article{Koga:2018ybs,
    author = "Koga, Yasutaka and Harada, Tomohiro",
    title = "Rotating accretion flows in {$D$} dimensions: Sonic points, critical points, and photon spheres",
    reportNumber = "RUP-18-9",
    doi = "10.1103/PhysRevD.98.024018",
    journal = "Phys. Rev. D",
    volume = "98",
    number = "2",
    pages = "024018",
    year = "2018"
}

@ARTICLE{fluid,
   author       = "Y. Koga",
   title        = "Photon surfaces in spherically, planar, and hyperbolically symmetric spacetimes in {$D$} dimensions: Sonic point/photon sphere correspondence",
   year         = "2019",
   journal      = "Phys. Rev. D",
   volume       = "99",
   pages        = "064034",
   doi          = "10.1103/PhysRevD.99.064034"
}

@ARTICLE{Michel,
   author       = "F. C. Michel",
   title        = "Accretion of matter by condensed objects",
   year         = "1972",
   journal      = "Astrophys. Space Sci.",
   volume       = "15",
   pages        = "153",
   doi          = "10.1007/BF00649949"
}

@ARTICLE{Bondi,
   author       = "H. Bondi",
   title        = "On Spherically Symmetrical Accretion",
   year         = "1952",
   journal      = "Mon. Not. R. Astron. Soc.",
   volume       = "112",
   pages        = "195",
   doi          = "10.1093/mnras/112.2.195"
}

@ARTICLE{DNF1,
    author = "C. Gundlach and R. H. Price and J. Pullin",
    title = "Late-time behavior of stellar collapse and explosions. {I}. Linearized perturbations",
    journal = "Phys. Rev. D",
    year = "1994",
    volume = "49",
    pages = "883",
    doi = "10.1103/PhysRevD.49.883"
}

@ARTICLE{DNFproblem,
   author       = "E. Eilon and A. Ori",
   title        = "Adaptive gauge method for long-time double-null simulations of spherical black-hole spacetimes",
   year         = "2016",
   journal      = "Phys. Rev. D",
   volume       = "93",
   pages         = "024016",
   doi          = "10.1103/PhysRevD.93.024016"
}

@article{Pricelaw,
    author      = "R. H. Price",
    title       = "Nonspherical Perturbations of Relativistic Gravitational Collapse. {I}. Scalar and Gravitational Perturbations",
    journal     = "Phys. Rev. D",
    volume      = "5",
    pages       = "2419",
    year        = "1972",
    doi         = "10.1103/PhysRevD.5.2419"
}

@ARTICLE{VaidyaPL,
   author       = "Chul-Moon Yoo and Masashi Kimura and Akihiro Ishibashi and Rikuto Ohashi",
   title        = "Ringdown in {V}aidya spacetimes: Time-dependent frequencies, {P}enrose limit, and time-domain analyses",
   year         = "2026",
   journal      = "Phys. Rev. D",
   volume       = "113",
   pages        = "044058",
   doi          = "10.1103/r2vm-zgqn"
}

@article{Lousto:2010qx,
    author = "Lousto, Carlos O. and Nakano, Hiroyuki and Zlochower, Yosef and Campanelli, Manuela",
    title = "Intermediate-mass-ratio black hole binaries: Intertwining numerical and perturbative techniques",
    year = "2010",
    journal = "Phys. Rev. D",
    volume = "82",
    pages = "104057",
    doi = "10.1103/PhysRevD.82.104057",
}

@ARTICLE{InfCorrection,
   author       = "Hiroyuki Nakano",
   title        = "A note on gravitational wave extraction from binary simulations",
   year         = "2015",
   journal      = "Classical Quantum Gravity",
   volume       = "32",
   pages        = "177002",
   doi          = "10.1088/0264-9381/32/17/177002"
}

@ARTICLE{GundlachMG,
   author  = "C. Gundlach and J. M. Martín-García",
   title   = "Gauge-invariant and coordinate-independent perturbations of stellar collapse: The interior",
   year    = "2000",
   journal = "Phys. Rev. D",
   volume  = "61",
   pages   = "084024",
   doi     = "10.1103/PhysRevD.61.084024"
}

@ARTICLE{LIGOGW150914,
   author  = "{B. P. Abbott {\it et al.} (LIGO Scientific Collaboration and Virgo Collaboration)}",
   title   = "Observation of Gravitational Waves from a Binary Black Hole Merger",
   year    = "2016",
   journal = "Phys. Rev. Lett.",
   volume  = "116",
   pages   = "061102",
   doi     = "10.1103/PhysRevLett.116.061102"
}

@ARTICLE{LIGOGW170817,
   author  = "{B. P. Abbott {\it et al.} (LIGO Scientific Collaboration and Virgo Collaboration)}",
   title   = "{GW170817}: Observation of Gravitational Waves from a Binary Neutron Star Inspiral",
   year    = "2017",
   journal = "Phys. Rev. Lett.",
   volume  = "119",
   pages   = "161101",
   doi     = "10.1103/PhysRevLett.119.161101"
}

@article{LIGOScientific:2025wao,
    author = "{A. G. Abac {\it et al.} (LIGO Scientific Collaboration, Virgo Collaboration, and KAGRA Collaboration)}",
    title = "Black Hole Spectroscopy and Tests of General Relativity with {GW250114}",
    year = "2026",
    journal = "Phys. Rev. Lett.",
    volume = "136",
    pages = "041403",
    doi = "10.1103/6c61-fm1n"
}

@article{LIGOScientific:2026wpt,
    author = "{A. G. Abac {\it et al.} (LIGO Scientific Collaboration, Virgo Collaboration, and KAGRA Collaboration)}",
    title = "{GWTC-4.0}: Tests of General Relativity. {III}. Tests of the Remnants",
    eprint = "2603.19021",
    archivePrefix = "arXiv",
    reportNumber = "LIGO-P2500067",
    journal = "",
    year = ""
}

@ARTICLE{KokkotasSchmidt,
   author  = "K. D. Kokkotas and B. G. Schmidt",
   title   = "Quasi-Normal Modes of Stars and Black Holes",
   year    = "1999",
   journal = "Living Rev. Relativity",
   volume  = "2",
   pages   = "2",
   doi     = "10.12942/lrr-1999-2"
}

@ARTICLE{BertiCardosoStarinets,
   author  = "E. Berti and V. Cardoso and A. O. Starinets",
   title   = "Quasinormal modes of black holes and black branes",
   year    = "2009",
   journal = "Classical Quantum Gravity",
   volume  = "26",
   pages   = "163001",
   doi     = "10.1088/0264-9381/26/16/163001"
}

@ARTICLE{KonoplyaZhidenko,
   author  = "R. A. Konoplya and A. Zhidenko",
   title   = "Quasinormal modes of black holes: From astrophysics to string theory",
   year    = "2011",
   journal = "Rev. Mod. Phys.",
   volume  = "83",
   pages   = "793",
   doi     = "10.1103/RevModPhys.83.793"
}

@ARTICLE{Dreyer,
   author  = "Olaf Dreyer and Bernard Kelly and Badri Krishnan and Lee Samuel Finn and David Garrison and Ramon Lopez-Aleman",
   title   = "Black-hole spectroscopy: testing general relativity through gravitational-wave observations",
   year    = "2004",
   journal = "Classical Quantum Gravity",
   volume  = "21",
   pages   = "787",
   doi     = "10.1088/0264-9381/21/4/003"
}

@ARTICLE{BertiSpec,
   author  = "E. Berti and A. Sesana and E. Barausse and V. Cardoso and K. Belczynski",
   title   = "Spectroscopy of {K}err Black Holes with {E}arth- and Space-Based Interferometers",
   year    = "2016",
   journal = "Phys. Rev. Lett.",
   volume  = "117",
   pages   = "101102",
   doi     = "10.1103/PhysRevLett.117.101102"
}

@ARTICLE{Barausse,
   author  = "E. Barausse and V. Cardoso and P. Pani",
   title   = "Can environmental effects spoil precision gravitational-wave astrophysics?",
   year    = "2014",
   journal = "Phys. Rev. D",
   volume  = "89",
   pages   = "104059",
   doi     = "10.1103/PhysRevD.89.104059"
}

@ARTICLE{CardosoMaselli,
   author  = "V. Cardoso and K. Destounis and F. Duque and R. Panosso Macedo and A. Maselli",
   title   = "Black holes in galaxies: Environmental impact on gravitational-wave generation and propagation",
   year    = "2022",
   journal = "Phys. Rev. D",
   volume  = "105",
   pages   = "L061501",
   doi     = "10.1103/PhysRevD.105.L061501"
}

@ARTICLE{VaidyaOriginal,
   author  = "P. C. Vaidya",
   title   = "The gravitational field of a radiating star",
   year    = "1951",
   journal = "Proc. Indian Acad. Sci. A",
   volume  = "33",
   pages   = "264",
   doi     = "10.1007/BF03173260"
}

@ARTICLE{MGtestI,
   author  = "E. Berti and K. Yagi and N. Yunes",
   title   = "Extreme gravity tests with gravitational waves from compact binary coalescences: ({I}) inspiral--merger",
   year    = "2018",
   journal = "Gen. Relativ. Gravit.",
   volume  = "50",
   pages   = "46",
   doi     = "10.1007/s10714-018-2362-8"
}

@ARTICLE{MGtestII,
   author  = "E. Berti and K. Yagi and H. Yang and N. Yunes",
   title   = "Extreme gravity tests with gravitational waves from compact binary coalescences: ({II}) ringdown",
   year    = "2018",
   journal = "Gen. Relativ. Gravit.",
   volume  = "50",
   pages   = "49",
   doi     = "10.1007/s10714-018-2372-6"
}

@ARTICLE{Clifton,
   author  = "T. Clifton and P. G. Ferreira and A. Padilla and C. Skordis",
   title   = "Modified gravity and cosmology",
   year    = "2012",
   journal = "Phys. Rep.",
   volume  = "513",
   pages   = "1",
   doi     = "10.1016/j.physrep.2012.01.001"
}

@article{Berti:2025hly,
    author = "Berti, Emanuele and others",
    title = "Black hole spectroscopy: from theory to experiment",
    year = "2026",
    journal = "Classical Quantum Gravity",
    volume = "43",
    pages = "123001",
    doi = "10.1088/1361-6382/ae59e2"
}

@ARTICLE{NakanoIoka,
   author  = "H. Nakano and K. Ioka",
   title   = "Second-order quasinormal mode of the {S}chwarzschild black hole",
   year    = "2007",
   journal = "Phys. Rev. D",
   volume  = "76",
   pages   = "084007",
   doi     = "10.1103/PhysRevD.76.084007"
}

@article{Ioka:2007ak,
    author = "Ioka, Kunihito and Nakano, Hiroyuki",
    title = "Second- and higher-order quasinormal modes in binary black-hole mergers",
    year = "2007",
    journal = "Phys. Rev. D",
    volume = "76",
    pages = "061503",
    doi = "10.1103/PhysRevD.76.061503"
}

@article{Xue:2003vs,
    author = "Xue, Li-Hui and Shen, Zai-Xiong and Wang, Bin and Su, Ru-Keng",
    title = "Numerical simulation of quasi-normal modes in time-dependent background",
    year = "2004",
    journal = "Mod. Phys. Lett. A",
    volume = "19",
    pages = "239",
    doi = "10.1142/S021773230401240X"
}

@article{Shao:2004ws,
    author = "Shao, Cheng-Gang and Wang, Bin and Abdalla, Elcio and Su, Ru-Keng",
    title = "Quasinormal modes in a time-dependent black hole background",
    doi = "10.1103/PhysRevD.71.044003",
    journal = "Phys. Rev. D",
    volume = "71",
    pages = "044003",
    year = "2005"
}

@article{Abdalla:2006vb,
    author = "Abdalla, Elcio and Chirenti, Cecilia B. M. H. and Saa, Alberto",
    title = "Quasinormal modes for the {V}aidya metric",
    doi = "10.1103/PhysRevD.74.084029",
    journal = "Phys. Rev. D",
    volume = "74",
    pages = "084029",
    year = "2006"
}

@article{Lin:2021fth,
    author = "Lin, Kai and Sun, Yang-Yi and Zhang, Hongsheng",
    title = "Quasinormal modes for dynamical black holes",
    doi = "10.1103/PhysRevD.103.084015",
    journal = "Phys. Rev. D",
    volume = "103",
    pages = "084015",
    year = "2021"
}

@article{Capuano:2026tjy,
    author = "Capuano, Lodovico and Lovo, Thomas and Prieto-Varela, Gorka and Sarkar, Subhodeep and Kuntz, Adrien and Barausse, Enrico and Kothawala, Dawood",
    title = "When the Ringing Stops: Purely Imaginary Modes in the Ringdown Spectrum of Dynamical Black Holes",
    eprint = "2605.28951",
    archivePrefix = "arXiv",
    journal = "",
    year = ""
}

@article{ZhaoPani,
    author = "Yu-Qian Zhao and Paolo Pani",
    title = "Quasinormal modes and tidal responses of black holes in generic anisotropic matter environments",
    eprint = "2606.11380",
    archivePrefix = "arXiv",
    journal = "",
    year = ""
}

@article{Ariadna2026,
    author = "Ariadna Uxue Palomino Ylla and Kosuke Makino and Akane Tanaka and Akihiro Ishibashi and Chul-Moon Yoo",
    title = "Ringdown waves from hairy black holes",
    journal = "J. Cosmol. Astropart. Phys.",
    volume = "09",
    number = "2026",
    pages = "046",
    doi = "10.1088/1475-7516/2026/09/046"
}

\end{document}